\documentclass[onecolumn,extra]{gji}

\usepackage{timet}
\usepackage{amsmath}
\usepackage{amsfonts}
\usepackage{mathrsfs}
\usepackage{graphicx}
\usepackage{bm}
\usepackage{caption}
\usepackage{subcaption}
\usepackage{booktabs}
\usepackage{tikz}
\usepackage{hyperref}

\newcommand{\dS}{\,\mathrm{d}S}

\newcommand{\cbraket}[2]{({#1},{#2})}

\newcommand{\trace}[1]{\mathrm{tr}(#1)}

\newcommand{\ee}{\mathrm{e}}

\newcommand{\Gauss}[2]{\mathcal{N}\!\left(#1, #2\right)}
\newcommand{\Sph}[1]{\mathbb{S}^{2}_{#1}}
\newcommand{\LebSph}[2]{L^{#2}(\Sph{#1})}
\newcommand{\SobSph}[2]{H^{#2}(\Sph{#1})}

\title[Bayesian sea-level inference]
  {A scalable Bayesian framework for modern sea-level inference}
  \author[Heathcote \emph{et al.}]
  {D.A. Heathcote$^{1}$, T. Holland$^{1}$, A.M. Mag$^{2}$, M.E. Tamisiea$^{3}$, 
  S. Coulson$^{4}$, \and S. Dangendorf$^{5}$, A.J. Lloyd$^{6}$, A. Mashayek$^{1}$, 
  J.X. Mitrovica$^{7}$, \& D. Al-Attar$^{1}$ \\
  $^{1}$Bullard Laboratories, University of Cambridge, Madingley Road, Cambridge CB3 0EZ, UK. E-mail: dah94@cam.ac.uk\\
  $^{2}$Department of Earth Sciences, University of Oxford, Oxford OX1 3AN, UK.\\
  $^{3}$Center for Space Research, University of Texas, Austin, TX 78759, USA.\\
  $^{4}$Department of Earth Sciences, University of New Hampshire, Durham, NH 03824, USA. \\
  $^{5}$Department of River-Coastal Science \& Engineering, Tulane University, New Orleans, LA 70118, USA.\\
  $^{6}$Lamont-Doherty Earth Observatory, Columbia University, Palisades, NY 10964, USA.\\
  $^{7}$Department of Earth and Planetary Sciences, Harvard University, 20 Oxford Street, Cambridge, MA 02138, USA.
  }

\date{}
\pagerange{\pageref{firstpage}--\pageref{lastpage}}
\volume{200}
\pubyear{{\the\year{}}}

\begin{document}
\label{firstpage}
\maketitle

\begin{summary}

    This paper presents a scalable Bayesian framework for the inference of modern-day sea-level change and surface mass redistribution. To motivate 
    this approach, we first review and quantify the performance of some standard methods for the analysis of satellite gravity and ocean altimetry observations. 
    We find that these methods substantially underestimate uncertainties and are subject to systematic biases, deficiencies that stem from their incomplete 
    treatment of sea-level physics and from uncertainty estimates based solely on the propagation of observational noise. To address these limitations, our 
    approach combines three advances. First, we use recent developments in adjoint sea-level theory to embed the full physics into the forward and inverse 
    modelling. Second, we formulate the Bayesian inverse problem in an infinite-dimensional setting, thereby avoiding discretisation artefacts and the 
    underestimation of uncertainties inherent in truncated model spaces. Finally, our computational methods render such inversions tractable at full observational 
    resolution while supporting joint model spaces and multiple data types. By employing a matrix-free approach with iterative solvers and randomised low-rank 
    decompositions -- implemented in the linked open-source libraries \texttt{pygeoinf} and \texttt{pyslfp} -- basic calculations are possible on a single laptop, 
    with the most intensive tasks parallelising trivially across available cores. We demonstrate the methodology through a series of synthetic experiments, 
    culminating in a joint inversion of satellite gravity and ocean altimetry data that decomposes regionally averaged sea-level change into steric and
    manometric components with quantified uncertainties, including the degeneracies that remain.

\end{summary}


\section{Introduction}

Accurate inference of modern-day sea-level change is essential for understanding the impacts of a warming climate and informing global coastal adaptation strategies 
\citep[e.g.][]{nicholls2011planning,larour2017should,dangendorf2026human}. 
A critical component of this effort is quantifying surface mass redistribution within the ice sheets and oceans \citep[e.g.][]{mitrovica2001recent, tamisiea2011moving,
shepherd2012reconciled, hay2015probabilistic, horton2018mapping, frederikse2020causes, cazenave2022contemporary, wang2024improved}. Although satellite gravity and ocean altimetry 
observations (from GRACE/FO and Sentinel-6, for example) provide critical constraints on these mass redistributions, standard estimators often fail to fully incorporate gravitationally self-consistent physics, resulting in systematic biases and artificially 
narrow uncertainties \citep[e.g.][]{clarke2005effect,riva2010sea, sterenborg2013bias, blazquez2018exploring, lickley2018bias, al2024reciprocity}.
To progress beyond these standard estimators, the community has increasingly explored joint inversions using predefined sea-level fingerprints
\citep[e.g.][]{rietbroek2016revisiting, uebbing2019processing, willen2026improving} and spatial Bayesian hierarchical models 
\citep[e.g.][]{zammit2014resolving, bamber2018land, calafat2022sources}. While these approaches represent major steps forward in 
decomposing sea-level contributions and rigorously evaluating regional budgets, they remain fundamentally constrained by their reliance on 
finite-dimensional spatial bases \citep[e.g.][]{rietbroek2016revisiting} or incomplete physical modelling
that ignores gravitationally self-consistent sea-level change \citep[e.g.][]{zammit2014resolving}. 
This paper presents a scalable, infinite-dimensional Bayesian framework 
designed to address these limitations whilst providing robust, physically grounded uncertainty quantification. Though the focus of this work is on satellite gravity and ocean altimetry, 
our approach constitutes a coherent framework in which any combination of other relevant observations (e.g., tide gauge records, GPS displacements, ice altimetry) can be used to 
quantitatively estimate changes in ice sheets, the oceans, or  hydrology.

Our method advances previous studies by combining improvements in three key areas. First, we take full account of the physics of gravitationally self-consistent sea-level change 
within both forward and inverse modelling, the latter being possible through the application of recent work on adjoint sea-level theory by \cite{al2024reciprocity}. Second, we 
formulate and solve the Bayesian inverse problems in an infinite-dimensional setting \citep[e.g.][]{stuart2010inverse, bui2013computational, petra2014computational, schillings2017analysis}. 
This directly complements the recent proof-of-concept by \cite{coulson2026inverting}, who successfully inverted altimetry observations for basin-scale ice mass changes using a finite-dimensional 
Bayesian formalism coupled with adjoint techniques. However, limiting the dimension of the model space in an ad hoc manner -- such as projecting the physics onto a  set of
predefined fingerprint patterns -- is equivalent to choosing a prior whose support is restricted to a finite-dimensional subset of the true model space, which generically
leads to uncertainties being underestimated. By maintaining 
an infinite-dimensional setting and utilising modern matrix-free numerical techniques \citep[e.g.][]{isaac2015scalable}, we ensure accurate uncertainty
 quantification without sacrificing computational feasibility and without the introduction of discretisation artefacts. 
Finally, Bayesian methods are inherently applicable to problems with multiple data types or composite model spaces.
Our computational framework, built on the open-source libraries \texttt{pygeoinf} and \texttt{pyslfp} \citep[][]{al_attar_pygeoinf_2026,al_attar_pyslfp_2026}, allows such joint 
inversions to be easily assembled from their component parts and evaluated
efficiently. Indeed, all results shown in this paper were generated on a single laptop. The basic
operations of the framework -- posterior expectations, actions of the posterior covariance,
individual posterior samples -- each
take on the order of a few minutes, while more intensive tasks, such as the generation of
posterior standard deviation maps, comprise many independent operations of this kind: these
take several hours in serial, but parallelise trivially, with run times falling broadly in
proportion to the number of cores available.

To demonstrate the practical application of this methodology, the paper is structured around three synthetic examples:
\begin{enumerate}
    \item \textbf{Satellite gravity:} We utilise satellite gravity observations to estimate loads and load averages, comparing our Bayesian approach to 
    the widely used method of \cite{wahr1998time}.
    \item \textbf{Ocean altimetry:} We estimate global-mean sea-level rise (GMSLR) from ocean altimetry data, comparing the results to the 
    standard averaging method \citep[e.g.][]{nerem2010estimating, ablain2015improved}. We also consider the decomposition 
    of global and regional sea-level change into its different physical components. 
    \item \textbf{Joint inversion:} We combine satellite gravity and ocean altimetry observations within a joint inversion, 
    and quantify the gain in information for quantities of interest such as  GMSLR.
    
\end{enumerate}
For these inversions, we deliberately choose simple prior and noise distributions, though we ensure their amplitudes and spatial scales remain representative. Our methods do assume these distributions are Gaussian; however, the theory is not restricted to the simple parametric covariances used here, nor does computational efficiency depend on these simplifications. In future application-focused studies, we will incorporate physically informed prior distributions and more comprehensive representations of observational uncertainties.

Although our idealised models limit the immediate real-world interpretation of these results, we outline general methods to quantify how sensitive the posterior distribution is to the choice of prior. Ultimately, we conclude that for certain quantities of interest -- such as regional averages of surface loads or GMSLR estimates -- this sensitivity is relatively mild. Therefore, the strong performance of our Bayesian methods in these simplified examples is likely indicative of their broader potential.

It is necessary here to outline the intended scope of this work.  We assume that the response of the Earth to surface loading 
is elastic over the relevant timescales, which is generally thought to be appropriate in the study of modern-day sea-level change. It follows that observations over a given time-period 
are determined entirely by concurrent loading, and hence we can focus on a series of time-independent inverse problems. In doing this, we  assume that contributions 
from ongoing processes such as glacial isostatic adjustment have been corrected for, with the uncertainties in these corrections incorporated in the observational errors. 
Solutions to the inverse problem at different times can be combined in a sequential manner using Kalman filters or smoothers \citep[e.g.][]{wunsch2006discrete, law2015data}, 
an approach which generalises the methods introduced into sea-level research by \cite{hay2013estimating} and \cite{hay2015probabilistic}.

It would be possible to adapt the methods in this paper to account for viscoelastic deformation using the adjoint theory of \cite{crawford2018quantifying} 
and \cite{yu2025application}. Such an extension is likely to be important for some modern-day applications due to the existence of areas of dramatically 
lower viscosity within the upper mantle such as under West Antarctica \citep[e.g.][]{kaufmann2005lateral, ivins2023antarctic}. However, doing this would substantially
raise the computational costs due to the need for calculations within viscoelastic earth models with strong lateral viscosity variations. 
It would also then be necessary to consider time-dependent loads, though the problem could still be usefully posed within the framework of sequential 
data-assimilation to limit the computational costs.

\section{Gravitationally self-consistent sea-level theory}

Modelling sea-level change requires an understanding of how the Earth's surface deforms in response to the redistribution of surface mass. 
In this work, we restrict our attention to 
processes that are rapid relative to viscoelastic relaxation in 
the mantle, but slow compared to the barotropic adjustment timescale for the ocean. As a result, 
we assume that the Earth responds elastically, while ocean dynamics are
decoupled from self-gravitation and loading effects \citep[e.g.][]{vinogradova2015dynamic}. Such assumptions are standard within work on modern 
sea-level, though, as noted above, the likely existence of strong lateral 
viscosity variations within the mantle may limit the validity of the elastic assumption. Here one might also comment 
that, as is usual within modern sea-level work, we apply a quasi-static theory for rotational feedbacks that 
was developed originally for deglacial sea-level change \citep[e.g.][]{sabadini1982polar,milne1998postglacial}. Because the 
timescales considered are not necessarily long relative to that for the Chandler wobble, this approach 
may be reasonably questioned, though the extent to which this matters practically has not, to our knowledge, been quantified.
Throughout this paper terminology related to sea-level follows \cite{gregory2019concepts}, though it has not  been possible to be
fully consistent with their mathematical notations.

\subsection{The elastic sea-level equation}

\label{sec:SLE}

We consider a self-gravitating elastic earth model which, in its equilibrium state, occupies a volume $M$ with boundary $\partial M$.
The equilibrium density is $\rho$, with $\Phi$ the corresponding gravitational potential. It is assumed that the equilibrium earth 
model is in hydrostatic equilibrium. Following the application of a load, $\sigma$, the deformation of the model is expressed in 
terms of a displacement vector, $\mathbf{u}$, a perturbation to the gravitational potential, $\phi$, and a perturbation, $\bm{\omega}$,
to the model's angular velocity. The rotational perturbation gives rise to a corresponding variation in the centrifugal potential, denoted by $\psi$.

Within hydrostatic sea-level theory, the mean sea surface sits on an equipotential of the gravity potential (i.e., the sum of the gravitational 
and centrifugal potentials) known as the geoid. Relative sea-level is then defined to be the signed distance from the sea floor to the sea surface measured along 
the local vertical \citep[e.g.][]{mitrovica2003post,gregory2019concepts}. This is precisely the quantity whose change is registered by a tide gauge.  Note that the 
above definition can be extended to the land, with the relative sea-level then generally being negative. 
Following the application of the load, $\sigma$, and working to first-order accuracy, the relative sea-level 
change (RSLC)  can be written
\begin{equation}
    \label{eq:deltaSL}
     \xi = -\frac{1}{g} \left( g u + \phi + \psi \right) + \frac{\Phi_g}{g}, 
\end{equation}
where $g$ is the surface acceleration due to gravity in the equilibrium state, $u$ is the vertical displacement, and $\Phi_{g}$ is a constant potential shift 
that corresponds to a spatially uniform change in sea-level.

Since the oceans  have mass, their redistribution on the surface of the Earth contributes to the load. We can, therefore, express the  total 
change in load, $\sigma$,  as the sum of a \emph{direct load}, $\zeta$, and an \emph{induced water load}:
\begin{equation}
    \label{eq:inducedWaterLoads}
    \sigma = \zeta + \rho_w 1_{\mathcal{O}} \,\xi.
\end{equation}
 Here $\rho_{w}$ is the  water density and $1_{\mathcal{O}}$ the ocean function that equals one in the oceans and zero on land (suitably adjusted to 
 account for ice shelves). Note that we treat the ocean function as constant throughout this study; this approximation is highly suitable for modern-day sea-level, where small signal amplitudes render shoreline migration a negligible, second-order effect \citep[e.g.][]{crawford2018quantifying}. This relation is important because it is only the direct part of the load, $\zeta$, that can be freely specified within modelling.  Finally,   conservation of mass at 
 the surface requires
 \begin{equation}
 \label{eq:masscon}
     \int_{\partial M} \sigma \dS = 0.
 \end{equation}
 By coupling the expressions for the relative sea-level change to those for elastic loading and mass conservation, one arrives at the 
 elastic sea-level equation introduced by \cite{farrell1976postglacial}, which was later refined to 
 include shoreline migration and rotational feedbacks \citep[e.g.][]{sabadini1982polar, milne1998postglacial,mitrovica2003post,kendall2005post}.
Given that the changes are sufficiently small for shoreline migration to be neglected, 
the sea-level equation is linear, and hence its solution defines  a linear operator that 
maps the direct load, $\zeta$, to the response $(\xi, u, \phi,\bm{\omega})$. In the terminology of \cite{gregory2019concepts}, the perturbations to 
the gravitational potential, $\phi$, and to rotation, $\bm{\omega}$ (and hence $\psi$), together with the deformation,  $u$, comprise the gravitational, rotational 
and deformational (GRD) response to the load; the sea-level equation is thus the mapping from a surface mass redistribution to its GRD effects and the associated 
change in relative sea-level.

\subsection{Ocean dynamics}

\label{sec:ocean_dynamics}

Due to ocean dynamics, the mean sea surface does not coincide with an equipotential of the gravity potential. Since the sea surface no longer selects 
a particular equipotential, the geoid must instead be fixed by convention: following \cite{gregory2019concepts}, we define it as the equipotential such 
that the volume enclosed between it and the sea floor equals the time-mean volume of sea water in the oceans, adjusted to include equivalent volumes for 
any floating ice. The height of the mean sea surface above the geoid, measured along the local vertical, is known as the mean ocean dynamic sea-level and, 
by construction of the geoid, its average over the oceans vanishes.

The ocean circulation redistributes mass. This redistribution perturbs the gravitational field directly and, through the associated changes in ocean bottom pressure, 
loads and deforms the solid Earth; that is, it generates its own GRD response, a process referred to within oceanography as self-attraction and loading (SAL). 
The resulting perturbations to the geoid and sea floor are, in turn, felt by the circulation -- as are those generated by any external loading -- and so the full problem is 
two-way coupled. On timescales significantly longer than those of the ocean's barotropic adjustment, however, the dynamical response to such perturbations is quasi-static \citep[e.g.][]{vinogradova2015dynamic}. It is therefore physically 
justifiable to decouple the ocean circulation from the SAL feedback. This decoupling allows us to take the sterodynamic sea-level change -- the change in ocean dynamic sea 
level plus a spatially uniform global-mean thermosteric contribution \citep[][]{gregory2019concepts} -- as a prescribed field with the associated SAL effects recovered through 
a post-hoc calculation using the sea-level equation.

A sterodynamic sea-level change does not, however, by itself determine the associated load. Its steric part reflects an expansion or contraction of the water column and 
carries no mass; the load exerted on the sea floor is instead set by the change in ocean bottom pressure. Specifying this load requires, alongside the sterodynamic change, 
knowledge of the density perturbation within the water column (in practice, its vertical average) and it is through these two prescribed fields that ocean dynamics enter the 
 sea-level equation. These ideas are  developed quantitatively within Section~\ref{sec:AltimetryBias}.

\subsection{Numerical implementation}

\label{sec:NumSLE}

The sea-level equation is valid within a laterally heterogeneous elastic earth model, 
but in the present work we restrict our numerical calculations to those that are spherically symmetric. 
This is a reasonable approximation because \cite{mitrovica2011robustness} have shown that the effects
of laterally varying elastic structure on sea-level change are comparatively small. 
Within a spherically symmetric earth model, the solution of the elastostatic loading 
problem can be expressed very simply using spherical harmonic expansions and pre-computed Love numbers. 
Following the (non-standard) notations of \cite{al2024reciprocity}, the spherical harmonic 
expansion coefficients of the vertical displacement, $u_{lm}$, and 
gravitational potential, $\phi_{lm}$, can be written in terms of that, $\sigma_{lm}$, for the 
total load by
\begin{equation}
    u_{lm} = h_{l} \sigma_{lm}, \quad \phi_{lm} = k_{l}\sigma_{lm}, 
\end{equation}
where $h_{l}$ and $k_{l}$ are loading Love numbers.
The Love numbers used in our numerical examples were calculated within the model  PREM of \cite{dziewonski1981preliminary} using 
the method of \cite{al2014sensitivity}.

Within  \texttt{pyslfp}, we have implemented the  standard pseudo-spectral  method  to iteratively solve 
the sea-level problem, including the effect of rotational feedbacks
\citep[e.g.][]{mitrovica1991postglacial,milne1998postglacial, mitrovica2003post,kendall2005post}.
Fast spherical harmonic transformations are performed using the \texttt{pyshtools}
library \citep[][]{wieczorek2018shtools}. Except where otherwise noted, within the calculations shown
below the  spatial fields have been expanded up to degree 256. 
The current implementation of rotational feedbacks neglects the component of the
perturbed angular velocity along the reference rotation axis because this term is smaller 
than the equatorial components by around two orders of magnitude; 
such an approximation is not, however, inherent to our methods. 

\section{Quantifying errors associated with standard estimators}

\label{sec:standard}

Within this section we examine the statistical performance of two standard methods for analysing satellite gravity and ocean altimetry observations. Doing this should be instructive in motivating our new inference framework, while also introducing  some of the 
probabilistic methods we require. Indeed, the forward models and the probability distributions defined within this section are 
precisely those used within our later synthetic inversions. 

\subsection{Satellite gravity}

\subsubsection{Satellite gravity observations}

\label{sec:WMBBias}

A notable source of gravity data is from the GRACE/FO satellites, which provide global measurements of time-variable changes to the Earth's  gravitational potential 
\citep[e.g.][]{tapley2004grace, landerer2020extending}. These results are  usually expressed in terms of monthly-averaged 
perturbations to the gravitational potential's  spherical harmonic coefficients in a range $2\le l \le l_{\max}$. The potential 
coefficients can be estimated directly from the inter-satellite distances or inferred indirectly  using  mascons, with 
the latter method providing an effective means for suppressing longitudinal striping  associated with the orbital 
paths \citep[e.g.][]{watkins2015improved}. For the purposes of this study, it does not matter how the coefficients have
been obtained, but only that they have been provided along with appropriate uncertainties. We assume that these measurement 
errors are Gaussian with zero-expectation and with a specified covariance; these requirements being  consistent with 
standard data products \citep[e.g.][]{loomis2012simulation}. The gravitational potential coefficients start at $l=2$ because the degree-zero coefficient
is proportional to the Earth's total mass (which does not change), and the degree-one coefficients cannot be measured accurately 
by satellite gravity. This is because these  coefficients are linearly related to the components of geocentre 
motion to which the satellites' orbits are insensitive. The upper limit for the observed coefficients, $l_{\max}$, is set by the 
obtainable spatial resolution from the satellite data. In practice, uncertainties tend to increase with the degree, 
and so it is common for the coefficients used to be truncated at some lower degree, with 
$l_{\max} = 100$ being a representative value used in the examples below. Within many 
studies it is common to replace the degree-two coefficients with more accurate values obtained using satellite laser 
ranging \citep[e.g.][]{loomis2020replacing}.  Within the present context 
these details are not important; if missing or improved coefficients for certain degrees are available along with 
uncertainties, then they can be included within the analysis or inversions. 

Monthly-averaged GRACE/FO coefficients represent the cumulative effect of several physical processes. In particular, 
due to the viscoelastic nature of the Earth's mantle, there will be contributions to the gravitational potential change
associated with ongoing glacial isostatic adjustment (GIA) from the last ice age. To focus on modern-day loading, this 
GIA signal must be modelled using an assumed viscoelastic earth model and ice sheet reconstruction, and its contribution 
to modern-day observations removed \citep[e.g.][]{van2015effect, caron2018gia, Vishwakarma}. Similar corrections can be made 
for other processes such as tectonic deformation and subsidence. Of course, these corrections are never perfect, and so they contribute 
to the uncertainty within the resulting coefficients. Without essential loss of generality, we assume here that all  the necessary 
corrections to the data have been made, and that errors associated with each correction  are Gaussian with zero-expectation 
and a known covariance. Under the reasonable assumption that the various corrections are statistically independent, the total 
covariance for the data errors is, therefore,  just the sum of these individual covariances. 

\begin{figure*}
    \centering    
    \begin{subfigure}[b]{0.48\textwidth}
        \centering
        \includegraphics[width=\textwidth]{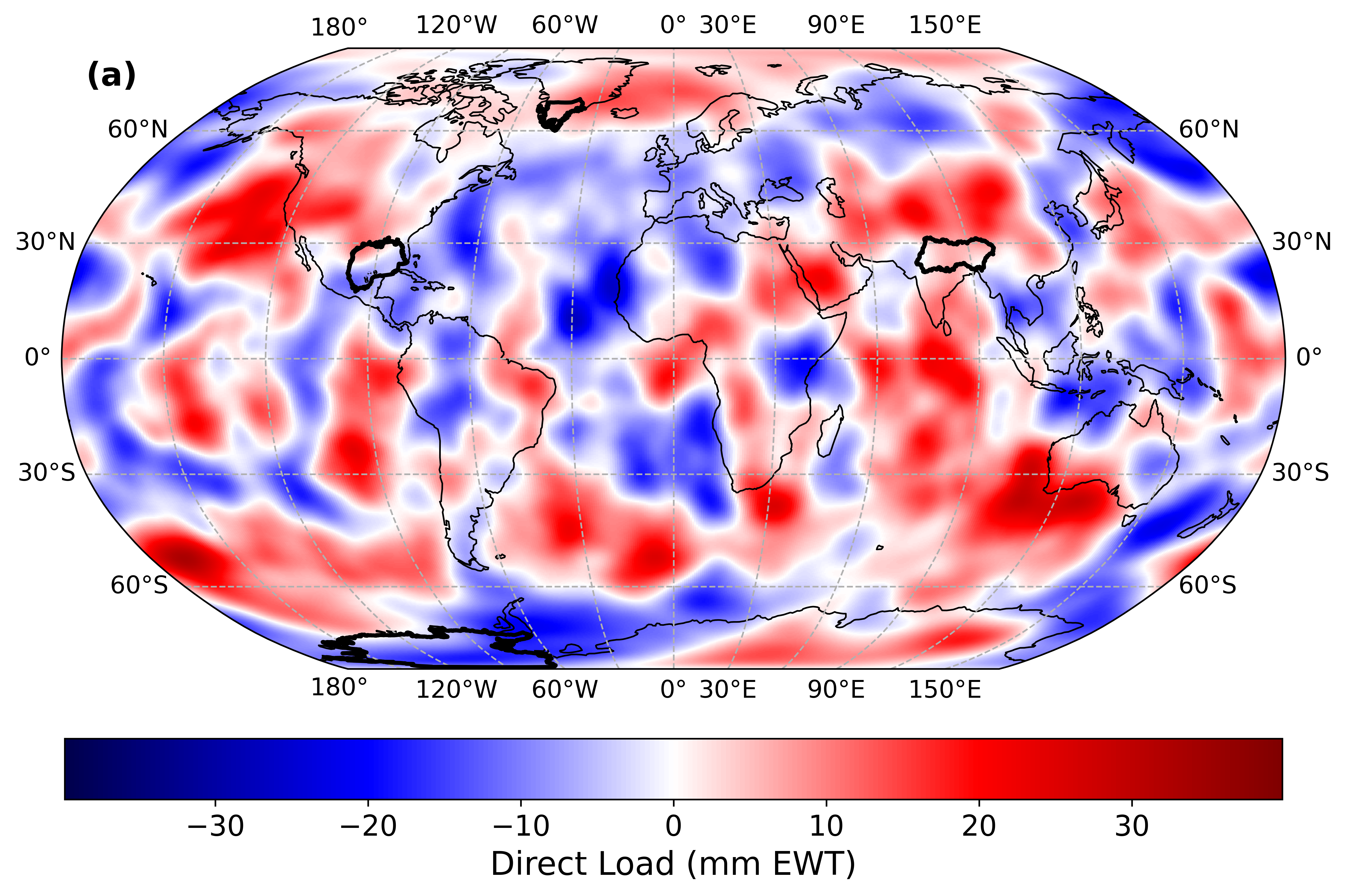}
        \label{fig:direct_load}
    \end{subfigure}
    \hfill 
    \begin{subfigure}[b]{0.48\textwidth}
        \centering
        \includegraphics[width=\textwidth]{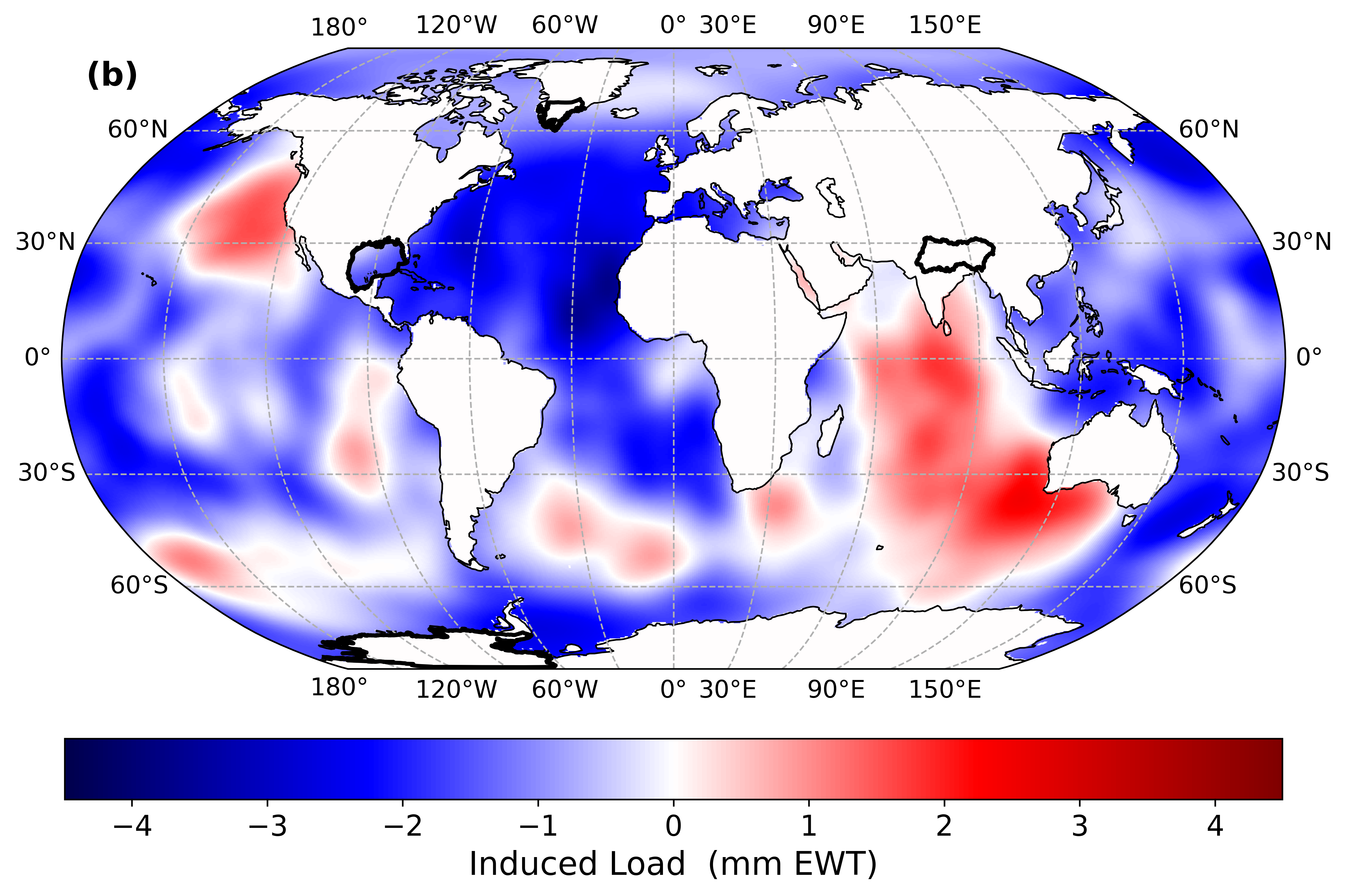}
        \label{fig:induced_load}
    \end{subfigure}
    \caption{Spatial distribution of a synthetic direct load and its gravitationally self-consistent sea-level response. 
    (a) A direct load drawn from a Gaussian prior distribution with a characteristic length scale of 250\,km and a pointwise
     standard deviation of 10\,mm. (b) The corresponding induced water load, obtained by solving the elastic sea-level equation.
      Both fields are expressed in Equivalent Water Thickness (EWT). Black outlines denote the four target regions
      analysed in Fig.~\ref{fig:grace_bias}: Greenland South (GRL S)
       Basin, West Antarctic Ice Sheet (WAIS), Gulf of Mexico, and the Ganges--Brahmaputra--Meghna (GBM) Basin.}
    \label{fig:grace_loads}
\end{figure*}

\subsubsection{A summary of the WMB method}

Within \cite{wahr1998time}, a simple approach was described for estimating surface loads from 
satellite gravity data that we term the WMB method for convenience. While extensively applied and refined 
\citep[e.g.][]{swenson2002methods, velicogna2005greenland, harig2012mapping}, the WMB method has a known 
limitation: it does not fully incorporate the physics of gravitationally self-consistent sea-level change \citep[e.g.][]{clarke2005effect}. 
Previous studies have attempted to quantify or correct the resulting bias. \cite{riva2010sea} employed an 
iterative correction scheme for land-based loads, suggesting WMB estimates might be underestimated by approximately 
16\%. \cite{sterenborg2013bias}  used synthetic examples to demonstrate systematic biases of around 10\%.
Finally, \cite{al2024reciprocity} calculated sensitivity kernels for load averages computed with the WMB method,
compared them to  kernels for the true load averages, and so highlighted 
semi-quantitatively the potential biases due to the neglect of induced water loads. However, these existing 
critiques rely either on heuristic corrections, a limited number of synthetic 
tests, or qualitative visual comparisons.

The starting point for the WMB method is the relation
\begin{equation}
\label{eq:wahr1}
    \phi_{lm} = k_{l} \sigma_{lm}, 
\end{equation}
between the  gravitational potential and the total load in the spherical harmonic domain, where we recall that the  $k_{l}$ 
are the appropriate degree-dependent Love numbers; up to notation, this relation is equivalent to eqs~(11) and (12) of \cite{wahr1998time}.
Physically, eq.~(\ref{eq:wahr1}) captures both the direct gravitational effect of the surface load, and the 
gravitational signal created by the resulting deformation of the solid Earth. Rearrangement of eq.~(\ref{eq:wahr1}) for the GRACE/FO case
leads trivially to the following expression
\begin{equation}
\label{eq:wahr2}
    \sigma \approx \sum_{l=2}^{l_{\max}}\sum_{m=-l}^{l} k_{l}^{-1} \phi_{lm} Y_{lm}, 
\end{equation}
which can be used to reconstruct the surface load from observed gravitational potential coefficients. 
Moreover, for any weighting function, $w$, we readily find that
\begin{equation}
\label{eq:wahr3}
    \int_{\partial M} w \sigma  \dS \approx \sum_{l=2}^{l_{\max}}\sum_{m=-l}^{l} k_{l}^{-1} \phi_{lm} w_{lm}, 
\end{equation}
where $w_{lm}$ are the spherical harmonic coefficients of $w$, and hence load averages over regions of interest can be 
estimated directly and easily from the observed potential coefficients. It is a 
simple matter to propagate observational uncertainties in the data through to
estimates of the load or of load averages.  Methods for reducing the effect of such data errors on load averages were 
discussed by \cite{swenson2002methods}, using optimal filters, and by \cite{harig2012mapping}, using Slepian functions. Other 
studies have considered how estimates can be improved by recovering the missing degree-one coefficients 
through the combination of GRACE data with constraints from ocean bottom pressure models
\citep[e.g.][]{chambers2004preliminary,swenson2008estimating,sun2016optimizing}.

\subsubsection{Quantifying the errors in the WMB method}

Consider a set of $p$ regions over which we wish to estimate averages of the total load, $\sigma$.
For a given direct load, $\zeta$, the elastic sea-level equation can be solved to obtain the 
gravitational potential perturbation, $\phi$, which is then mapped to a set of error-free synthetic GRACE/FO coefficients.
This process defines a linear forward operator that we denote by $A$. To model observed GRACE/FO data, we then write 
\begin{equation}
    d = A \zeta + z, 
\end{equation}
where $z$ is a random vector of observational errors drawn from a Gaussian distribution,  $\Gauss{0}{R}$, 
with zero expectation and covariance operator, $R$. From the  direct load, $\zeta$, we can also  obtain the 
relative sea-level change, $\xi$,  use this to determine the total load,
and then form the desired regional averages of the $p$ regions. This latter process defines a linear operator, $B$, and we
write $q = B \zeta$ for the vector of  true averages of the total load. 
Finally, the WMB method defines a linear operator, $C$, that maps GRACE/FO coefficients into the $p$-dimensional space 
of load averages, with the resulting estimate given by 
\begin{equation}
    \hat{q} = C d = C(A\zeta + z).
\end{equation}
It follows that the error associated with the WMB method takes the simple form
\begin{equation}
     \hat{q} -q = (CA-B)\zeta + C z. 
\end{equation}
This error comprises two parts. The first, $(CA-B)\zeta$,  depends on the direct load, and reflects both the 
physical approximations of the WMB method and resolution limits inherent to GRACE/FO observations. The second, 
$C z$, quantifies the propagation of observational noise into the estimated load averages. Applications of the WMB 
method have, to our knowledge, considered only the second contribution.

To analyse the  WMB method further, we suppose that the direct load 
is a random variable with distribution, $\Gauss{\overline{\zeta}}{Q}$. The true load averages, $q$, 
the WMB estimates, $\hat{q}$, and the estimator error, $\hat{q} -q$, then become 
random variables whose distributions are determined from those for the direct load and the observational 
noise. Because all mappings are linear and the underlying distributions are Gaussian, we can express the resulting distributions analytically
 \citep[e.g.][]{stuart2010inverse}:
\begin{align}
    q &\sim \Gauss{B\overline{\zeta}} {BQB^{*}}, \\
    \hat{q} &\sim \Gauss{CA\overline{\zeta}} {CAQA^{*}C^{*} + CRC^{*}}, \\
    \label{eq:wmb_error}
     \hat{q} -q &\sim \Gauss{(CA-B)\overline{\zeta}} {(CA-B)Q(CA-B)^{*} + CRC^{*}},
\end{align}
where the superscript $*$ is used to denote the adjoint of a linear operator. The distribution for the error 
term, $\hat{q} -q$, shows two main things. The first is that 
its expectation is non-zero unless $(CA-B)\overline{\zeta}=0$. This condition 
can hold for all possible distributions for the direct load if and only if $CA = B$, 
which is to say that the WMB method produces exact results in the absence of 
observational errors. Clearly this is not the case, and hence we expect that load 
averages obtained using the WMB method will be systematically  offset from 
their true value. These offsets depend on both the regions
chosen and the geographic distribution of the unknown direct load, and hence 
there is no simple method by which they can be corrected.
The second point is that the covariance for the error distribution is a sum of two parts.
 The term $CRC^{*}$  is the familiar propagation of 
observational errors that is accounted for within  applications of the WMB method. The other,
 $(CA-B)Q(CA-B)^{*}$, represents the intrinsic resolution of the WMB method due to 
the band-limited  potential coefficients and the neglect of gravitationally self-consistent sea-level 
change. The upshot of this analysis is that applications of the WMB method will produce load 
average estimates that are  subject to systematic offsets, and whose uncertainties will be
underestimated.

\begin{figure*}
    \centering        
    \includegraphics[width=\textwidth]{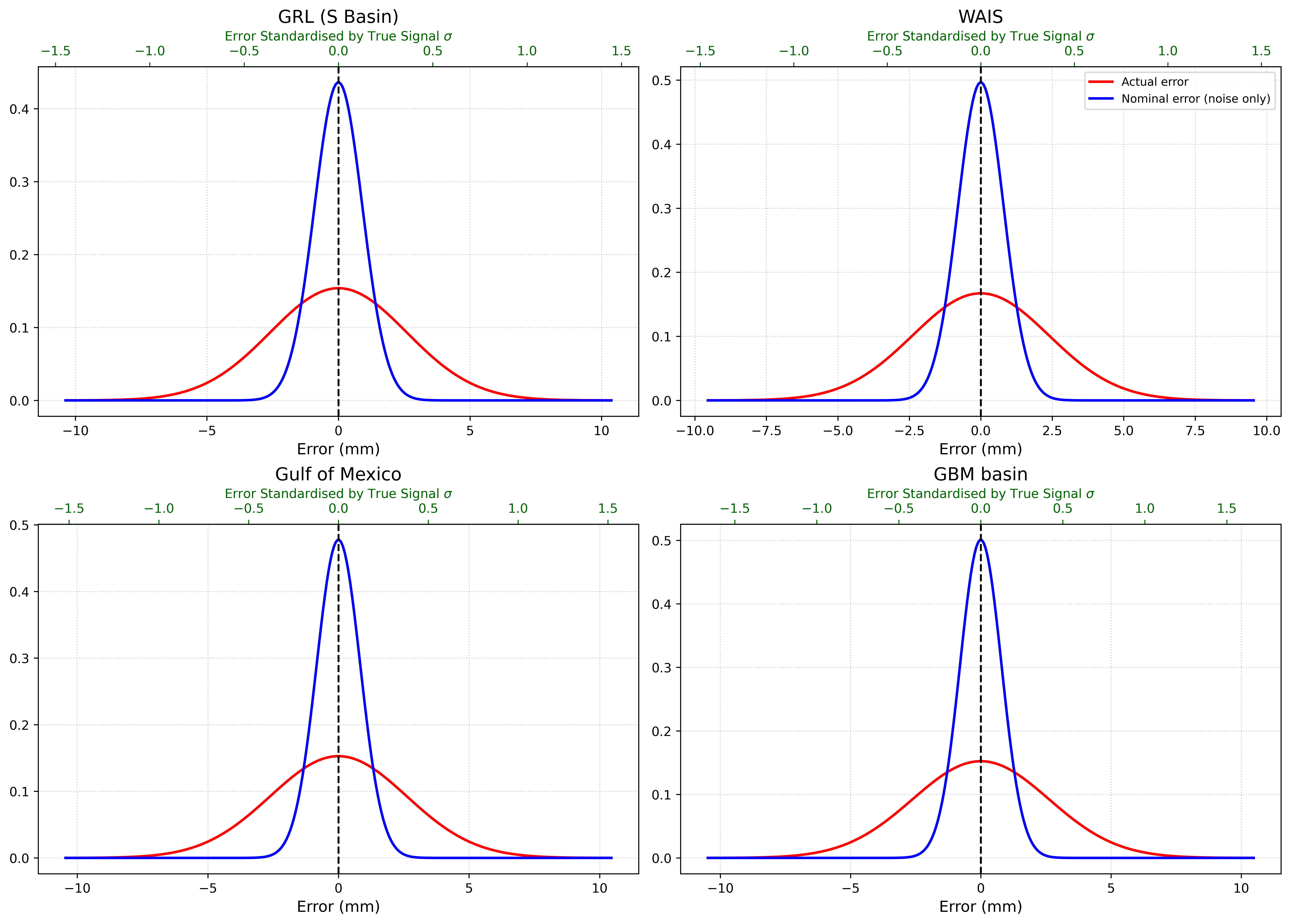}        
    \caption{Comparison of standard and true error distributions for regional averages of the total 
    load obtained using the WMB method. The blue curves represent the nominal WMB error 
    probability density functions (PDFs), which are derived solely by propagating observational noise. 
    The red curves show the actual, analytically derived error distributions which account for incomplete physical modelling 
    and the spatial resolution limits associated with the GRACE/FO data. The dashed black line indicates zero error. A secondary top axis (green) expresses 
    the error magnitude normalised relative to the standard deviation of the true load average at each location.}
    \label{fig:grace_bias}
\end{figure*}

\subsubsection{Numerical implementation}

\label{sec:WMBNum}

To apply these ideas numerically, we must first set up the various operators and  
distributions. 
For the direct loads we use the Sobolev space, $\SobSph{a}{2}$, which is a Hilbert space of 
continuous functions on the sphere of radius $a$ whose elements have square-integrable 
derivatives up to second order \citep[e.g.][]{taylor1996partial}. With this choice, 
the operators $A$, $B$, and $C$ are all continuous from their respective domains, while 
their adjoints can be determined by combining the results of \cite{al2024reciprocity}
with standard methods. A numerical approximation 
to $\SobSph{a}{2}$ is implemented within \texttt{pygeoinf} as
 part of its \texttt{symmetric\_space} sub-package using fast spherical harmonic transformations. 
 Relative to this discretisation, approximations to the necessary operators have been developed such 
 that their actions converge to their exact continuous forms as the truncation degree is increased.

To define the probability distribution for the direct load, 
we consider a Gaussian random field having zero expectation and whose covariance takes the simple 
form  \citep[e.g.][]{lindgren2011explicit}
\begin{equation}
\label{eq:sobolev_kernel}
    Q = \alpha (1 + \lambda^{2} \Delta)^{-q}, 
\end{equation}
where $\alpha, \lambda > 0$, $q > 1$ are scalar parameters and $\Delta$ is the Laplace--Beltrami operator (note 
we use the sign convention such that this operator is non-negative). The condition 
on $q$ is required for the covariance to be trace-class, and hence for the distribution to be well-defined. 
This covariance operator commutes with rotations, and is associated with random fields 
whose statistical properties are rotationally invariant. The amplitude term, $\alpha$, 
is related to the pointwise standard deviation, while $\lambda$ defines 
a characteristic length scale. The exponent $q$ sets the regularity of the field, 
with higher values corresponding, roughly, to greater degrees of differentiability.
Within the spherical harmonic domain, 
this covariance is diagonal, and hence samples from the distribution can 
be efficiently generated using the Karhunen--Lo\`{e}ve expansion \citep[e.g.][]{stuart2010inverse}.
 Full support is provided within \texttt{pygeoinf} for working with Gaussian 
distributions on function spaces. This includes algebraic operations 
such as affine transformations and conditioning on a subspace,  along with an optimised class for rotationally invariant 
distributions on symmetric spaces. A key detail of the implementation
is that it is matrix-free by default,  meaning that it is not necessary to
form or work with large dense matrices for covariance operators.

In our numerical examples, we consider load distributions with a vanishing expectation; consequently, the bias term in eq.~(\ref{eq:wmb_error}) 
is strictly zero. We can, however, analytically estimate the potential magnitude of this bias. Suppose we assign a non-zero expectation by 
drawing a sample from our original prior distribution and translating the field. This procedure is readily implemented in practice and is 
provided as an option within our supplementary Python scripts. The resulting error PDFs are subsequently shifted away from zero, with the magnitude 
of the bias dependent on the specific sample chosen to define the expectation. If we consider the ensemble average of these shifts across all
possible expectation fields drawn from the covariance $Q$, the variance of this bias precisely evaluates to the resolution term, $(CA-B)Q(CA-B)^{*}$. 
It follows that by estimating the full error covariance from a zero-expectation distribution, we inherently quantify the anticipated magnitude 
of the bias. This is observed by taking the difference between the full error variance and the nominal variance obtained through standard error 
propagation. While this quantification remains tied to our simplified load distribution, it represents the theoretical limit of the analysis without 
postulating a physically informed, non-zero expectation directly.

To proceed, we set the exponent $q$ to $3$, the length scale for the load distribution to 250\,km, and the
amplitude parameter such that
the pointwise standard deviation is equal to 10\,mm of equivalent water
thickness. These values are broadly representative, in both amplitude and
horizontal scale, of modern-day load variations associated with ice sheets,
the oceans, and land water storage. Figure~\ref{fig:grace_loads} shows a sample from the
resulting distribution along with the corresponding induced water load
obtained through solution of the elastic sea-level equation. We emphasise
that, in considering GRACE data, only the direct load as a whole need be
specified: the induced water load follows from the sea-level equation, while
the manner in which the direct load is built up from ice, hydrological, and
oceanic sources is immaterial, and hence none of the finer sea-level
decompositions of Section~\ref{sec:ocean_dynamics} enter at this stage.

Samples from this distribution are, of course, not fully realistic.
In particular, a rotationally invariant covariance imposes statistically homogeneous correlations. 
Consequently, loads over an ice sheet become unrealistically correlated with those in the surrounding oceans.
Such limitations are not of serious consequence for our present
purposes, with the calculations intended only to be illustrative of the 
theoretical analysis of the WMB errors expressed within eq.~(\ref{eq:wmb_error}).
The more detailed model of Section~\ref{sec:AltimetryBias}, in which the ice and
ocean-dynamic contributions are modelled using a joint distribution,
addresses directly the question of correlations between components; and both the methodology and 
its implementation within \texttt{pygeoinf} carry over immediately to more realistic Gaussian
random fields, a refinement we deliberately defer to applications involving
real data.

As discussed above, uncertainties within GRACE/FO data are due to a combination of 
(i) observational errors associated with the measurement and analysis of the satellite data 
and (ii) uncertainties within corrections for processes such as GIA. Within 
applications, it is typical for the latter source of uncertainty to be dominant 
\citep[e.g.][]{van2015effect, caron2018gia}, and hence  we focus on contributions of this form
through a simplified noise model. The idea is to define a distribution for an effective ``noise load'' 
which we can push forward onto the GRACE/FO coefficients using the simple Love number 
scaling. The noise load is designed to have a lower pointwise standard deviation than 
that for the direct load, but to be rougher so that beyond a 
specified degree the noise overtakes the signal. For the noise load's covariance 
we take the same functional form
\begin{equation}
\label{eq:noise_kernel}
    Q_n = \alpha_{n} (1 + \lambda_{n}^{2} \Delta)^{-q_{n}}, 
\end{equation}
where it is assumed that $1 < q_{n} < q$. For simplicity, we fix the length scale for the noise 
load to be half that for the direct load, reflecting the idea that uncertainties within the various corrections 
are likely to be larger at shorter length scales. To determine the remaining two parameters we then 
consider the ratio of the spectral variances between the direct signal and the noise load at degree $l$, which is given by
\begin{equation}
    s_{l} = \frac{\alpha}{\alpha_{n}}\left[1 + \frac{\lambda^2}{a^2}l(l+1)\right]^{-q}\left[1 + \frac{\lambda_{n}^2}{a^2}l(l+1)\right]^{q_{n}}, 
\end{equation}
where $a$ is the radius of the Earth model's surface. By fixing the value of the amplitude signal-to-noise ratio ($\sqrt{s_l}$) 
at degree two and setting it equal to one at a specified crossover degree, we can then solve for $\alpha_{n}$
and $q_{n}$. Within the examples below, we choose $\sqrt{s_2} = 10$ and take the crossover degree to be $64$. 
The resulting pointwise standard deviation for the
noise load is $3.76$\,mm while at the maximum observed degree of 100 the amplitude signal-to-noise ratio falls to 0.75. It should be emphasised that solution of the sea-level equation leads 
to coupling between the potential coefficients and the load at different orders and degrees, and 
hence the values above merely provide an estimate (but a reasonably good one, since the couplings are 
typically only around 10\% of the signal).

Having now set up the necessary distributions for the direct load and the noise, 
we can apply eq.~(\ref{eq:wmb_error}) to determine the distribution for the errors associated
with the WMB method. To evaluate this error distribution numerically without relying on computationally
expensive Monte Carlo simulations, we utilise the matrix-free architecture of the \texttt{pygeoinf}  library. 
Within this framework, linear operators such as $A$ are defined 
in a matrix-free manner through their action and that of their adjoints. Algebraic combinations of operators 
are then implemented lazily -- that is, a term such as $AB$ is not evaluated when formed, but only through its action on a 
given vector -- while the adjoints for such combinations are automatically generated using algebraic rules.
The benefit of this approach is that  explicit dense matrix 
representations of the operators are never formed, these being expensive to compute and difficult 
to store.  For example, within a typical situation, we might truncate spherical harmonic expansions 
within the sea-level code at degree 256 and use GRACE/FO coefficients up to degree 100. 
A dense representation of the forward operator would then have dimension 10,197$\times$ 66,049, require  
10,197 solutions of the (generalised) sea-level equation to compute, and need around 5.5\,GB of RAM for storage using double precision.  
While representing the degree 256 case as a dense matrix is technically feasible on modern hardware, it 
remains a highly inefficient use of resources. Yea, as spatial resolution requirements increase, the 
limitations of the dense approach become prohibitive; for instance, at degree 512, the storage requirements 
quadruple to over 21\,GB and the computational overhead grows accordingly. In contrast, the matrix-free architecture 
of \texttt{pygeoinf} handles such high-resolution problems with ease and without substantial memory requirements.

Our specific focus is on  load averages for the four regions  shown within Fig.~\ref{fig:grace_loads}. 
Two of the regions selected are linked to ice sheets \citep[][]{rignot2011ice, mouginot2019forty}, 
one lies within the oceans \citep[][]{contarinis2020value}, and the last is on land
within a region of interest to studies of the Indian Monsoon \citep[][]{lehner2013global}. For each location, we define the desired average
using a weighting function, $w$, that is initially chosen to be constant over the region, zero elsewhere, 
and to integrate to one. An integral-preserving spatial smoothing is then applied to each of the weighting functions 
using a characteristic length scale of 500\,km, this being done to reduce spectral leakage effects within the 
application of the WMB method \citep[e.g.][]{swenson2002methods}. By defining a joint Gaussian distribution for 
the direct load and the observational noise, we can  form the WMB error through the linear transformations
\begin{equation}
    \label{eq:wmb_error_block}
\hat{q} -q = \left(
    \begin{array}{cc}
        CA - B & C
    \end{array}
\right) 
 \left(
    \begin{array}{c}
        \zeta \\ z
    \end{array}
\right),
\end{equation}
whose implementation is trivial due to \texttt{pygeoinf}'s support for direct sums and block operators. 
As there are four regions, the error distribution is defined
on $\mathbb{R}^{4}$, and so its lazily-evaluated covariance operator can be expressed as a dense matrix at a cost 
of only four actions. It should be 
noted that the actual numerical calculation of the WMB error distribution is not 
based on eq.~(\ref{eq:wmb_error_block}), but instead uses a more elaborate block-factorisation 
that removes redundant solutions of the sea-level equation linked to the actions of both $A$ and $B$;
full details can be seen within the supplied codes.  The  result is that the  calculation 
of the error distribution requires four  solutions of the sea-level equation,  and four 
solutions of the generalised sea-level equation introduced by \cite{al2024reciprocity} to determine the necessary adjoint actions. 

Figure~\ref{fig:grace_bias} shows the probability density functions (PDFs) for load average errors in each region. In each case, the blue 
PDFs reflect the nominal WMB errors obtained through the propagation of observational uncertainties, while the red PDFs show the actual errors. 
Consistent with the earlier theoretical discussion, we see that the WMB method underestimates uncertainties, with marginal standard deviations being 
too small by factors of between two and three. Importantly, the standard deviations for the true errors are approximately 50\% of the signal standard deviation, and 
hence these are differences that would matter practically. As noted earlier, because the direct load distribution has a zero expectation, the true error PDFs are 
all centred on zero. However, the plausible size of the estimator bias can be evaluated by considering the difference between the true and nominal 
variances, which suggests bias magnitudes of around 50\% of the signal standard deviation. These results are broadly consistent with those from earlier modelling studies
\citep[e.g.][]{clarke2005effect,riva2010sea, sterenborg2013bias, al2024reciprocity} as well as empirical ensemble-based assessments \citep[e.g.][]{blazquez2018exploring}, 
although making definitive statements about WMB errors would require  more realistic prior and noise distributions.

\begin{figure*}
    \centering    
    \begin{subfigure}[b]{0.48\textwidth}
        \centering
        \includegraphics[width=\textwidth]{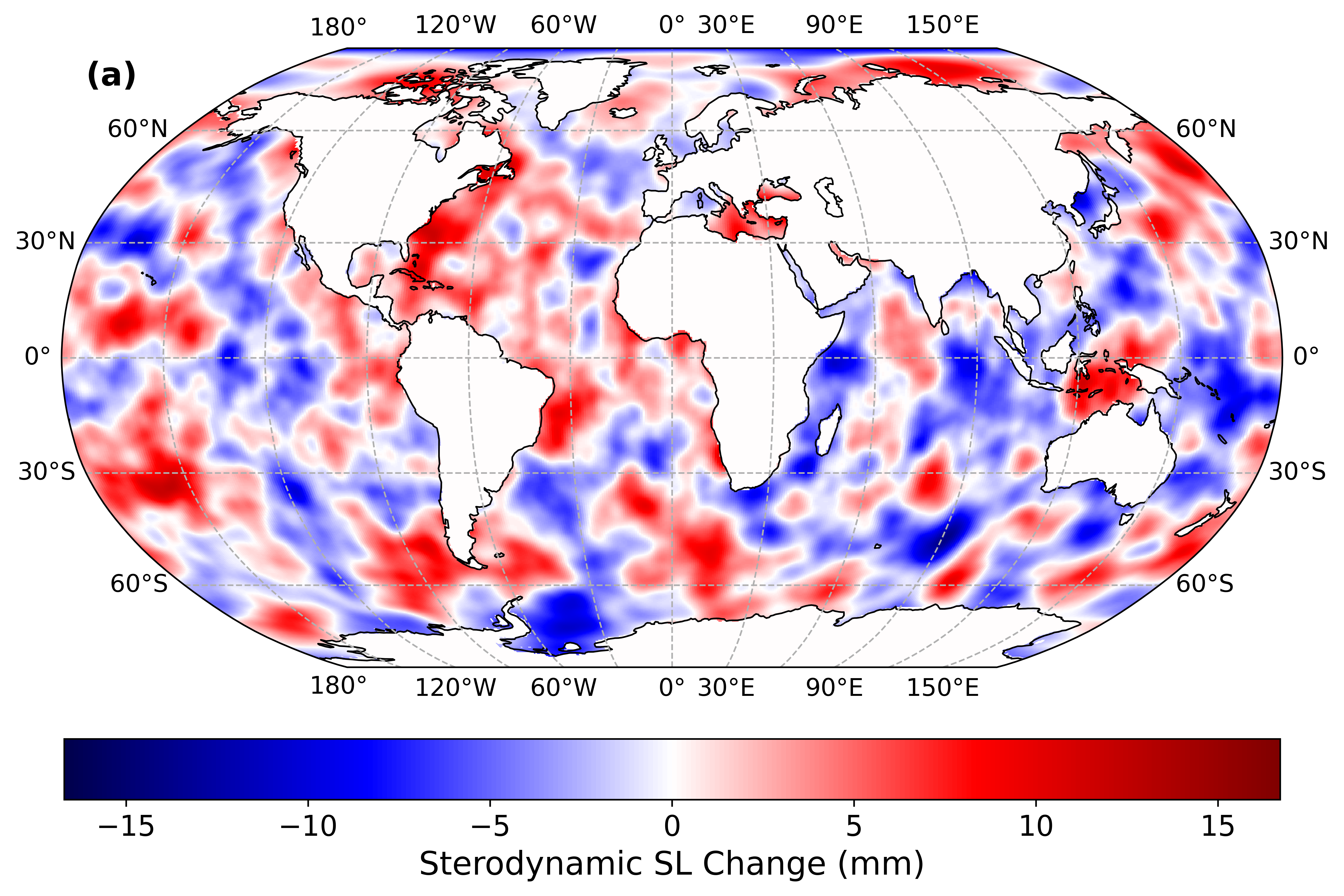}
        \label{fig:dynamic_ssh}
    \end{subfigure}
    \hfill 
    \begin{subfigure}[b]{0.48\textwidth}
        \centering
        \includegraphics[width=\textwidth]{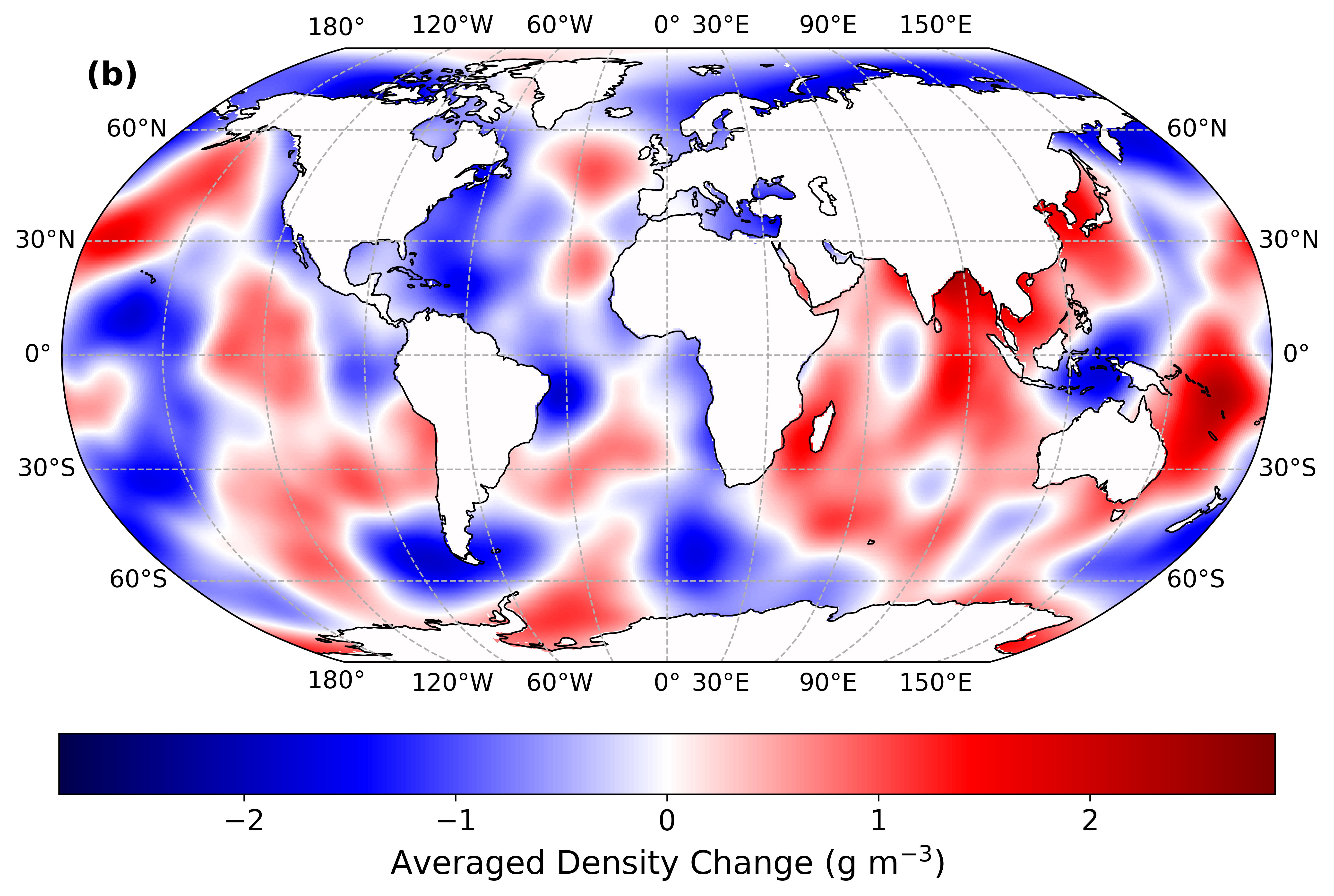}
        \label{fig:dynamic_density}
    \end{subfigure}
    \begin{subfigure}[b]{0.48\textwidth}
        \centering
        \includegraphics[width=\textwidth]{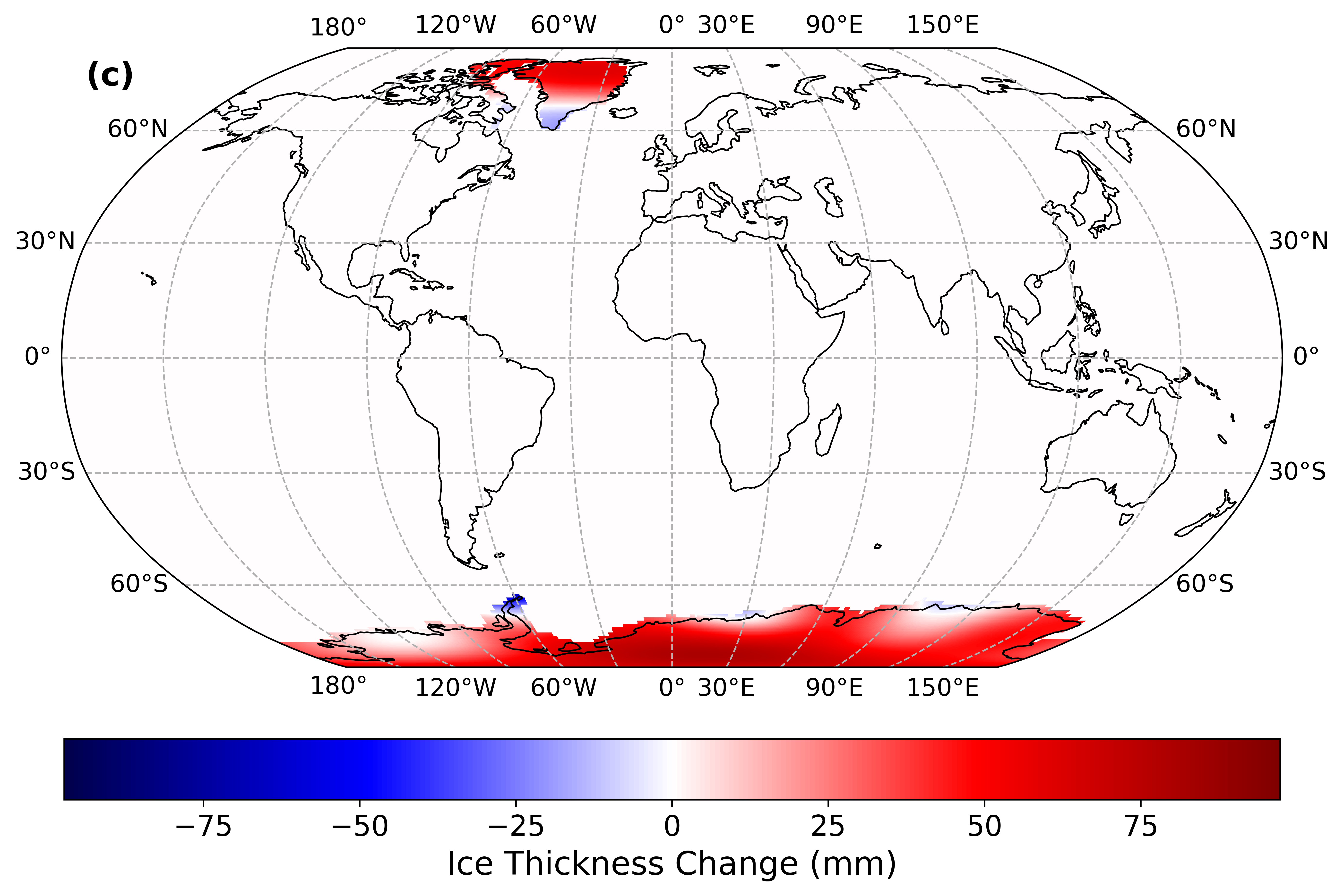}
        \label{fig:ice_thickness}
    \end{subfigure}
    \hfill 
    \begin{subfigure}[b]{0.48\textwidth}
        \centering
        \includegraphics[width=\textwidth]{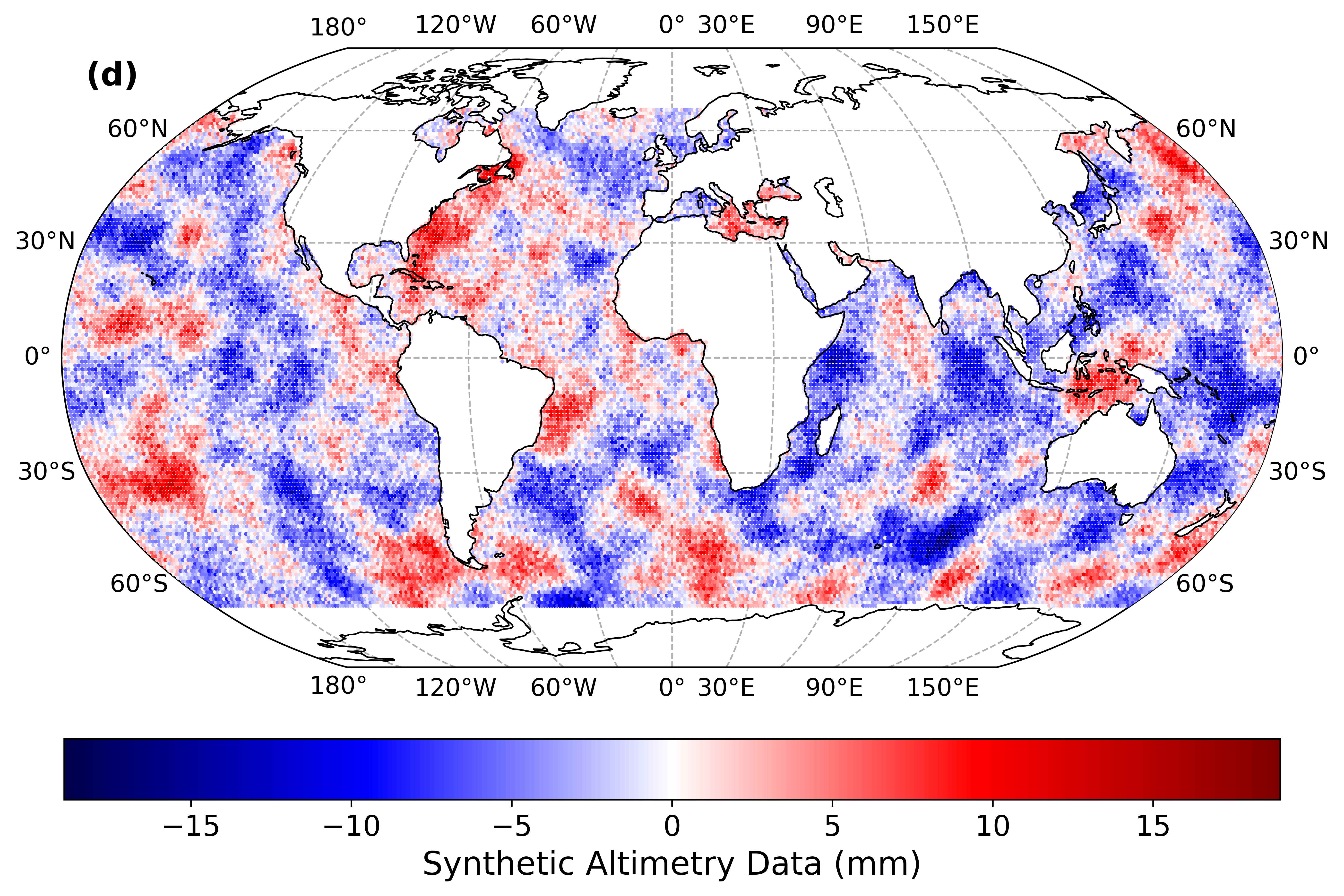}
        \label{fig:ssh_points}
    \end{subfigure}
    \caption{Simulated ocean altimetry observations associated with a sample from the three-component model distribution 
    described in the main text. Ocean dynamics are represented through the sterodynamic sea-level change in (a) and the 
    vertically averaged density change in (b). The joint covariance for these fields builds in a strong negative correlation 
    at long wavelengths, reflecting the dominant steric contribution to large-scale sea-level change. At shorter wavelengths, 
    the fields decorrelate to account for meso-scale ocean dynamic perturbations to the mean sea surface. The two ocean dynamic terms 
    are further coupled by requiring that they cause no overall mass change. The ice thickness change is shown in (c), with 
    the covariance for this field being uncorrelated with the others. Finally, (d) displays the resulting point observations sampled 
    on a $1^{\circ} \times 1^{\circ}$ grid between $\pm66^{\circ}$ latitude. In addition to the modelled geocentric sea-level change, these 
    simulated values include a lower-amplitude, large-scale correlated signal, reflecting systematic errors such as orbital uncertainties,
    and spatially uncorrelated noise that models a host of noise sources that fall below the $1^{\circ}$ resolution.
    }
    \label{fig:altimetry_maps}
\end{figure*}

\subsection{Ocean altimetry}

\label{sec:AltimetryBias}

\subsubsection{Ocean altimetry observations}

Ocean altimetry measurements provide 
high-resolution maps of geocentric sea-level  change -- variations in the height of the mean sea surface relative to the 
terrestrial reference frame -- over the entire ocean, excepting polar regions where measurements 
are complicated by orbital paths and sea ice cover  \citep[e.g.][]{wunsch1998satellite,stammer2017satellite,srinivasan2023satellite}.
Errors in altimetric measurements exhibit markedly different correlation structures. At the shortest scales, speckle
and thermal noise propagated through waveform re-tracking produce a quasi-white noise floor \citep[e.g.][]{sandwell2005, zaron2016}, 
while the measurement process also generates errors that are spatially coherent over roughly 5--100\,km, arising from backscatter inhomogeneities within the altimeter footprint \citep[e.g.][]{dibarboure2014, dufau2016} and from the mutually correlated retrieval errors induced by a common waveform re-tracking \citep[e.g.][]{zaron2016, tran2021}. Once measurements are aggregated onto coarser grids
typical of GMSLR studies, these short-range errors are largely subsumed, together with the white noise, into a component that is effectively uncorrelated between grid cells. By contrast, errors in the orbit solution and in the atmospheric and geophysical corrections 
\citep[e.g.][]{chelton2001} are coherent over much larger spatial scales -- up to basin scale for the orbit \citep[e.g.][]{tapley1985,esselborn2018} -- and are correlated in time over periods of up to several years, making them the dominant contribution to uncertainty in global-mean sea-level trends  \citep[e.g.][]{ablain2019, prandi2021, guerou2023}. Finally, recovering a geocentric water-volume signal requires further corrections, most notably for glacial isostatic adjustment \citep[e.g.][]{kopp2015geographic,spada2017,melini2019some}; the uncertainties in such corrections are readily incorporated within statistical noise models
and we assume this has been done in what follows. 

\subsubsection{Summary of the spatial averaging method}

By definition, GMSLR measures the change in ocean volume divided by its area, which is mathematically equivalent to the spatial average of relative sea-level change across the oceans \citep[e.g.][eq.42]{gregory2019concepts}. Fundamentally, these changes in ocean volume are driven by variations in either ocean mass or density. The mass component, known as barystatic sea-level rise, reflects the transfer of water to or from the land (e.g., from ice sheets or land water storage). Meanwhile, the density component varies due to changes in temperature and salinity, which combine to drive steric sea-level change. Because the contribution from salinity variations can be shown to be globally negligible, total GMSLR is effectively the sum of barystatic and global-mean thermosteric sea-level rise \citep[e.g.][eq.43]{gregory2019concepts};
for simplicity, we will hereafter refer to this latter thermosteric quantity simply as the global-mean \emph{steric} sea-level rise.

To determine GMSLR from ocean altimetry data, it is standard practice to perform a simple latitude-weighted spatial average \citep[e.g.][]{nerem2010estimating, ablain2015improved}. The values obtained
are subject to uncertainty due to observational errors and incomplete coverage at high latitudes. More fundamentally, this process
 approximates the global-mean geocentric sea-level rise, which differs from GMSLR by the average vertical motion of the sea floor \citep[e.g.][]{tamisiea2011ongoing, kopp2015geographic,lickley2018bias, gregory2019concepts}. Building on the study of \cite{lickley2018bias}, \cite{al2024reciprocity} assessed the bias in GMSLR estimates arising from these factors using a small number of forward calculations along with a visual examination of the associated sensitivity kernels. The aim of this section is to extend these arguments using statistical techniques, as has been done for the WMB method.

\subsubsection{Quantifying errors in the spatial averaging method}

To analyse the performance of the standard spatial averaging method, we consider a physical model in which geocentric sea-level varies 
in response to ice-sheet mass flux and ocean dynamics.
While we focus on these two mechanisms, the framework can readily accommodate additional mass sources, such as mountain glaciers or land water storage. 
Following Section~\ref{sec:ocean_dynamics}, the ocean dynamic contribution is prescribed through two 
fields: the sterodynamic sea-level change, $\eta_d$, and the vertically  averaged change in ocean density, $\delta\rho_w$.  
Because ocean dynamics cannot change the mass of the oceans, we impose the model constraint
\begin{equation}
    \label{eq:ocean_mass}
  \int_{\partial M} 1_{\mathcal{O}} \left(\rho_{w}\, \eta_{d}
        +  \delta \rho_{w}\, \eta_{0}\right) \dS = 0, 
\end{equation}
where $\eta_{0}$ is the background sea-level.
The direct load associated with the ice sheet and ocean dynamic perturbations is
\begin{equation}
    \label{eq:alt_load}
    \zeta = \rho_{i} (1-1_{\mathcal{O}})\, \eta_{i}
        + 1_{\mathcal{O}} \left(\rho_{w}\, \eta_{d}
        + \delta \rho_{w}\,\eta_{0}\right),
\end{equation}
where $\rho_{i}$ is a constant ice density; the factor
$1-1_{\mathcal{O}}$ multiplying the ice thickness change accounts for
ice shelves, which we assume to be in hydrostatic balance with the
underlying oceans. 
Solving the sea-level equation given this direct load, we obtain the
resulting relative sea-level change, $\xi$, along with the vertical
displacement, $u$, and the centrifugal potential change, $\psi$.
Following \cite{lickley2018bias}, the geocentric sea-level change,
$\eta$, can then be written
\begin{equation}
    \eta = \eta_{d} + \xi + u + \frac{\psi}{g}.
\end{equation}

Within this expression, the sum $\eta_d + \xi + u$ is the change in the height of the sea surface relative 
to a fixed reference frame. Because changes in Earth's rotation axis occur slowly relative to satellite orbital periods, altimetry measurements do not naturally capture them; the centrifugal term, $\psi/g$, is therefore included to mirror standard processing conventions that remove these rotational effects. While practical implementations of these corrections vary \citep{lickley2018bias}, our framework can evaluate $\eta$ both with and without this term. We present results including the rotational correction, having confirmed through sensitivity testing that its exclusion does not alter our conclusions.

To simulate the observational process, we generate synthetic altimetry data by adding random noise to this calculated geocentric sea-level field. Applying the standard spatial averaging method to this synthetic dataset yields an observation-based estimate for the GMSLR. We can then compare this estimate directly against the true GMSLR, which is defined analytically from the underlying model state by the integral
\begin{equation}
\frac{1}{A_{\mathcal{O}}}\int_{\partial M}1_{\mathcal{O}}(\eta_{d} + \xi) \dS,
\end{equation}
with $A_{\mathcal{O}}$ denoting the total area of the oceans \citep[e.g.][eq.42]{gregory2019concepts}.

To summarise the above ideas, let us combine the three inputs into a model vector
\begin{equation}
    m = \left(
        \begin{array}{c}
             \eta_{d} \\ \delta \rho_{w} \\ \eta_{i} 
        \end{array}
    \right). 
\end{equation}
We can then express the synthetic ocean altimetry observations in the form
\begin{equation}
    d = A m +z, 
\end{equation}
where $A$ is an appropriate linear operator and $z$ a vector of random errors with distribution $\Gauss{0}{R}$.
Of course, the operators $A$ and $R$ are not the same as in the satellite gravity example, but  context 
will be sufficient to clarify which is meant. Similarly, we can introduce a linear operator, $B$, that 
maps a model vector to the true GMSLR change, and an operator, $C$, that maps ocean altimetry data
to the estimated GMSLR change via the usual spatial averaging.  Writing $q= Bm$ and $\hat{q} = Cd$, it 
follows immediately that the error in the GMSLR estimate  has distribution
\begin{equation}
    \hat{q} -q \sim \Gauss{(CA-B)\overline{m}} {(CA-B)Q(CA-B)^{*} + CRC^{*}},
\end{equation}
where we have assumed that the model is a random variable with distribution $\Gauss{\overline{m}}{Q}$.
The interpretation of this result is identical to that for the satellite gravity problem.

\begin{figure*}
    \centering        
    \includegraphics[width=0.75\textwidth]{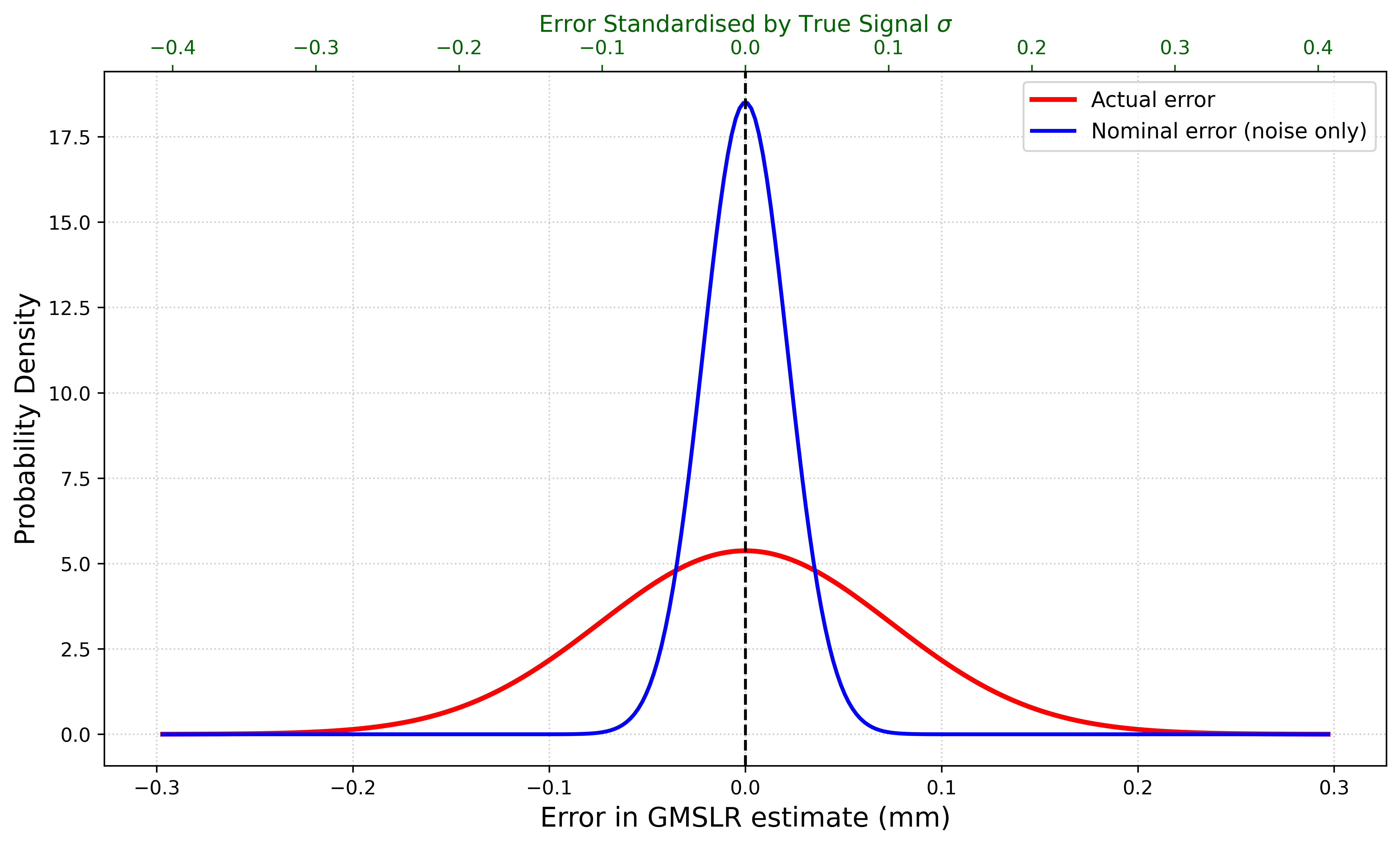}     
    \caption{Error probability density functions (PDFs) for the global-mean Sea Level (GMSL) change estimated 
    via spatial averaging of altimetry data. The blue curve represents the nominal error distribution, 
    relying solely on the propagation of assumed noise. The red curve displays the actual, analytically 
    derived error distribution that accounts for spatial sampling biases (e.g., the lack of data 
    within polar regions) and the physical distinction between relative and geocentric sea-level change.}       
    \label{fig:altimetry_bias}
\end{figure*}

\subsubsection{Numerical implementation}

\label{sec:AltimetryNum}

Each  field  $\eta_{i}$, $\eta_{d}$,  and $\delta \rho_{w}$ is assumed to be an element of
  $\SobSph{a}{2}$,  and hence the model space is the direct sum 
\begin{equation}
    \mathcal{M}  =  \SobSph{a}{2} \oplus \SobSph{a}{2} \oplus \SobSph{a}{2}.
\end{equation}
Note that, for convenience, these fields are globally defined over the surface, though only their restrictions to 
the appropriate geographic regions are physically significant. These necessary restrictions are built into the 
various operators; additionally, when plotting the fields, we zero out the non-physical parts of their respective domains.
The action of the forward operator, $A$, involves the solution of the sea-level equation, while 
its adjoint action requires solution  of a generalised sea-level equation following \cite{al2024reciprocity}.
The application of discontinuous ocean and ice masks reduces the global regularity of the direct load  to 
$\LebSph{a}{\infty}$, but the regularising nature of the sea-level equation ensures that $\xi$ resides in the $L^{p}$ 
Sobolev space $W^{1,p}(\Sph{a})$ for any finite $p$. Because $W^{1,p}(\Sph{a})$ embeds continuously into 
$C^{0}(\Sph{a})$ for $p>2$ \citep[e.g.][]{taylor2010partial}, point evaluation is globally well-defined, 
and hence the forward operator is a continuous linear mapping.

To construct a suitable distribution for the model vector, we start with a zero-expectation Gaussian distribution with covariance:
\begin{equation}
Q = 
\left(\begin{array}{ccc}
\alpha_{d} (1 + \lambda_{d}^2 \Delta)^{-q} & -\beta \sqrt{\alpha_{d} \alpha_{w}} \ee^{-\mu^2 \Delta}(1 + \lambda_{d}^2 \Delta)^{-q/2} (1 + \lambda_{w}^2 \Delta)^{-q/2} & 0 \\
-\beta \sqrt{\alpha_{d} \alpha_{w}} \ee^{-\mu^2 \Delta}(1 + \lambda_{d}^2 \Delta)^{-q/2} (1 + \lambda_{w}^2 \Delta)^{-q/2} & \alpha_{w} (1 + \lambda_{w}^2 \Delta)^{-q} & 0 \\
0 & 0 & \alpha_{i} (1 + \lambda_{i}^2 \Delta)^{-q}
\end{array}\right).
\end{equation}
Here, the diagonal terms have the same form as eq.~(\ref{eq:sobolev_kernel}), but with different amplitude and length-scale parameters for each of the three fields; 
for simplicity we have chosen the same exponent, $q$, for each of the fields, though this could readily be varied.
The off-diagonal blocks couple the two ocean dynamic components, with the magnitude of the correlation set by $0 \le \beta < 1$, and its decay at higher degrees controlled by the additional length scale $\mu$ \citep[e.g.][]{kleiber2017coherence}. This choice reflects the idea that at 
low degrees, sterodynamic sea-level changes are dominated by steric contributions, and hence that over these scales there should be a strong negative correlation 
between this field and vertically averaged density changes. Moving to shorter wavelengths, this correlation weakens as meso-scale ocean dynamic perturbations to 
the mean sea surface become more important.

The resulting joint distribution is rotationally invariant, with each field initially supported over the whole sphere. Geometry is imposed by pushing this distribution forward through a block-diagonal operator, $P$, formed from spatial masks, with the ice thickness multiplied by a mask equal to one over the (grounded) ice sheets and zero elsewhere, and the two ocean fields by the ocean function. Under this mapping, the covariance transforms as $Q \mapsto PQP^*$. The result is a Gaussian random field whose support is strictly limited to the areas where each component is physically meaningful, while pointwise standard deviations within those regions remain unchanged. 

Next, because dynamic changes cannot alter the total mass of the oceans, this physical requirement must be explicitly built into the distribution. Let $D$ be the linear operator that maps a model vector in $\mathcal{M}$ to the mass change for the oceans. We can condition the covariance subject to the exact constraint $Dm = 0$ (as a special case of the Bayesian methods discussed in Section~\ref{sec:bayes}) via the transformation:
\begin{equation}
Q \mapsto Q - QD^*(DQD^*)^{-1}DQ.
\end{equation}

To set the parameters for the prior distribution, we first fixed the various length scales to broadly sensible physical values: $\lambda_{i} = 500$\,km
for the ice sheet thickness change, $\lambda_{d} = 100$\,km for the sterodynamic sea-level change, and $\lambda_{w} = 500$\,km for the vertically
averaged density change. The magnitude of the anti-correlation, $\beta$, between $\eta_{d}$ and $\delta \rho_{w}$ was set to a high value of $0.9$, while
the length scale over which this anti-correlation weakens was taken to be $200$\,km.
To determine the three amplitude parameters, we proceeded as follows. First, we specified a value of $4$\,mm for the pointwise standard deviation of the sterodynamic
sea-level, with this choice fixing the value of $\alpha_{d}$. The pointwise standard deviation for $\delta \rho_{w}$ was scaled by the average depth of the oceans
to yield a mean steric sea-level change, which we set to $0.75$ times the sterodynamic value. This ensures that the bulk of the sterodynamic signal is
steric in origin, consistent with modern-day conditions.
Having determined the ocean dynamic components of the distribution, we then calculated the standard deviation for the global-mean steric sea-level rise. This
value was scaled by $1.7$ to obtain the associated barystatic sea-level rise, from which $\alpha_{i}$ was determined.
The values stated here are the defaults within the provided scripts and were chosen to be broadly typical of modern-day sea-level budgets. However,
this parameterisation -- relying on a single dimensioned amplitude alongside a set of physically meaningful ratios -- means that the calculations can
be readily adapted for other scenarios as desired.

Figure~\ref{fig:altimetry_maps}(a--c) shows a sample drawn from the resulting distribution. While these results -- which are fundamentally based on rotationally invariant Gaussian 
random fields -- are not fully realistic, they demonstrate how more complex physical knowledge can be incorporated into the choice of distributions. Indeed, 
we noted earlier that the distribution used in the assessment of the WMB method led to a strong correlation between the direct load over an ice sheet and that 
within the surrounding oceans. This artefact has now been eliminated in the more sophisticated model through the use of a composite model space. More generally, 
by introducing suitable off-diagonal blocks within the joint covariance, physically motivated correlations between different model components can be imposed 
in a statistical sense.

To represent the complex error structures inherent in ocean altimetry measurements, we model the noise covariance as the sum of two distinct components. The first 
term accounts for short-range errors -- such as speckle and thermal noise -- which are effectively subsumed into a spatially uncorrelated white-noise floor when aggregated 
onto a coarser grid. We set the constant standard deviation of this uncorrelated component to $0.5$ times the pointwise standard deviation of the sterodynamic sea-level 
change. To this, we add a spatially correlated term to capture large-scale errors, such as those arising from orbital uncertainties and broad-scale atmospheric corrections. 
For this component, we employ the same Sobolev--Mat\'{e}rn covariance defined in eq.~(\ref{eq:sobolev_kernel}), assigning it a length scale of $2000$\,km and a pointwise amplitude 
equal to $0.01$ times that of the sterodynamic signal. Although the absolute amplitude of this correlated error is relatively small, its spatial coherence means that it is 
not readily removed by spatial averaging; consequently, it contributes disproportionately to the overall uncertainty in GMSLR estimates. Fig.~\ref{fig:altimetry_maps}d shows the synthetic altimetry data, calculated from the sampled model shown in panels (a--c) and corrupted with noise drawn from the chosen distribution.

Having set up the various operators and distributions, it is straightforward to use \texttt{pygeoinf} to determine the nominal and true error distributions for the spatial 
averaging method. The results are shown in Fig.~\ref{fig:altimetry_bias}. We see that the nominal error obtained through propagation 
of noise substantially underestimates the true errors, in this case by a factor of around three.

\begin{figure*}
    \centering        
    \includegraphics[width=\textwidth]{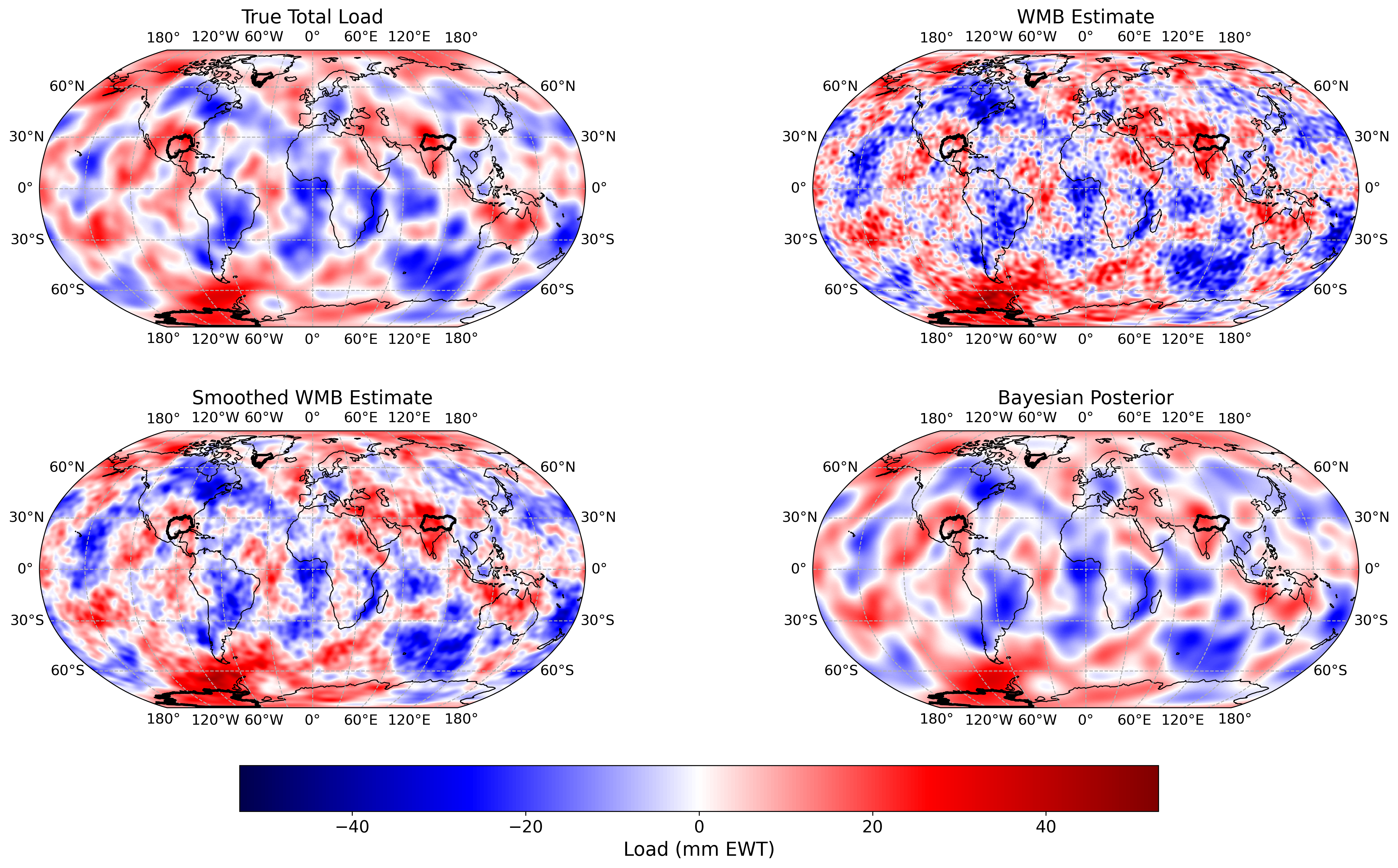}      
    \caption{Comparison of a true total load with various estimates derived from synthetic GRACE/FO potential 
    coefficients. The top-left panel displays the true total load, sampled from its prior distribution. The 
    top-right and bottom-left panels show the standard WMB estimates, unsmoothed and smoothed (using a 500\,km 
    length scale), respectively. The bottom-right panel presents the expected value of the Bayesian posterior 
    distribution. The Bayesian inversion effectively suppresses spectral leakage and noise, providing a 
    higher-fidelity recovery of the true spatial load patterns.}    
    \label{fig:grace_maps}
\end{figure*}

\subsection{Summary and implications}

The preceding analyses of the WMB method for satellite gravity  and the standard GMSL estimator for
ocean altimetry   highlight shared limitations in terms of (i) failing to fully incorporate gravitationally self-consistent sea-level physics 
and (ii) restricting uncertainty estimation solely to the propagation of observational noise. The result is that both techniques 
suffer from  systematic biases and  underestimate uncertainties. It is, however, important to emphasise that the precise 
quantification of these errors depends  on the probability  distributions chosen for the underlying model parameters and the observational
noise. The  model and noise distributions employed in the examples above are idealised, but their characteristic length scales and 
amplitudes were selected to broadly represent  realistic modern-day  scenarios. This means that the quantitative results 
in this section should be seen as indicative of real and significant issues, but in need of refinement through the consideration 
of more realistic statistical distributions.

Crucially, the numerical framework utilised in this section
is not bound to any specific statistical choices beyond the Gaussian assumptions. The matrix-free architecture of the \texttt{pygeoinf} and \texttt{pyslfp} 
libraries allows for the  implementation and investigation of alternative,  physically informed priors or more complex, correlated 
noise models, should other researchers wish to explore them. Moreover, the forward operators, adjoints, and statistical structures 
introduced in this section to
analyse existing methods provide the exact mathematical and computational machinery required  
within the Bayesian framework we now turn to.

\section{Linear Bayesian inverse problems}
\label{sec:bayes}

For the remainder of this paper we will be concerned with the solution of 
linear inverse problems using Bayesian methods, under the assumption of 
Gaussian priors and noise. Here we  provide 
a  summary of the necessary theoretical results following 
\cite{stuart2010inverse}. We then discuss some general numerical aspects 
of the implementation, along with practical quantitative methods 
for working with the posterior distribution.
Finally, we consider the performance of the Bayesian solution from 
a frequentist perspective, this providing a means for understanding 
the sensitivity of the results to the choice of prior.

\subsection{Formulation and solution}

Building on and generalising the notations from the previous section, we consider 
an inverse problem in which the data generation is modelled through 
the random variable
\begin{equation}
    d = A m + z, 
\end{equation}
where $m \sim \Gauss{\overline{m}}{Q}$ is the model vector and $z \sim \Gauss{\overline{z}}{R}$
the observational noise. Both the model space, $\mathcal{M}$, and the data space, $\mathcal{D}$, 
are assumed to be real separable Hilbert spaces, with $\mathcal{D}$ being finite-dimensional. The forward 
operator, $A$, is a continuous linear mapping from $\mathcal{M}$ to $\mathcal{D}$.
The prior and data covariances are  self-adjoint, non-negative, and 
trace-class; the latter condition is trivially met on the data space, but 
it places a  strong  constraint when the model space is infinite-dimensional. 

The prior distribution, $\Gauss{\overline{m}}{Q}$, fully describes our knowledge of the model 
before any observations have been made. From the above relation, we see that the data itself has the prior distribution
\begin{equation}
    d \sim \Gauss{A\overline{m}+\overline{z}}{AQA^{*}+R}. 
\end{equation}
We assume that the prior data covariance, $AQA^{*}+R$, is 
invertible. This ensures that any point within the 
data space can be generated through some combination 
of a model state and random noise. The
condition can, for example, be met if $A$ is surjective while $R$ vanishes, 
and hence the formalism applies  to the error-free case
under appropriate circumstances. Alternatively, if the data covariance is
non-singular then there is no restriction on the image 
of the forward operator. 

If a value, $d_{\mathrm{obs}}$, for the data is observed then we can ask how our knowledge of the 
model should be updated. The answer is given through Bayes' theorem in terms 
of a posterior model distribution.
Within a  linear problem with Gaussian prior and noise, it is well known
that the posterior distribution is also Gaussian, denoted by $m_p \sim \Gauss{\overline{m}_p}{Q_p}$. The posterior expectation and covariance are given analytically by
\begin{align}
    \label{eq:postexp}
    \overline{m}_{p} &= \overline{m} + K(d_{\mathrm{obs}}-A\overline{m}-\overline{z}), \\
    \label{eq:postcov}
    Q_{p} &= (1-KA)Q, 
\end{align}
where the Kalman operator is
\begin{equation}
    \label{eq:kalman}
    K= QA^{*}(AQA^{*}+R)^{-1}. 
\end{equation}
Furthermore, a simple calculation shows that the random variable
\begin{equation}
    \label{eq:rto}
    m_{p} = m + K(d_{\mathrm{obs}}-Am-z), 
\end{equation}
with $m \sim \Gauss{\overline{m}}{Q}$ and  $z \sim \Gauss{\overline{z}}{R}$, exactly follows the posterior distribution. This formulation allows 
for samples of the posterior to be generated by drawing from the prior and noise distributions and transforming the results accordingly. This approach underlies the randomise-then-optimise method for sampling from the posterior distribution which we discuss further below
\citep[e.g.][]{bardsley2014randomize, schillings2017analysis}.

\subsection{Numerical considerations}
\label{sec:NumCon}

\subsubsection{Model discretisation and convergence}

Within numerical work we cannot deal directly with infinite-dimensional 
spaces. Nevertheless, because the Bayesian posterior is well-defined 
mathematically, it is possible to design numerical approximations 
that converge to any desired level of accuracy 
\citep[e.g.][]{bui2013computational,petra2014computational}.
This situation contrasts  with the ``discretise first'' approach 
that is more common within solid Earth geophysics \citep[e.g.][]{tarantola2005inverse, 
wunsch2006discrete, menke2018geophysical} but which  leads to a sequence
of finite-dimensional inverse problems that generally do not have a  well-defined limit, 
and hence whose solutions always depend on the discretisation scheme
\citep[e.g.][]{stuart2010inverse}.

The key points within the implementation of  infinite-dimensional 
Bayesian methods are (i) choosing a finite-dimensional
approximation to the model space that accurately represents samples
from the prior distribution, and (ii) forming an associated approximation to
the forward operator whose discretisation error is small relative to
the observational noise. While formal error analyses are possible, a pragmatic 
approach can also be taken where a discretisation is  progressively refined until 
the resulting posterior distribution does not  change appreciably. Within
the applications considered below, model spaces are discretised using 
spherical harmonic expansions truncated at degree 256. This value 
is more than sufficient to represent samples from the prior distributions, 
being instead selected through the need to accurately solve the 
sea-level equation. Key calculations have been repeated at degree 512 
to check that convergence has been achieved at a practical level.

\begin{figure*}
    \centering        
    \includegraphics[width=\textwidth]{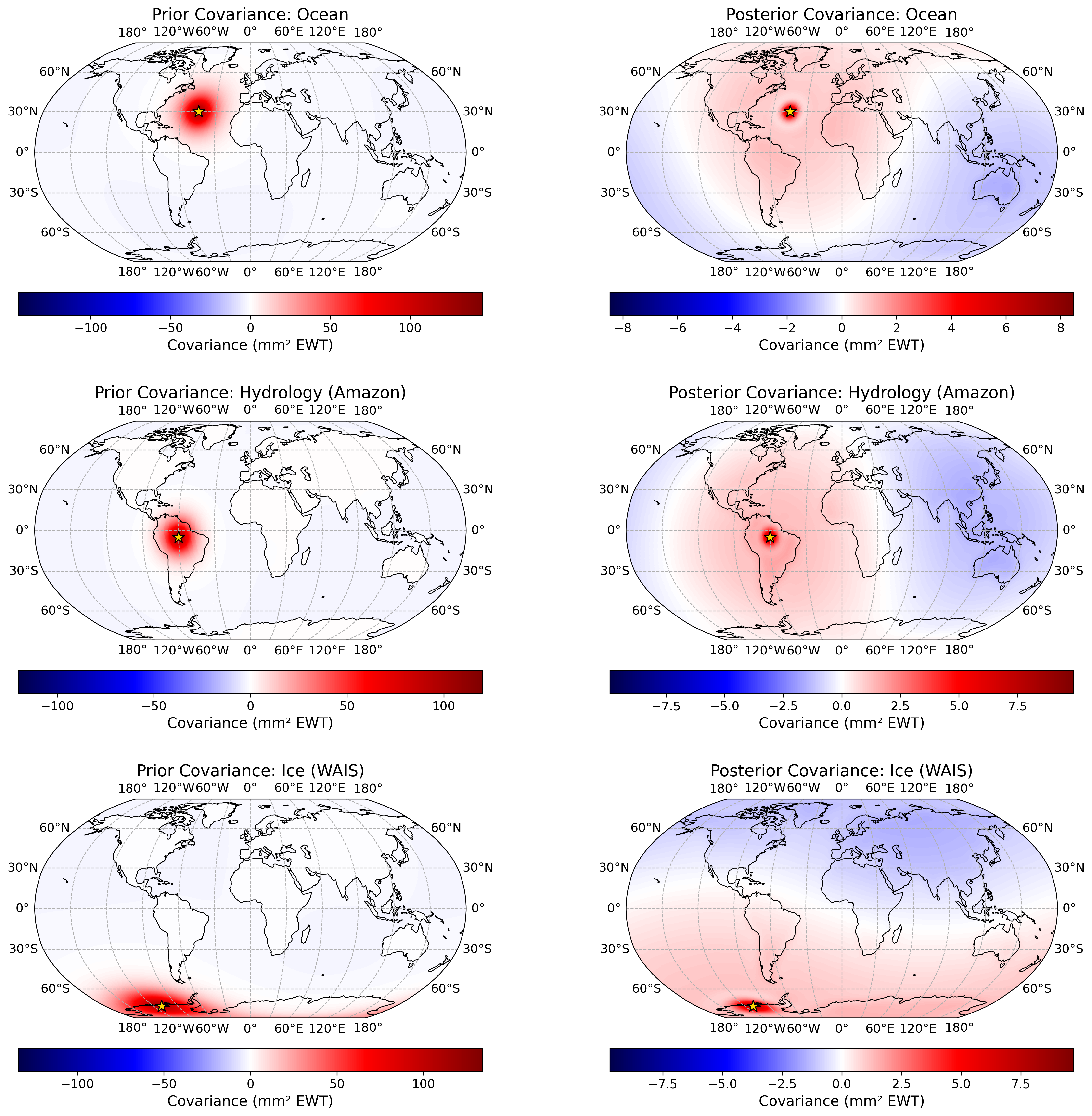}        
    \caption{Prior (left column) and posterior (right column) spatial covariance functions evaluated at three 
    representative locations (marked by gold stars): the North Atlantic Ocean, the Amazon Basin, and the West 
    Antarctic Ice Sheet (WAIS). Values are expressed in $\mathrm{mm}^2$ Equivalent Water Thickness (EWT). The 
    posterior maps demonstrate a substantial reduction in local variance (the value at the reference point) 
    compared to the prior, reflecting the strong  constraints provided by the synthetic GRACE/FO observations. 
    Additionally, the posterior covariances exhibit prominent long-range correlation structures not seen in the 
    prior, primarily reflecting the absence of degree-one observations.}
    \label{fig:grace_cov}
\end{figure*}

\subsubsection{Preconditioned iterative linear solvers}

Let us suppose that we have a suitable discretisation for the model space along with 
corresponding approximations for the action of the forward operator $A$, 
its adjoint $A^{*}$, the prior covariance $Q$, and the noise covariance $R$. 
Numerical calculation of the posterior distribution is then reduced to 
the repeated solution of the linear system associated with the 
prior data covariance, $AQA^{*}+R$. One such solution is required to 
determine the posterior expectation, while one further solution 
is needed per action of the posterior covariance or per sample from 
the posterior distribution using eq.~(\ref{eq:rto}). This 
linear system is posed on the finite-dimensional data space and is  self-adjoint and positive 
definite. For the applications considered within this paper, the 
data spaces have dimensions in the range 10,000 to 50,000. In fact, 
the upper value here is associated with ocean altimetry data sampled on 
a $1^{\circ} \times 1^{\circ}$ grid, whereas typical global data products 
are sampled on $0.25^{\circ} \times 0.25^{\circ}$ grids which would 
push the data space dimensions to around 500,000.

Each action  of the prior data covariance requires one solution of the sea-level equation 
and one of the generalised sea-level equation.  At truncation degree 256, the solution of either 
problem takes around one second using \texttt{pyslfp}; here, and throughout the paper, quoted timings 
refer to a laptop with 16 CPU cores and 64\,GB of RAM which acts as our reference machine.
At the lower end of the data dimensions, construction of the prior data 
covariance, $AQA^*+R$, as a dense matrix would take about six hours in serial, and require about 800\,MB 
for storage in double precision. For the largest data spaces considered, the time 
rises to around 28 hours with  20\,GB for storage. Both calculations could, however, be performed in 
an embarrassingly parallel manner, and hence the times reduced substantially 
given suitable resources.  Having formed the dense matrix, solution
of the linear system could be determined  using Cholesky factorisation and 
back-substitution. Indeed, using a standard linear algebra library 
like \texttt{scipy} \citep[][]{2020SciPy-NMeth}, the necessary factorisation would take only a few seconds 
for problems with dimension 10,000. However, extending this direct approach to data spaces of 50,000 is 
practically prohibitive. While the raw storage of the matrix requires 20\,GB, intermediate allocations and 
memory layout requirements during the assembly and factorisation phases routinely exceed standard memory 
limits. Furthermore, the 28-hour computational 
cost of merely assembling the dense matrix at this scale renders the approach intractable without access 
to specialised computational resources.
For the inverse problems considered it is therefore possible to use direct linear solvers  for the lower-dimensional 
cases, but the larger data spaces necessitate the use of matrix-free iterative methods instead.
Indeed, even when direct solvers are computationally viable, iterative methods 
can be more efficient by avoiding the construction of the prior data covariance as 
a dense matrix. Consequently, we focus on determining the action of the \textit{inverse} prior data 
covariance using the preconditioned conjugate gradient method \citep[e.g.][]{saad2003iterative}.

Without preconditioning, the number of iterations required to solve a linear system 
can be comparable to the dimension of the data space. This makes the construction of efficient
and effective 
preconditioners, which can reduce this number substantially, a central concern within this paper. Most standard preconditioning methods are designed for 
very large, sparse linear systems arising from the solution of partial differential equations 
\citep[e.g.][]{saad2003iterative, chen2005matrix}.  Here, by contrast, the sizes of the linear systems are 
comparatively  modest, but the operators are dense and defined only through their action; hence there is no simple 
means for determining any particular elements. It is therefore necessary to consider fully 
matrix-free preconditioning schemes, a topic that has received comparatively little attention in the
literature \citep[e.g.][]{bellavia2013matrix, de2014matrix,benner2018low,ambartsumyan2020hierarchical,
subrahmanya2025randomized}. To address this, the \texttt{pygeoinf} library 
implements a range of generic matrix-free preconditioners. It is also, as we will see, sometimes possible and 
advantageous to design bespoke, 
physically motivated  preconditioners tailored for a given problem.  

\subsubsection{Exploring the posterior distribution}

The central challenge within the Bayesian solution of inverse 
problems in infinite-dimensional spaces is the 
practical assessment of uncertainty. Numerically, 
we can approximate the action of the posterior covariance
to a desired level of accuracy. This could be 
used to explicitly construct the posterior covariance
as a dense matrix relative to the numerical
discretisation used for the model space. In 
practice, however, the discretised spaces are so large 
as to make such calculations impossible. Within the present application, 
for example, the discretised model space for a single scalar field expanded to degree 256 has  
dimension 66,049. In double precision, this equates to 
roughly 35\,GB of memory just for storage. More prohibitively, constructing this matrix 
would necessitate evaluating the inverse prior data covariance 66,049 times -- a 
computationally intractable task even with highly optimised solvers.
To circumvent this bottleneck, we  abandon dense matrix calculations and instead leverage 
the matrix-free architecture of  \texttt{pygeoinf}   to assess uncertainty through alternative, 
computationally tractable methods.

Using eq.~(\ref{eq:rto}) it is possible to draw samples from the posterior distribution. 
Each sample requires one action of the inverse prior data covariance, while the 
generation of a set of samples can be trivially parallelised given suitable computational
resources. A simple application  is to generate maps of the pointwise 
standard deviation of a scalar field such as the direct load within the inversion of GRACE/FO data. 
The relative error of the standard deviation estimated from  $N$ samples is $1/\sqrt{2N}$, meaning that visually useful maps can be generated from a few hundred samples. 

Such maps of pointwise standard deviation do not, of course, quantify spatially correlated errors. To look at this aspect, 
we suppose that the model space consists of continuous scalar fields, but the arguments generalise
to suitable vector fields also. The value of the model at a point, $x$,
can then be written 
\begin{equation}
    m(x) = \cbraket{\hat{\delta}_{x}}{m}_{\mathcal{M}}, 
\end{equation}
where $\hat{\delta}_{x}$ is the representation of the Dirac measure based at $x$, this being an element
of the model space whose existence is guaranteed by the Riesz representation theorem \citep[e.g.][]{schechter2001principles}. If the model is 
distributed according to the posterior distribution, then its values at two points, $x$ and $x'$, 
are jointly Gaussian distributed with covariance
\begin{equation}
    \mathrm{Cov}[m(x),m(x')] = \cbraket{Q_{p}\hat{\delta}_{x}}{\hat{\delta}_{x'}}_{\mathcal{M}}. 
\end{equation}
This leads to the introduction of a two-point posterior covariance function 
\begin{equation}
    k_{p}(x,x') = (Q_{p}\hat{\delta}_{x})(x'),
\end{equation}
which for fixed $x$ can be generated at the cost of one evaluation of the posterior covariance. A
two-point prior covariance function can be similarly defined, with visual comparison of the two functions providing 
insight into the spatial scales resolved by the inversion.

\begin{figure*}
    \centering        
    \includegraphics[width=\textwidth]{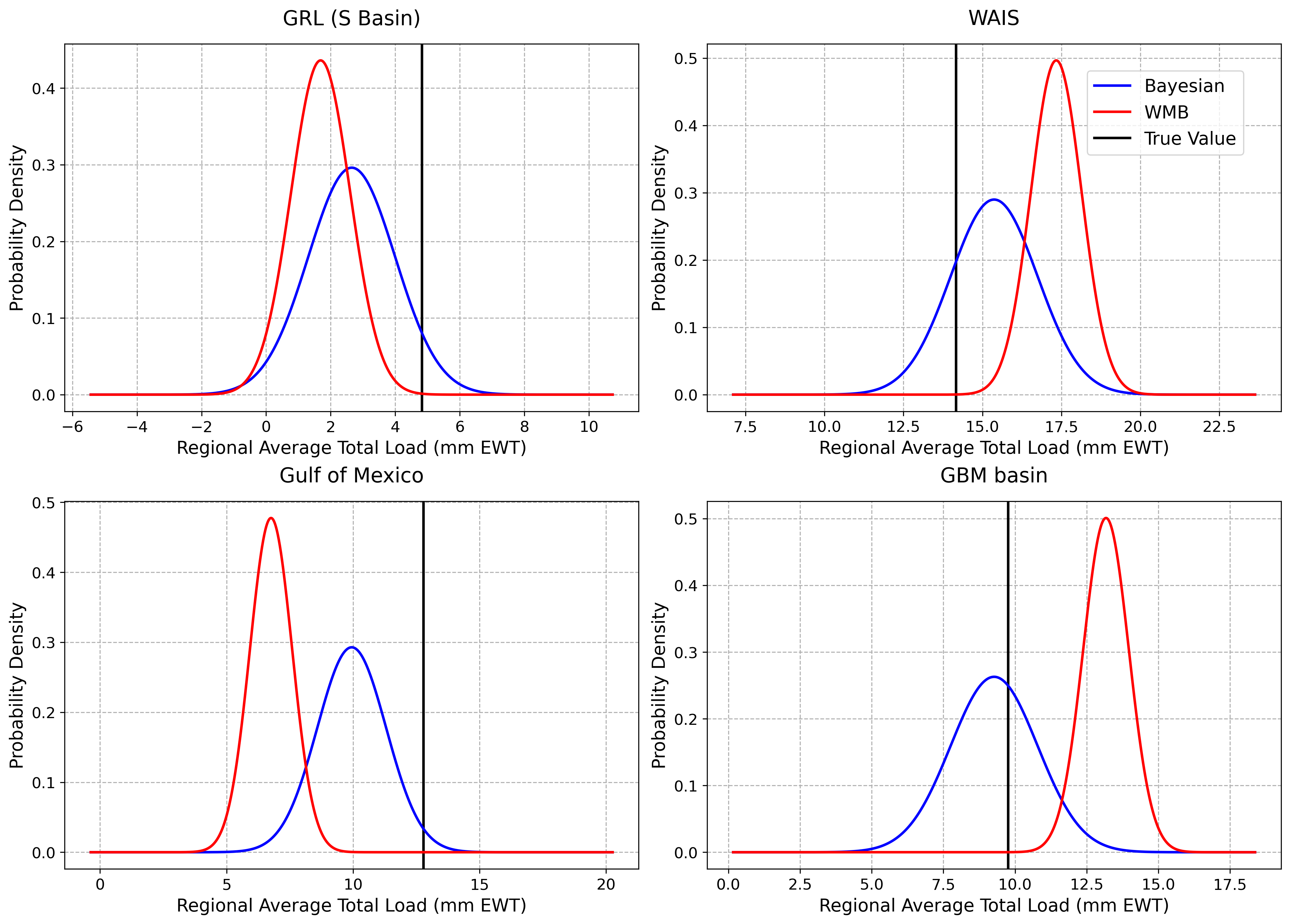}     
    \caption{Marginal posterior probability density functions (PDFs) for the regional average of the total 
    load (expressed in mm Equivalent Water Thickness, EWT) across four target areas. The blue curves 
    represent the Bayesian posterior distributions, while the red curves show the error distributions 
    associated with the standard WMB method. Vertical black lines denote the true synthetic values in each region. 
    As expected when inverting data generated from the prior and noise distributions, the Bayesian framework yields 
    estimates that consistently bound the true values while providing broader, more realistic uncertainty quantification 
    compared to the WMB method.}
    \label{fig:grace_pdfs}
\end{figure*}

Building on the above idea,  it is not always necessary to 
consider the full posterior distribution. Rather, the scientific questions may 
concern only values within some low-dimensional space.
Indeed, the WMB method is primarily used in the determination of load averages
over a targeted set of regions, while we have seen that an important application of ocean altimetry 
observations is the estimation  of global-mean sea-level rise. Let  
$B$ then denote a linear mapping from the model space to some $p$-dimensional
property space, $\mathcal{P}$, whose components are the quantities of interest. 
We write $q = Bm$ for the property vector corresponding to the model $m$.
The posterior distribution for the model, $m_{p} \sim \Gauss{\overline{m}_{p}}{Q_{p}}$, 
can be readily pushed forward under this mapping to obtain
\begin{equation}
    q_{p} \sim \Gauss{B\overline{m}_{p}}{BQ_{p}B^{*}}.
\end{equation}
The expected value 
for this latter distribution can be obtained immediately from that for the model. 
Its covariance, $BQ_{p}B^{*}$, is defined on a $p$-dimensional space, and can be 
constructed as a dense matrix by acting the operator on a basis for $\mathcal{P}$. 
This in turn requires $p$ actions of the posterior model covariance, $Q_{p}$, and hence 
the same number of solutions of the underlying linear system. For property
spaces that are not too large (e.g., dimensions of order tens), the necessary calculations can be readily performed 
when coupled with efficient preconditioned methods for solving the linear systems, while the 
whole process can be trivially parallelised.

As a final example, a further useful method for analysis of the posterior covariance is based on 
randomised low-rank factorisation \citep[e.g.][]{halko2011finding}. The intuition behind this method is that while the covariance operator acts 
on an infinite-dimensional space, its most important features are often concentrated on a finite-dimensional subspace of modest size. This means that by probing 
the posterior covariance with a relatively small set of random vectors, we are able to capture a sufficiently accurate idea of the covariance structure. 
This allows us to, for example, form a low-rank Cholesky factorisation
\begin{equation}
    Q_{p} = L_{p}L_{p}^{*}, 
\end{equation}
where $L_{p}$ maps from a much smaller finite-dimensional Euclidean space (the rank of the decomposition) back to the model space. In a similar 
manner, approximate eigendecompositions can be generated, or singular value decompositions for non-self-adjoint operators.
Because each random vector can be acted on independently, this process can be 
readily parallelised. Such decompositions can be formed using a fixed rank, or allowing a variable rank 
determined through a probabilistic convergence criterion. Within \texttt{pygeoinf}, we support these randomised decompositions following \cite{halko2011finding}, with 
refinements including the parallelisation of variable-rank decompositions.

The advantage of forming a low-rank factorisation of the posterior covariance is that the cost of its action is negligible compared to that of the original operator. Moreover, samples from the approximate posterior distribution can be generated at a similarly low cost through a single action of the Cholesky factor, $L_{p}$, on a random vector. Such computational savings are vital when solving sequential assimilation problems using, for example, a Kalman filter \citep[e.g.][]{law2015data}. In this context, the posterior distribution at one time is used to form the prior at a later time by pushing it forward under the system's dynamics. Because the exact posterior covariance is represented implicitly (relying on lazy evaluation), passing it directly to the next time step would result in nested iterative solves. This would cause the computational cost to balloon as the assimilation progresses. By expending the upfront cost to compute a low-rank factorisation of the posterior at each time step, however, this chain of dependency is broken, ensuring that all steps within the assimilation process maintain a constant and tractable computational cost.

\subsubsection{Summary}

It is useful to draw together the computational threads of this section. Within the
matrix-free framework, the basic unit of computation is a single preconditioned solution of
the linear system associated with the prior data covariance, and each of the tasks discussed
above can be costed by the number of such solves it requires. The posterior expectation, an
action of the posterior covariance, and an individual posterior sample each require one; the
 push-forward of the posterior onto a $p$-dimensional property space requires $p$; while maps
 of pointwise standard deviations and low-rank factorisations require, respectively, one to
 two hundred solves and slightly more than the rank of the decomposition. Crucially, wherever
 multiple solves are needed they are mutually independent, and hence such tasks are
 embarrassingly parallel. Moreover, because the matrix-free architecture never forms large
 dense matrices, each solve carries only a modest memory footprint, and so many can proceed
 concurrently on a single machine; this stands in contrast to dense formulations, whose memory
 demands become prohibitive well before their arithmetic costs, and it is in this sense that
 the framework scales down to minimal hardware just as readily as it scales up across hundreds 
 or thousands of cores. 
 
 In practice, therefore, the framework operates within two regimes. Calculations
 requiring a handful of solves can be performed interactively on very modest hardware, each
 solve taking on the order of a few minutes for the problems considered within this paper. Tasks
 requiring hundreds of solves take correspondingly longer in serial, but their wall-clock
 times fall broadly in proportion to the number of cores available, and this without
 modification to the underlying algorithms. It should be noted that parallelisation of this kind improves 
 throughput (the total volume of tasks completed over time) rather than latency (the 
 execution speed of a single task): the iterations within a single solve remain sequential, 
 and hence the cost of one solve sets a floor on the time required for any
 calculation. The throughput-oriented regime can be expected to dominate within sequential
 assimilation problems where, as noted above, a low-rank factorisation of the posterior must
 be formed at every time step. In such applications, the practical viability of the method
 rests not upon the cost of any single solve, but upon the efficiency with which the many
 independent solves within each factorisation can be distributed across the available cores.

\subsection{Frequentist assessment of the Bayesian solution}
\label{sec:freq}

\subsubsection{Bounding prior sensitivity}

\label{sec:priorsense}

Within the context of Bayesian statistics, the posterior distribution represents the complete solution of the 
inference problem. The prior  quantifies our initial beliefs, while Bayes' theorem tells
us how to coherently update these beliefs in light of the available observations \citep[][]{ramsey1926truth, 
savage1954foundations,lindley1972bayesian, freedman1995some, stark2015constraints}. What the 
theory cannot do is tell us what our initial beliefs \emph{should} be, nor  is there a way to  
meaningfully ask from \emph{within} the theory whether one choice of prior is better than another.

In order for Bayesian inference to be useful, it is therefore necessary within each application to carefully 
and convincingly justify the choice of prior with reference to appropriate physical ideas. Nevertheless, it 
is typically still the case that some range of priors will be seen as acceptable, and hence one should ask 
the extent to which the resulting posterior distributions vary. Within favourable circumstances, 
the available data might be sufficient that any plausible prior leads to practically the same 
posterior. But such cases where the ``data swamps the prior'' can   occur only within 
relatively simple problems that would have unique solutions in the absence of observational noise. 
Indeed, within the context of an under-determined inverse problem there must necessarily be directions  within 
the model space that have no effect on the data, and hence along which the posterior is entirely
determined by the prior. These ideas are sketched quantitatively below for the special case of linear problems 
with Gaussian priors and noise, while   more 
general results can be found in \cite{stuart2010inverse}.

To proceed, we follow \cite{diaconis1986consistency} by analysing the performance 
of the Bayesian solution from a frequentist perspective. We take the mapping 
from the data to the posterior expectation within eq.~(\ref{eq:postexp}) to 
define a point estimator, and  ask how well it performs under 
hypothetical repetitions. We  suppose that the data are generated using a \emph{different} prior 
distribution to that assumed within the inversion (the noise distribution could be 
varied similarly). Specifically, 
we replace the observed data in eq.~(\ref{eq:postexp}) with 
the random variable $d = Am+z$ where the model distribution
takes the perturbed form
\begin{equation}
    m \sim \Gauss{\overline{m}+\Delta \overline{m}}{Q + \Delta Q},
\end{equation}
while the noise distribution is unchanged. Writing $\hat{m}$
for the resulting  estimator, a simple calculation shows 
that the associated error is
\begin{equation}
    \hat{m}-m = (1-KA)(\overline{m}-m) + K(z-\overline{z}), 
\end{equation}
which has expected value
\begin{equation}
    \mathrm{E}[\hat{m}-m] = (KA-1)\Delta \overline{m},
\end{equation}
and covariance
\begin{equation}
    \mathrm{Cov}[\hat{m}-m] = Q_{p} + (KA-1)\Delta Q(KA-1)^{*},
\end{equation}
with $Q_{p}$ the posterior covariance operator defined in eq.~(\ref{eq:postcov}).
We see that when the model and data are drawn from 
the same distributions used within the inversion,  
the Bayesian estimator is unbiased while the covariance 
for its error equals the posterior as defined in eq.~(\ref{eq:postcov}).
This is the ideal situation for which the Bayesian method provides 
an optimal estimator. More generally, the  occurrence of the operator $KA-1$ in both 
the bias and covariance expressions formalises the limitations 
of infinite-dimensional inference discussed earlier. Because the  data space is strictly finite-dimensional, 
the Kalman operator $K$ necessarily has a finite-dimensional image. Consequently, $KA-1$ can never vanish 
globally across the model space. In the infinite number of directions where the data provide no constraint, the action of $A$ 
is zero, and $KA-1$ is proportional to the identity operator. In these unconstrained directions, the error properties of our 
estimator are dictated entirely by the discrepancy between the true model distribution and our chosen prior. 

\begin{figure*}
    \centering        
    \includegraphics[width=0.8\textwidth]{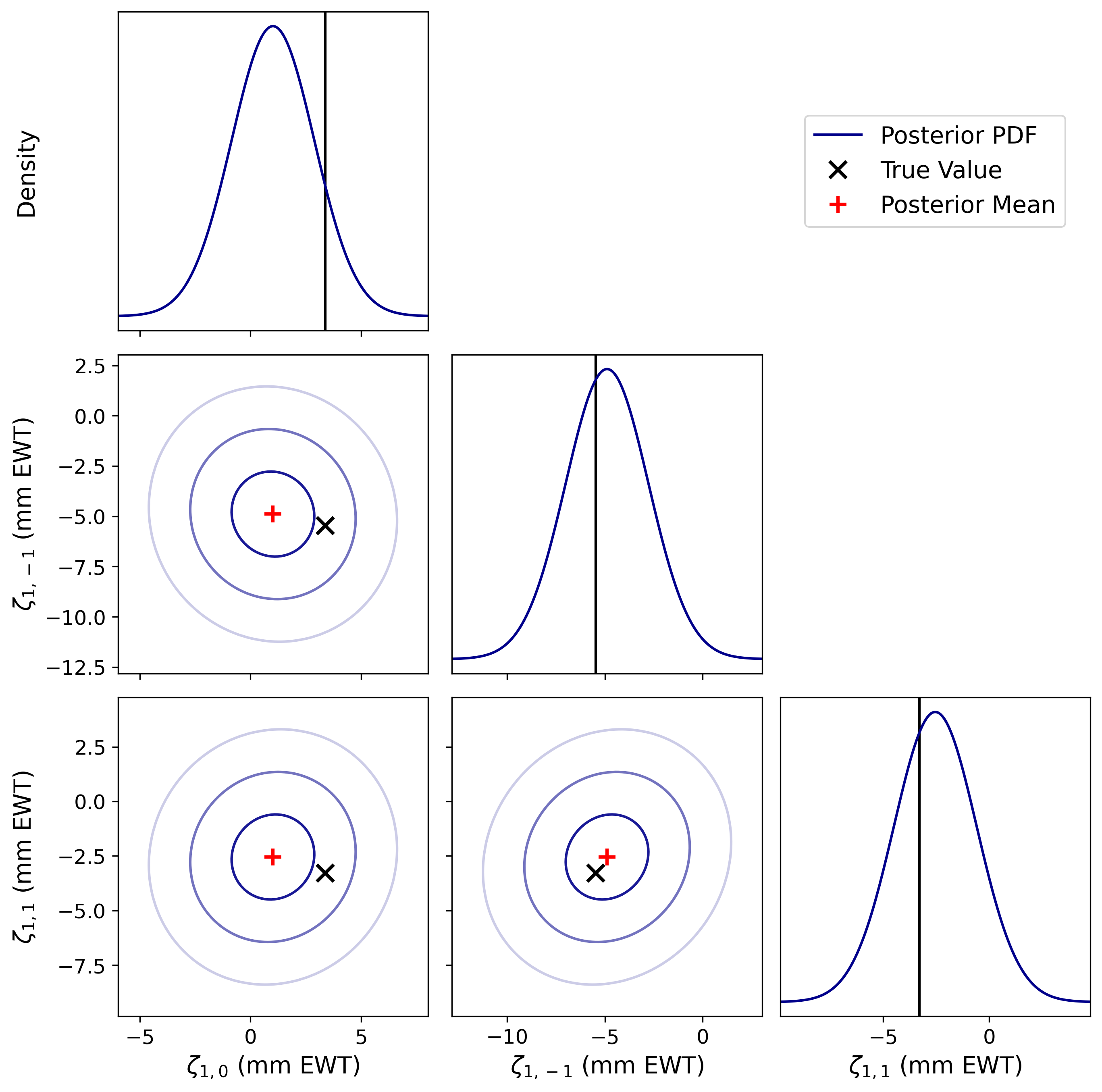}        
    \caption{Evaluation of the Bayesian framework's recovery of the degree-one spherical harmonic coefficients 
    ($\zeta_{1-1}$, $\zeta_{10}$, and $\zeta_{11}$) of the direct load from synthetic GRACE/FO observations. 
    These coefficients are directly proportional to geocentre motion and are not directly observable by satellite gravity.
     The diagonal panels display the 1D marginal posterior density functions for each coefficient; secondary top axes 
     indicate the distance from the prior mean, normalised by the prior standard deviation. The off-diagonal panels 
     show the 2D joint posterior probability contours (at integer standard deviations) for coefficient pairs. 
     Black crosses mark the true synthetic values, and red crosses indicate the posterior expectations. By 
     embedding gravitationally self-consistent sea-level physics into the inversion, the framework provides 
     indirect, quantitative constraints on these degree-one loads.}    
    \label{fig:grace_degree_one}
\end{figure*}

We have already noted that in practice we need not be concerned with the full posterior distribution, but 
only its push-forward to a low-dimensional property space. Writing again $q = Bm$ for the 
property vector, the error  associated with the resulting Bayesian estimator, $\hat{q} = B\hat{m}$, has expected value
\begin{equation}
    \label{eq:bayesbias}
    \mathrm{E}[\hat{q}-q] = B(KA-1)\Delta \overline{m},
\end{equation}
and covariance
\begin{equation}
    \label{eq:bayescov}
    \mathrm{Cov}[\hat{q}-q] = BQ_{p}B^{*} + B(KA-1)\Delta Q(KA-1)^{*}B^{*}.
\end{equation}
The effect of the prior on the property space is, therefore, controlled by the operator $B(KA-1)$
which  vanishes if the image of $KA-1$ is contained within the kernel of $B$. That this
condition might be met \textit{exactly} is highly unlikely, of course, with the quantities of interest
needing to be  tailored precisely to what the data can tell us. Nevertheless, in favourable
circumstances these subspaces might be sufficiently close that the posterior for the quantities of 
interest  is relatively insensitive to the choice of prior. To be concrete, we  write 
\begin{equation}
    B = \sum_{i=1}^{p} e_{i} \otimes t_{i}, 
\end{equation}
where $\{e_{i}\}_{i=1}^{p}$ is an orthonormal basis for the property space, the $\{t_{i}\}_{i=1}^{p}$
are vectors in the model space which we call target kernels, and $\otimes$ denotes a tensor product. It  follows from eq.~(\ref{eq:bayesbias})  that 
\begin{align}
    \label{eq:priorsense1}
    \|\mathrm{E}[\hat{q}-q]\|_{\mathcal{P}}^{2}  &= \sum_{i=1}^{p} \cbraket{(KA-1)^{*}t_{i}}{\Delta \overline{m}}_{\mathcal{M}}^{2} \\
    &\le \sum_{i=1}^{p} \|(KA-1)^{*}t_{i}\|_{\mathcal{M}}^{2} \|\Delta \overline{m}\|_{\mathcal{M}}^{2}, 
\end{align}
where we have used the Cauchy--Schwarz inequality. This shows that the relative magnitude of the estimator's bias
can be bounded by the norms of the vectors $\|(KA-1)^{*}t_{i}\|_{\mathcal{M}}^{2}$. Through a similar
argument, we find that 
\begin{equation}
    \label{eq:priorsense2}
    \|\mathrm{Cov}[\hat{q}-q] -BQ_{p}B^{*}\|_{\mathrm{HS}(\mathcal{P})} \le \|\Delta Q\|_{\mathrm{HS}(\mathcal{M})}
     \sum_{i=1}^{p} \|(KA-1)^{*}t_{i}\|_{\mathcal{M}}^{2}, 
\end{equation}
where $\|\cdot\|_{\mathrm{HS}(\mathcal{X})}$ denotes the Hilbert--Schmidt norm which is defined for a linear operator, $A$, 
on a space, $\mathcal{X}$, by 
\begin{equation}
    \|A\|_{\mathrm{HS}(\mathcal{X})}^{2} = \sum_{i}\|Ae_{i}\|^{2}_{\mathcal{X}}, 
\end{equation}
with $\{e_{i}\}$ an orthonormal basis \citep[e.g.][]{schechter2001principles}; in the above application, this norm is well-defined because
covariances are required to be trace-class.

It is important to emphasise that the bounds derived in eqs~(\ref{eq:priorsense1}) and (\ref{eq:priorsense2})
are conservative, representing a worst-case scenario that is independent of an assumed prior. Provided the prior 
expectation and its projection onto the property space are non-zero, we can express the bias bound in a relative form
\begin{equation}
    \label{eq:priorsense_bias_rel}
    \frac{\|\mathrm{E}[\hat{q}-q]\|_{\mathcal{P}}}{\|B\overline{m}\|_{\mathcal{P}}} 
    \le \left( \frac{\|\Delta \overline{m}\|_{\mathcal{M}}}{\|\overline{m}\|_{\mathcal{M}}} \right) 
    \left( \frac{\|\overline{m}\|_{\mathcal{M}}}{\|B\overline{m}\|_{\mathcal{P}}} \left( \sum_{i=1}^{p} \|(KA-1)^{*}t_{i}\|_{\mathcal{M}}^{2} \right)^{1/2} \right).
\end{equation}
This demonstrates that the fractional error in the expected property estimate is bounded by the fractional perturbation 
of the prior mean, scaled by a geometric amplification factor. Moreover, we can  write the bound on the 
covariance sensitivity  similarly as
\begin{equation}
    \label{eq:priorsense_cov_rel}
    \frac{\|\mathrm{Cov}[\hat{q}-q] -BQ_{p}B^{*}\|_{\mathrm{HS}(\mathcal{P})}}{
        \|BQ_{p}B^{*}\|_{\mathrm{HS}(\mathcal{P})}
    } \le \left(\frac{ \|\Delta Q\|_{\mathrm{HS}(\mathcal{M})}}{\|Q\|_{\mathrm{HS}(\mathcal{M})}}\right)
\left( \frac{\|Q\|_{\mathrm{HS}(\mathcal{M})}}{\|BQ_{p}B^{*}\|_{\mathrm{HS}(\mathcal{P})}}
     \sum_{i=1}^{p} \|(KA-1)^{*}t_{i}\|_{\mathcal{M}}^{2}\right).
\end{equation}
In both cases, the fractional perturbation of the estimator's error metrics is bounded by the product of the fractional 
error in the assumed prior (the first term on the right) and an amplification factor dictated by the problem's physics 
and geometry (the second term on the right). Within an application, we can calculate these amplification factors
at low cost, and based on their size, assess the robustness of our results to the choice of prior.

\begin{figure*}
    \centering        
    \includegraphics[width=\textwidth]{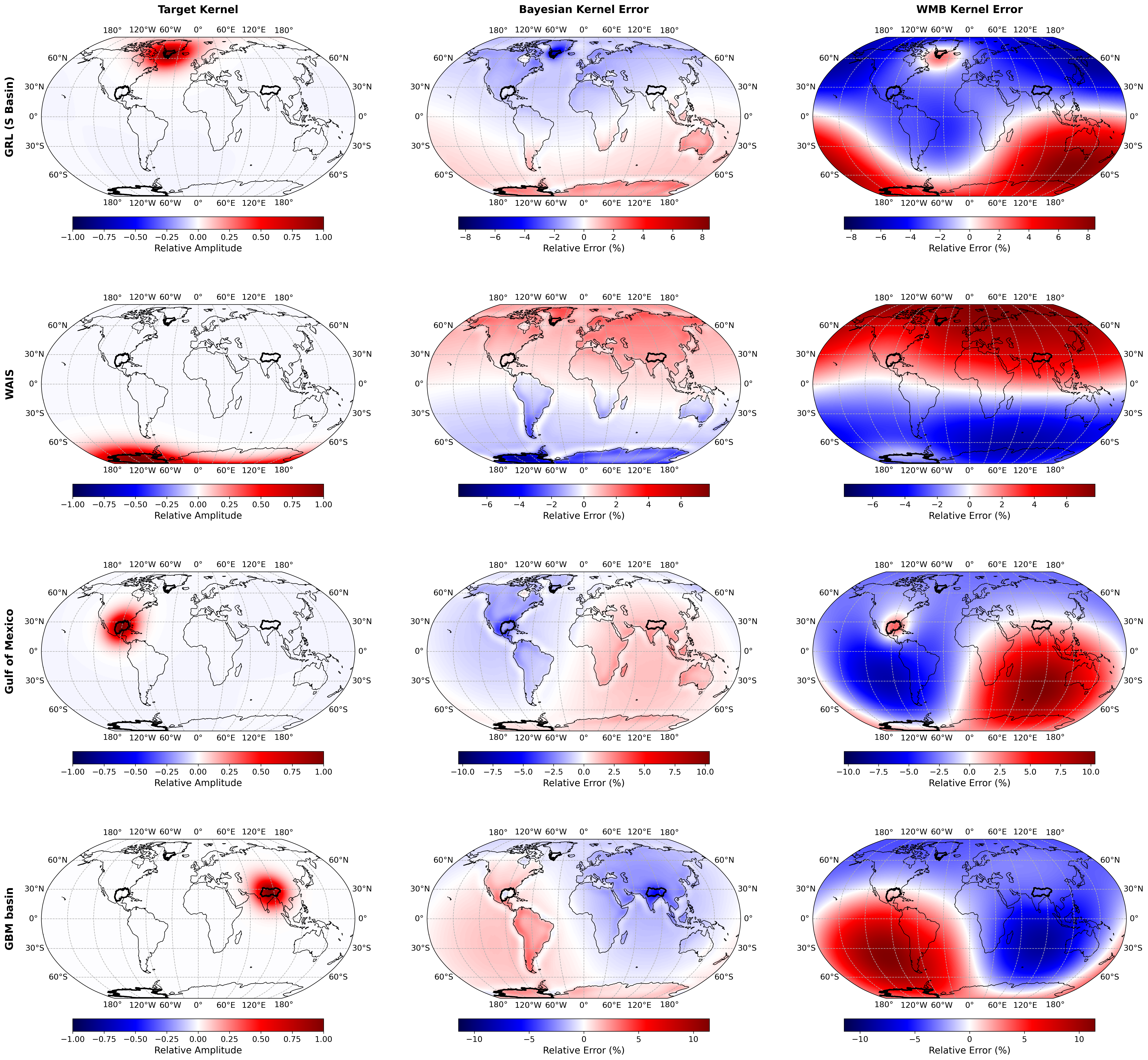}        
    \caption{Spatial comparison of the target and error estimator kernels for load averages of the four regions. 
    The left column displays the target kernels, $t_i$, whose 
    inner products with the direct load yield the regional averages of the total load; these fields are normalised 
    to a maximum pointwise amplitude of one. The middle and right columns show the corresponding error kernels 
    for the Bayesian estimator, $(KA - 1)^*t_i$, and the standard WMB estimator, $(CA - B)^*e_i$, respectively.
    Error kernel amplitudes are expressed as a percentage relative to the target kernel's maximum value.
    The WMB method exhibits pronounced global far-field leakage, which the Bayesian framework better suppresses 
    by fully embedding sea-level physics.}
    \label{fig:grace_kernels}
\end{figure*}

\subsubsection{Optimality of the Bayesian estimator}

\label{sec:optest}

It is worth comparing the term $B(KA-1)$ with the rather
similar operator $CA-B$ that arose within our discussion of the errors within
the WMB and altimetry averaging methods. Recalling that $C$ is the mapping from the data to the estimate,
we see that the Bayesian estimator takes the same basic form. Indeed, pushing eq.~(\ref{eq:postexp}) forward
under $B$, the estimate $\hat{q} = Bm_{p}$ is the affine function
\begin{equation}
    \label{eq:bayesest}
    \hat{q} = Cd + c, \qquad C = BQA^{*}(AQA^{*}+R)^{-1}, \qquad c = B(1-KA)\overline{m} - BK\overline{z},
\end{equation}
of the data, with $C$ providing the required mapping from the data into the property space and $c$ a fixed
translation within it. The standard estimators of Section~\ref{sec:standard} are of this same form, but with
$c = 0$. From this perspective,
the Bayesian approach is then just one of a range of possible affine estimators. It is, however,
notable that the Bayesian estimator depends on the forward operator, $A$, and  so takes account of the full physics of the problem,
while also accounting explicitly for the assumed noise through the covariance, $R$. 

In fact,
one can readily derive the Bayesian estimator by optimising the choice of $C$ and $c$ with respect to the
estimator's performance against samples from the  assumed prior. To make this precise, let $C$ be an arbitrary
continuous linear mapping from the data space into the property space, let $c$ be an arbitrary element of the
latter, and write $\hat{q}(C,c) = Cd + c$ for the associated estimator. On the assumption that the model and
noise are independent, its error, bias, and error covariance are given by
\begin{align}
    \label{eq:affine_error}
    \hat{q}(C,c) - q &= (CA-B)m + Cz + c, \\
    \label{eq:affine_bias}
    \mathrm{E}[\hat{q}(C,c) - q] &= (CA-B)\overline{m} + C\overline{z} + c, \\
    \label{eq:affine_cov}
    \mathrm{Cov}[\hat{q}(C,c) - q] &= (CA-B)Q(CA-B)^{*} + CRC^{*},
\end{align}
the last of these being the analogue of eq.~(\ref{eq:wmb_error}). The translation enters the bias but not
the covariance, of course. It follows that, whatever the choice of $C$, the estimator can be rendered unbiased by setting
$c = -(CA-B)\overline{m} - C\overline{z}$, and this costs nothing in variance. 

Adopting the mean-square error, $J$, as our measure of performance, and using the standard bias-variance decomposition, we can write
\begin{equation}
    \label{eq:mse}
    J(C,c) = \mathrm{E}\|\hat{q}(C,c) - q\|_{\mathcal{P}}^{2}
    = \|\mathrm{E}[\hat{q}(C,c) - q]\|_{\mathcal{P}}^{2} + \trace{\mathrm{Cov}[\hat{q}(C,c)-q]},
\end{equation}
and hence minimise first over $c$ (annihilating the bias term for any $C$), and then over $C$ alone.
Writing $S = AQA^{*}+R$ for the prior data covariance and completing the square in the operator sense, it
follows using eq.~(\ref{eq:kalman}) and eq.~(\ref{eq:postcov}) that
\begin{equation}
    \label{eq:sqcomp}
    \mathrm{Cov}[\hat{q}(C,c)-q] = (C-BK)S(C-BK)^{*} + BQ_{p}B^{*}.
\end{equation}
Because $S$ is positive-definite, the first term on the right is non-negative and vanishes precisely when
$C = BK$, and hence
\begin{equation}
    \label{eq:loewner}
    \mathrm{Cov}[\hat{q}(C,c)-q] \ge BQ_{p}B^{*}
\end{equation}
with the ordering meaning that their difference is a positive semi-definite operator. This is a stronger statement than the minimisation of the
mean-square error alone, which follows from taking the trace. The minimising pair is unique, with $C = BK$ because
$J$ is a strictly convex quadratic on the finite-dimensional space of such mappings. The accompanying
unbiased translation is then $c = B(1-KA)\overline{m} - BK\overline{z}$, thereby recovering eq.~(\ref{eq:bayesest})
in full and yielding $\min J = \trace{BQ_{p}B^{*}}$. It is worth noting that the optimal $C$ factorises as $B$
acting on $K$, so that the best estimator of $q$ is simply $B$ applied to the best estimator of $m$, irrespective
of the quantities of interest; this justifies our practice of pushing the posterior forward under $B$ rather
than designing a bespoke estimator in each case. Finally, note that the restriction to affine estimators entails
no loss of generality here, for $\hat{q}$ coincides with the conditional expectation $\mathrm{E}[q\,|\,d]$,
which minimises the mean-square error over all square-integrable estimators, linear or otherwise \citep[e.g.][]{kallenberg1997foundations}.

The above point of view provides an interesting link with the inference methods pioneered by Backus, Gilbert, and Parker 
\citep[e.g.][]{BackusGilbert1967, BackusGilbert1968,Backus1970a,Parker1977,stark2008generalizing, mag2025bridging}. Indeed, the bounds obtained earlier form a Bayesian 
counterpart to the results of \cite{Backus1970a}, who showed that without suitable prior information, any finite amount of data generically provides no information 
on the quantities of interest. Strictly speaking, the optimality of these Bayesian estimators is relative to the specific prior and noise distributions used to construct 
them. In practice, however, they perform well even when the prior is misspecified, provided it is chosen to span a suitably wide range of the model space so as not 
to artificially suppress the data. Furthermore, we are not required to guess at this sensitivity. The frequentist bounds in eqs~(\ref{eq:priorsense1}) and (\ref{eq:priorsense2}) 
serve as explicit diagnostic tools from which we can determine when a specific property is strongly constrained by the 
data and, consequently, robust against our subjective choice of prior.

\subsection{Empirical Bayes and prior construction}

\label{sec:empirical_bayes}

Within this paper, we take the strict Bayesian view that the prior distribution quantifies knowledge of the model before the data have been collected. 
From this perspective, the choice of prior is a subjective one that cannot be questioned from within the theory, though one can ask about the implications of
different choices. There is, however, the alternative approach of Empirical Bayes \citep[e.g.][]{morris1983parametric}. Within this paradigm, 
the prior is not fixed entirely but is instead chosen from a parameterised family by maximising the marginal likelihood (the Bayesian evidence) of the observed data. 
Evaluating this evidence can be computationally prohibitive for large-scale inverse problems due to the need to compute the log-determinant of the 
prior data covariance. 
To address this, the \texttt{pygeoinf} library natively supports efficient, matrix-free evidence estimation based on Stochastic Lanczos Quadrature (SLQ)
\citep[e.g.][]{ubaru2017fast}. In this manner, data-driven hyperparameter tuning is a practically viable strategy even for high-resolution, 
global-scale geophysical inversions. Beyond Empirical Bayes, this computational machinery can also be used to construct physically motivated priors using independent information, such as outputs from numerical climate models. By treating a set of such model outputs as samples from a parameterised Gaussian random field, one can employ evidence maximisation to rigorously determine suitable values for the underlying hyperparameters \citep[e.g.][]{kennedy2001bayesian, kaufman2010bayesian}.

\section{Synthetic inversions}

Before detailing the synthetic experiments, we emphasise the primary objective of this section. The following inversions of 
simulated satellite gravity and ocean altimetry data are presented to illustrate the computational viability of our infinite-dimensional 
Bayesian framework and to provide concrete examples of the types of analyses it enables. As the data are synthetic and the prior and noise 
distributions idealised, the numerical values obtained are not themselves of interest. Our focus is instead on more general features of the 
inference: the resolution of the associated averaging kernels, the degree of variance reduction achieved, and the sensitivity of the estimators 
to the assumed prior. Such indications are necessarily tentative, but they illustrate that the framework can be used not only to perform an inversion, 
but to assess  which conclusions are likely to be robust when subject to reasonable variations of the prior.

\subsection{Satellite gravity}

\subsubsection{Formulation of the problem}

We begin by considering the inversion of synthetic GRACE/FO observations to estimate surface loads, or 
averages thereof over targeted regions. The formulation of the forward problem and generation of synthetic data 
follows precisely the discussion within Section~\ref{sec:WMBBias}. The same load distribution is then used  as the prior 
within the solution of the Bayesian inverse problem. As noted, such a  choice is favourable to the 
performance of the Bayesian method, with the associated estimator then being unbiased and having an optimal covariance, 
but this circumstance is not relevant to our present aims.

\begin{figure*}
    \centering        
    \includegraphics[width=\textwidth]{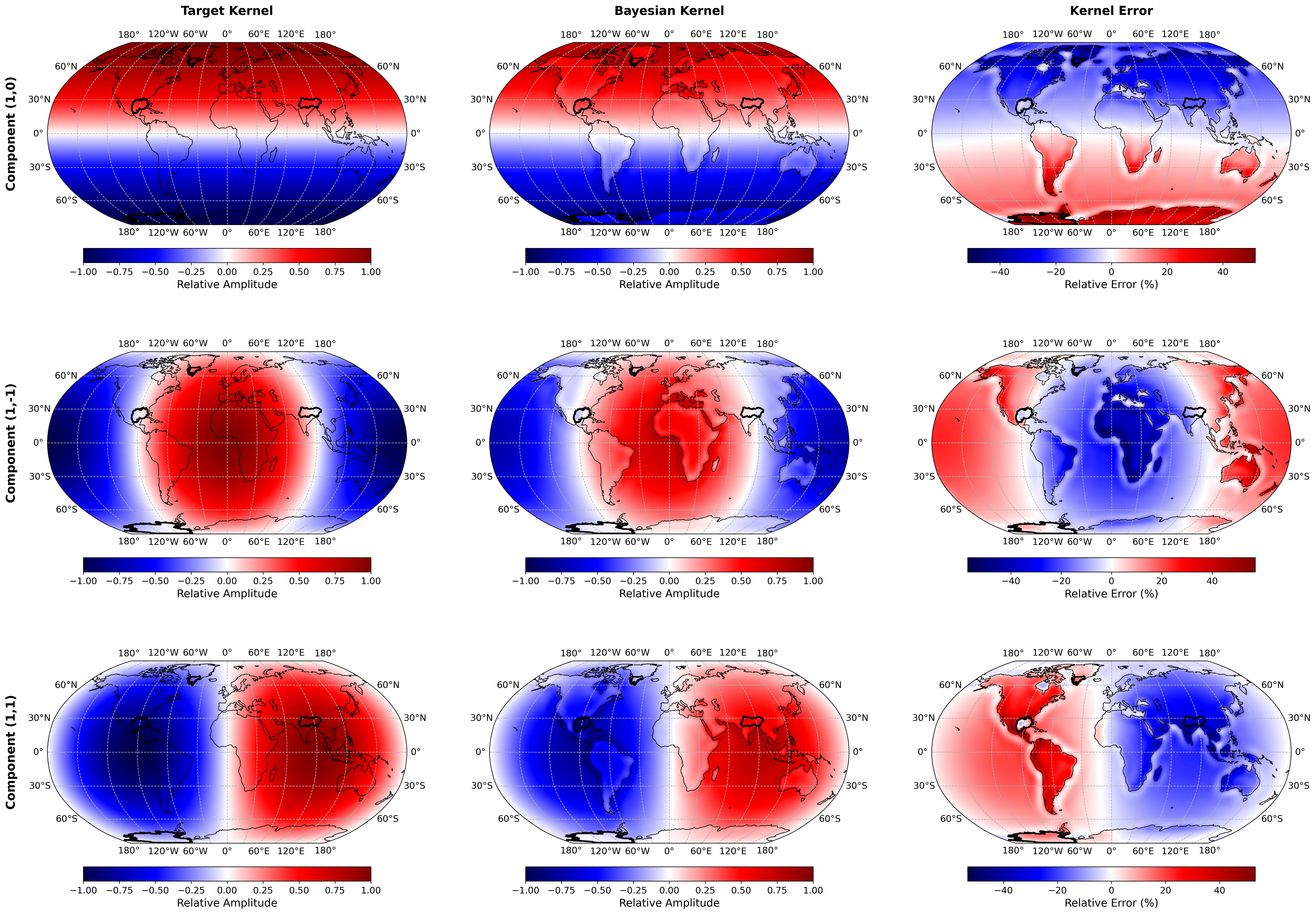} 
    \caption{
    Spatial comparison of the target kernels, Bayesian estimator kernels, and the corresponding error kernels for the 
    three degree-one spherical harmonic components $(1,0)$, $(1,-1)$, and $(1,1)$ of the direct load using synthetic GRACE/FO 
    observations. The left column displays the target kernels, which are proportional to the corresponding real spherical harmonics
    normalised to a maximum amplitude of one. The middle column displays the Bayesian estimator kernels that
    have been similarly normalised. Because satellite gravity 
    observations are inherently insensitive to degree-one loads, these estimator kernels are constructed from a linear combination 
    of the kernels at degrees $2 \le l \le l_{\max}$. The right column displays the error kernels associated 
    with the Bayesian estimators, with amplitudes expressed as a percentage relative to the target kernel's maximum absolute value. 
    The relatively large amplitude of these error kernels indicates that Bayesian estimates of degree-one coefficients are more 
    sensitive to the choice of prior distribution.}    
    \label{fig:grace_deg1_kernels}
\end{figure*}

\subsubsection{The WMB preconditioner}

Within this example, GRACE/FO coefficients are used in the range $2 \le l \le 100$, with the corresponding
data space having a dimension of just over ten thousand. As we have noted, the necessary linear systems are 
best solved using  iterative methods, with the development of a good preconditioner being key to 
the performance. A range of matrix-free preconditioners could be applied to this problem, generic in the sense that they do not depend on the specific forms of the operators involved. There is, however, a simple and effective way of preconditioning this problem inspired by 
the WMB method. 

Let us suppose that the covariance, $Q$, for the prior load distribution
is rotationally invariant, and hence in the spherical harmonic domain its action corresponds to 
a simple degree-dependent scaling, $q_{l}$. Similarly, we suppose that the noise distribution
has a covariance, $R$, whose action  is also given through a simple degree-dependent scaling, $r_{l}$. 
These conditions are met within the problem at hand, but in broader applications where they are not, \textit{approximations} of the true covariances naturally suffice for constructing preconditioners. We can then 
define the action of the WMB preconditioner  on the $(l,m)$th spherical harmonic by
\begin{equation}
    \phi_{lm} \mapsto (q_{l}k_{l}^{2}+r_{l})^{-1} \phi_{lm},
\end{equation}
where the $k_{l}$ are loading Love numbers as in eq.~(\ref{eq:wahr1}). Note that to the same 
level of approximation, the operator, $K = QA^{*}(AQA^{*}+R)^{-1}$, 
is given in the spherical harmonic domain by
\begin{equation}
    \phi_{lm} \mapsto \frac{q_{l}k_{l}}{q_{l}k_{l}^{2}+r_{l}} \phi_{lm}, 
\end{equation}
which reduces to the WMB method in the absence of data errors. The action 
of the WMB preconditioner is cheap to compute, while its setup costs 
are tiny. Nevertheless,  it allows for the 
 linear system to  be solved in about thirty iterations, which is excellent performance for 
 a problem of this size.
 For reference, within these conjugate gradient solutions, the iterations are stopped 
 when the relative norm of the residual is one hundred times smaller than the pointwise noise-to-signal ratio, 
 with the analogous criterion used in all examples below.

\subsubsection{Visual inspection of the posterior expectation}

In Fig.~\ref{fig:grace_maps}, we compare the Bayesian posterior expectation of the  load 
to  its true value and with estimates obtained using the WMB method.  The latter approach maps the 
data to an estimate of the total load, while the Bayesian method returns one for the
 direct load. The posterior expectation can, however, be converted into the corresponding
total load through solution of the sea-level equation, and we similarly 
plot the true total load so that all fields within the figure are directly comparable.
A known issue with the WMB method -- which can be seen prominently within the upper-right panel 
of the figure -- is that  short-wavelength noise is amplified due to the division in eq.~(\ref{eq:wahr1}) by the loading Love numbers, 
whose magnitudes decrease with degree. To address this point heuristically, we  
show also the result of smoothing the WMB estimate over a 500\,km length scale
which  suppresses much of this noise. 
A simple visual comparison of the true total load with the smoothed WMB estimate and the 
Bayesian posterior expectation suggests that both inversion methods  can recover broadly correct 
amplitudes and spatial patterns. In detail, however, the amplitudes and patterns within the posterior expectation are 
closer to the truth, while this estimate is less obviously affected by observational noise and does not 
require ad hoc smoothing.

From a computational perspective, the WMB estimate 
can be obtained in a fraction of a second, with its main cost being a single fast 
spherical harmonic transformation. Calculation of the Bayesian posterior expectation
takes a few minutes on a modern laptop  using the preconditioned iterative method
discussed above.  While the relative increase in computational time is considerable, 
practically speaking the Bayesian estimate can still be obtained quite
quickly.  The comparison  would be very  different had we relied on direct linear solvers, with 
assembly of the prior data covariance in dense form then requiring several hours of CPU time.

\begin{figure*}
    \centering        
    \includegraphics[width=0.8\textwidth]{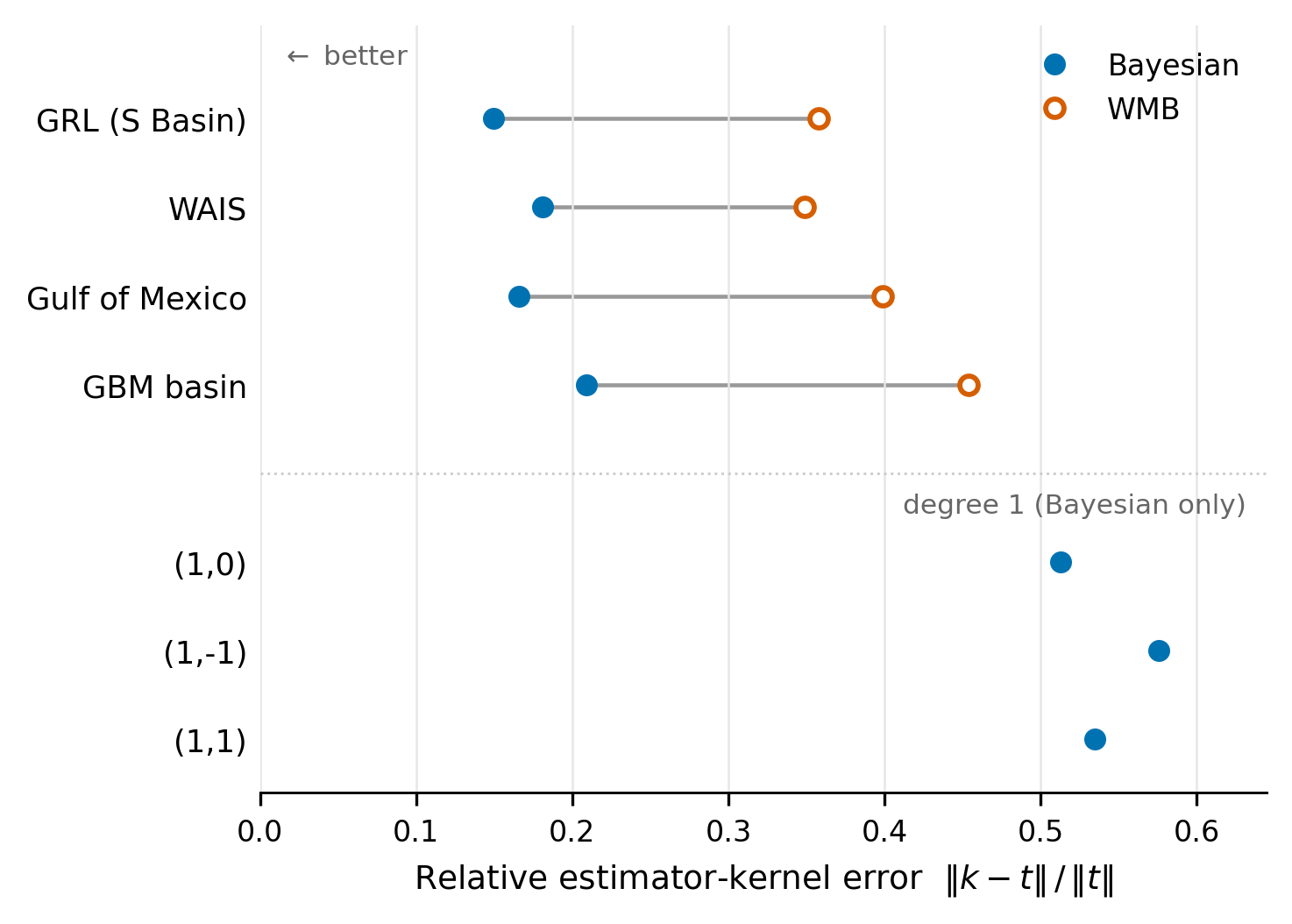} 
    \caption{Relative estimator-kernel error, $\|k-t\|_{\mathcal{M}}/\|t\|_{\mathcal{M}}$, for the quantities of interest considered
    within the inversion of synthetic GRACE/FO observations, 
    where $t$ denotes the target kernel and $k$ that associated with the estimator. Smaller values indicate a more faithful recovery of 
    the intended quantity, and through eqs~(\ref{eq:priorsense_bias_rel}) and (\ref{eq:priorsense_cov_rel}) also imply a weaker dependence of 
     Bayesian estimates on the assumed prior. The upper group shows the four regional load averages, with the Bayesian estimator (filled symbols) 
    and the WMB estimator (open symbols) joined for clarity; the WMB errors are larger by a factor of about two in every case. The lower group shows 
    the degree-one coefficients of the direct load, for which no WMB estimator is available, and for which the errors are markedly larger than for 
    any of the regional averages.}   
    \label{fig:grace_kernal_ratios}
\end{figure*}

\subsubsection{Pointwise variance and covariance functions}

Using the randomise-then-optimise approach within eq.~(\ref{eq:rto}), it is possible 
to produce maps of the pointwise standard deviation for the Bayesian estimate of either 
the direct load or the total load. As discussed, the accuracy of the maps depends on the number of 
samples used. In practice, we find that around one hundred samples are sufficient 
to produce  visually useful results.  With each sample taking a few minutes to generate, a map based on one hundred samples
requires several hours of computation in serial. The samples are, however, mutually
independent, and so this calculation falls within the throughput-oriented regime discussed
in Section~\ref{sec:NumCon}: distributed across the 16 cores of our reference machine, the same map is
produced in a few tens of minutes. We do not
show example standard deviation maps for the GRACE/FO problem because they are not 
visually interesting.  Indeed, the standard deviation is very nearly
constant, which reflects the global coverage of the GRACE/FO data
as well as the fact that the assumed prior and  noise distributions are spatially homogeneous.

A more interesting aspect of uncertainty quantification for this problem comes from a
comparison of the prior and posterior covariance functions for some representative
locations. This can be seen within Fig.~\ref{fig:grace_cov}, which includes an 
ocean location, one relevant to land water storage, and one for an ice sheet. 
The prior covariance functions are very similar relative to their reference points, 
though they are not precisely equal because they have been calculated for the total 
load, and so include the effects of gravitationally self-consistent
sea-level change. For each location  the posterior covariance functions 
 have much richer spatial forms.  In each case the 
value of the covariance at the reference point (i.e., the variance at these locations) 
has reduced by more than a factor of ten. This shows the significant information gain 
over the prior through the assimilation of the GRACE/FO observations. The posterior variances 
for each location are approximately equal, which is consistent 
with the sampled standard deviation maps discussed above. The peaks of the 
posterior covariance functions at their reference points are also notably sharper than those of the priors,
 showing that the inversion has been able to resolve spatial length scales
finer than those  assumed in the prior distribution.
Furthermore, the posterior covariance maps reveal 
prominent long-range spatial structures that are  absent in the prior. These far-field correlations are 
driven primarily by the lack of  degree-one observations. Indeed, the calculations can be repeated with 
the degree-one component removed from the prior, 
with the result that the posterior covariance functions are localised tightly about the chosen reference point. 

\begin{figure*}
    \centering
    \includegraphics[width=\textwidth]{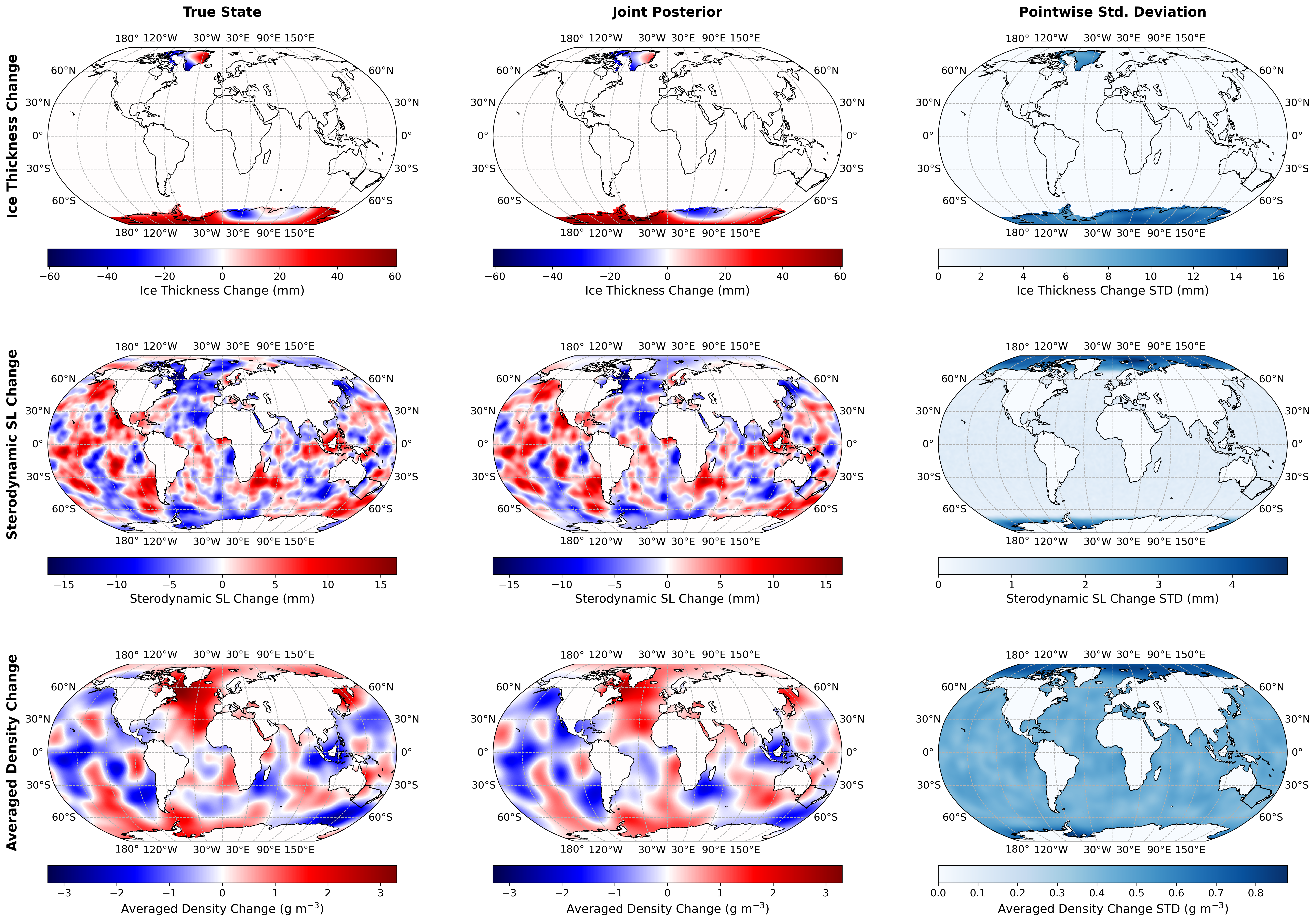}
    \caption{Comparison of the true synthetic model state (left column) with the posterior expectation of the
    joint inversion (middle column), for the ice thickness change (top), sterodynamic sea-level change (middle),
    and vertically averaged density change (bottom). The right column shows maps of the pointwise posterior
    standard deviation, obtained via the randomise-then-optimise method. The sterodynamic uncertainty rises
    towards its prior value poleward of the $\pm66^{\circ}$ altimetry limit, while the density change remains
    the least constrained of the three fields.}
    \label{fig:joint_posterior_maps}
\end{figure*}

\subsubsection{Regional load averages}

In Section~\ref{sec:WMBBias}, we introduced a linear operator, $B$, that maps the direct load, $\zeta$, 
to the averages of the total load over a set of $p$ geographic regions. The posterior distribution
can be pushed forward onto the associated property space, with the dense form of the 
resulting covariance, $BQ_{p}B^{*}$, obtained in just $p$ actions of $Q_{p}$. In 
Fig.~\ref{fig:grace_pdfs}, we show the marginal posterior PDFs for the regions  
considered earlier along with the corresponding WMB estimates. These results 
 are for a single synthetic inversion and so are not necessarily representative.
But consistent with Section~\ref{sec:WMBBias}, we  see that the 
WMB method does indeed suffer from significant biases while systematically underestimating uncertainties. By contrast, the 
Bayesian method arrives at larger but more realistic uncertainties so long as the choice of prior is 
justified; here, of course, we are generating the data using the prior and hence this 
condition is automatically satisfied.

For each region, we can determine the Kullback--Leibler (KL) divergence of the
posterior from the prior, which provides a measure of the information gained
through the inversion \citep[e.g.][]{sanz2023inverse}. For Gaussian
distributions on the real line this quantity takes a simple form which is helpful in interpreting
the results. Writing the prior and posterior marginals for a scalar property
$q$ as $\mathcal{N}(\mu_{0},\sigma^{2})$ and $\mathcal{N}(\mu_{1},\tau^{2})$,
respectively, and setting $r = \tau^{2}/\sigma^{2}$ for the ratio of the
posterior to the prior variance, we have
\begin{equation}
D_{\mathrm{KL}}\!\left(\mathrm{post}\,\|\,\mathrm{prior}\right)
= \tfrac{1}{2}\left(r - 1 - \ln r\right)
\; + \; \frac{(\mu_{1}-\mu_{0})^{2}}{2\sigma^{2}},
\label{eq:kl-gauss}
\end{equation}
in natural units of information (nats). The first term depends only on $r$ and measures the sharpening of the
distribution; the second measures the displacement of the posterior mean in
units of the prior standard deviation. The distinction matters because a distribution may be moved without being 
narrowed; while a shifted mean conveys information about the parameter's value, a lack of narrowing indicates that the 
posterior uncertainty is still entirely dictated by the prior.

\begin{figure*}
    \centering
    \includegraphics[width=\textwidth]{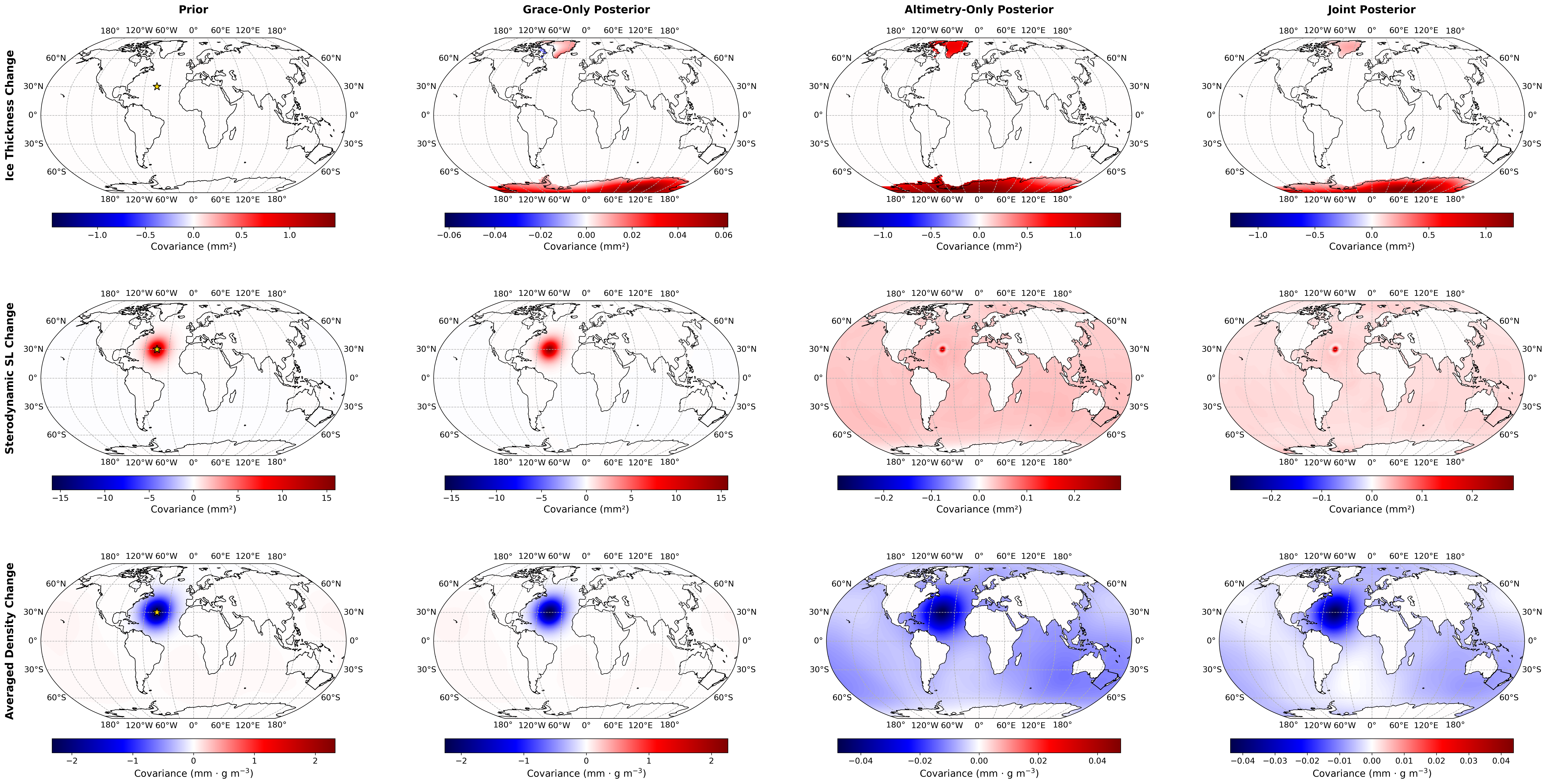}
    \caption{Spatial covariance functions evaluated for the sterodynamic sea-level change at a reference location in the North 
    Atlantic (marked by a star). Rows show the covariance of this quantity with the global ice thickness (top), sterodynamic sea-level 
    (middle), and vertically averaged density (bottom) fields. Columns compare the prior with the GRACE-only, altimetry-only, and joint 
    posteriors. Note the differing colour scales between panels.  
  }
    \label{fig:cov_ocean}
\end{figure*}

The values obtained lie between $1.26$ and $3.84$ nats, so the posterior differs
substantially from the prior in every case. More telling is that this is
accompanied by variance reductions in the range $94$--$96$\%, giving
$r \approx 0.05$. If the posterior uncertainty were still dominated by the prior, one would expect $r \approx 1$, meaning any KL divergence would have to arise almost entirely from a displaced mean (the second term in Eq.~\ref{eq:kl-gauss}). Instead, the fact that the data constrain these load averages to such a small fraction of their prior variance suggests that the final estimates are fundamentally driven by the data, although this remains a qualitative indication rather than a formal demonstration.

The main point of this example is not, however, to rigorously assess the performance of the 
Bayesian method for estimating load averages, with the full answer to this question requiring, at a minimum, 
the consideration of realistic  observational uncertainties. Rather, we primarily wish to demonstrate that   
load average estimates
with full uncertainty quantification can be obtained  from GRACE/FO observations at acceptable computational cost.
Indeed, within the present case, just five actions of the posterior covariance, $Q_{p}$, are 
required. Each action takes around one minute running 
in serial, while the four calculations needed to determine $BQ_{p}B^{*}$ in
dense form can be trivially parallelised, so that the whole calculation takes about two minutes
on our reference laptop.
As with the case of the posterior expectation, while the relative increase in computational cost  is large compared to 
the WMB method, the absolute costs remain manageable  for practical 
applications, for which a few tens of regions are likely to be analysed at any one time.

\subsubsection{Degree-one recovery}

As noted previously, satellite gravity observations are inherently insensitive to degree-one loads, which correspond to geocentre motion.
Standard estimators must either omit these terms entirely or rely on supplementary oceanographic and geophysical models. However, the 
integration of gravitationally self-consistent sea-level physics into the problem couples spherical harmonic degrees, and so provides an 
indirect constraint on these otherwise unobservable components \citep{riva2010sea}.

\begin{figure*}
    \centering
    \includegraphics[width=\textwidth]{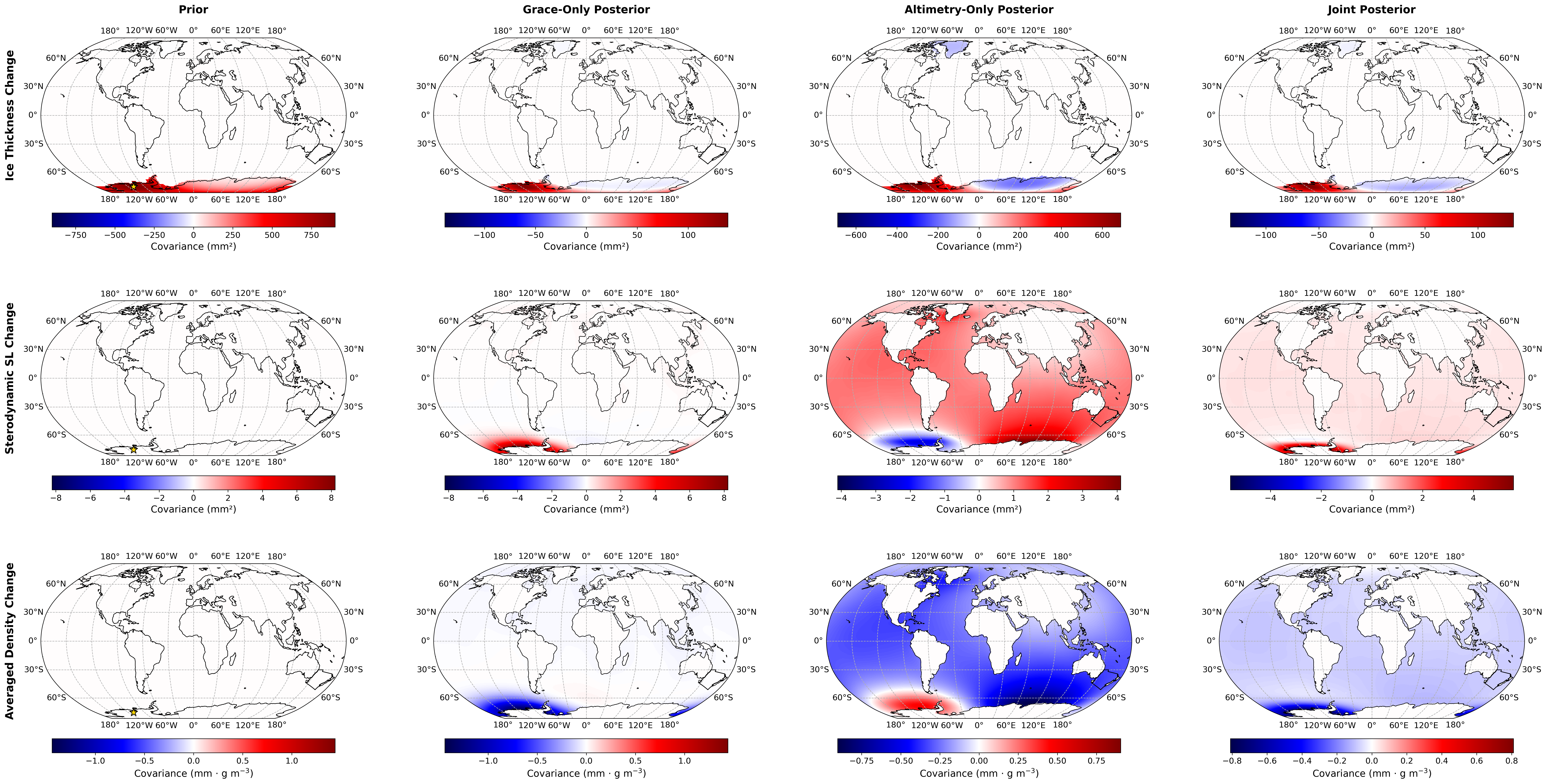}
    \caption{As Fig.~\ref{fig:cov_ocean}, but for the ice thickness change at a reference location on the West
    Antarctic Ice Sheet. }
    \label{fig:cov_ice}
\end{figure*}

Fig.~\ref{fig:grace_degree_one} illustrates this recovery, displaying the 1D marginal posterior density functions and 2D joint contours for 
the degree-one coefficients of the direct load ($\zeta_{1,-1}$, $\zeta_{1,0}$ and $\zeta_{1,1}$). In each case the posterior is consistent with 
the true synthetic value, though as the target field is itself a sample from the prior, this is to be expected of any correct implementation. It is 
the reduction in variance, though, that shows the data to have been informative. The individual reductions are 70.5\% for $\zeta_{1,-1}$, 76.9\% 
for $\zeta_{1,0}$ and 75.0\% for $\zeta_{1,1}$, with a joint Kullback--Leibler divergence of 1.97 nats. Measured against the rule of thumb of order 
$n$ nats in $n$ dimensions, this falls short of what would be regarded as a substantial information gain, and the constraints at degree one are 
accordingly weaker than those on the properties considered above. They are nonetheless quantitative estimates with attached uncertainties, which the WMB 
method cannot provide. Consistent with this, we noted earlier in Fig.~\ref{fig:grace_cov} a distinct degree-one signal within the posterior covariance 
function of the total load, which illustrates how uncertainties at degree one are mapped into other quantities of interest.

Our approach does not diminish the value of  independent constraints on degree-one loads, such as those derived from satellite laser ranging or 
global GPS networks; rather, it complements existing methods for degree-one recovery. Previous studies have inferred these otherwise unobservable coefficients 
by algebraically combining GRACE observations with external ocean bottom pressure models \citep[e.g.][]{chambers2004preliminary,swenson2008estimating,sun2016optimizing}. 
The present formalism offers a probabilistic statement of the same idea. Rather than combining these inputs in a deterministic post-processing step, the output of oceanographic 
models can be used to construct physically informed priors for ocean dynamics (cf. Section~\ref{sec:empirical_bayes}), so that the geophysical constraints introduced in these earlier studies enter the inversion directly 
and their uncertainties are propagated through to the degree-one estimates.

\subsubsection{Prior sensitivity}

We now consider the issue of prior sensitivity within the Bayesian inversion of GRACE/FO data. Due to the relatively low computational costs 
required to determine the posterior expectation or to push forward the posterior distribution to a low-dimensional property space, it is possible 
to repeat such calculations multiple times using a range of priors to gauge the effects on the posterior. We do not, however, present such results here, with such an
analysis being better placed within a real inversion. Instead, we illustrate practically the methods in Section~\ref{sec:freq} that regard the Bayesian 
posterior expectation as an estimator, and use this as a basis for assessing prior sensitivity.

We begin by considering the estimation of regional averages of the total load, recalling that the operator $B$ maps the direct load into a set of $p$ such averages. 
This operator takes precisely the form assumed within Section~\ref{sec:priorsense}, with the average over the $i$th region expressed through an inner product of the direct load with a 
target kernel, $t_{i}$. The kernels for the four regions considered are shown in the first column of Fig.~\ref{fig:grace_kernels}. The next column shows the difference, 
$(KA-1)^{*}t_{i}$, between the kernel associated with the Bayesian estimator and the target kernel, normalised by the maximum pointwise absolute value of the latter field. 
Through eqs~(\ref{eq:priorsense_bias_rel}) and (\ref{eq:priorsense_cov_rel}), we established that the norms $\|(KA-1)^{*}t_{i}\|_{\mathcal{M}}$ can be used to bound the sensitivity of 
the Bayesian solution to prior perturbations. The relative norms $\|(KA-1)^{*}t_{i}\|_{\mathcal{M}}/\|t_{i}\|_{\mathcal{M}}$ for the regions lie in the range 15--21\%, and are summarised 
alongside the corresponding values for the WMB estimator in Fig.~\ref{fig:grace_kernal_ratios}. As these bounds do not vanish, the load averages are not guaranteed to be independent of the prior, 
and this choice cannot be discounted. The bounds in eqs~(\ref{eq:priorsense_bias_rel}) and (\ref{eq:priorsense_cov_rel}) are, however, necessarily conservative, and values of this size suggest 
that the sensitivity is modest in practice. Repeating the calculations with a range of priors, in the manner discussed above, confirms that this is so.

For the WMB method, we can similarly associate a kernel with the estimator for each region \citep[cf.][]{al2024reciprocity}. Specifically, the difference between the WMB kernel and the target kernel 
for the $i$th region takes the form $(CA-B)^{*}e_{i} = A^{*}C^{*}e_{i} - t_{i}$, where $e_{i}$ is the $i$th vector within an orthonormal basis for the property space. The final column of 
Fig.~\ref{fig:grace_kernels} shows examples of this difference, while the relative norms for each of the regions considered span the range 35--40\%, roughly twice those of the corresponding Bayesian 
estimators. Visually, the WMB kernel differences are dominated by a degree-one signal and exhibit pronounced far-field leakage.

These norms admit a simple interpretation. Writing $k_{i}$ for the kernel associated with an estimator, the error in the $i$th regional average for a model $m$ is, in the absence of observational noise, 
equal to $(k_{i}-t_{i}, m)_{\mathcal{M}}$, and hence bounded by $\|k_{i}-t_{i}\|_{\mathcal{M}}\|m\|_{\mathcal{M}}$. This bound holds uniformly over all models of a given size, with equality only when $m$ 
is aligned with the kernel difference itself. It would therefore be possible to contrive a model that favoured one estimator over the other, but for sources not selected with this in mind the smaller 
kernel differences of the Bayesian estimator should translate into more accurate estimates. Note, finally, that these same norms appear within the bounds of eqs~(\ref{eq:priorsense_bias_rel}) 
and (\ref{eq:priorsense_cov_rel}). This is no coincidence: the operator $KA-1$ measures the extent to which the property of interest is not determined by the data alone, and what the data do not 
determine must be supplied by the prior. Fidelity of the estimator kernel and insensitivity to the choice of prior are, in this sense, two aspects of the same thing.

\begin{figure*}
    \centering
    \includegraphics[width=\textwidth]{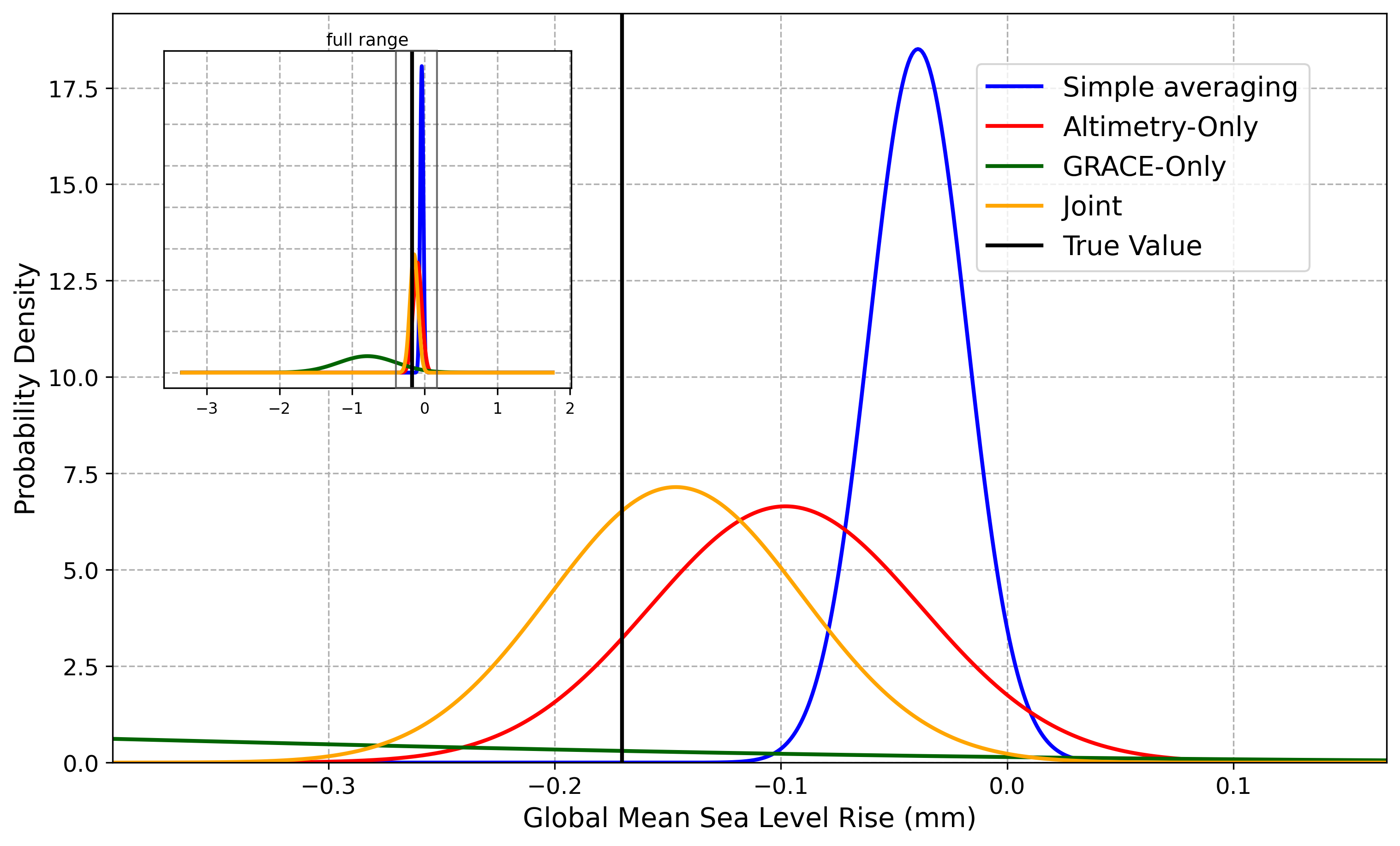}
    \caption{Probability density functions for GMSLR under the simple latitude-weighted average of the altimetry
    data and the altimetry-only, GRACE-only, and joint posteriors; the vertical line marks the true synthetic
    value, and the inset shows the full range including the broad GRACE-only distribution. The averaging method
    is overconfident, its nominal uncertainty excluding the truth. The altimetry-only and joint posteriors give
    variance reductions of 99.3 and 99.4\% relative to the prior, respectively, while GRACE/FO alone
    achieves 69.9\%: the gravity data constrain the barystatic part of GMSLR but not its steric part
    (cf. Fig.~\ref{fig:joint_gmslr_split}).}
    \label{fig:joint_gmsl}
\end{figure*}

Finally, we can apply this same analysis to the recovery of the degree-one coefficients. Fig.~\ref{fig:grace_deg1_kernels} shows the target kernels, the Bayesian estimator kernels, 
and the differences between them for all three degree-one components. As noted previously, satellite gravity observations are inherently insensitive to degree-one loads, and so the 
estimator kernels must be assembled from the kernels of the individual data, each of which acquires a degree-one component only through the physical effect of induced water loading. 
The resulting relative norms, summarised in Fig.~\ref{fig:grace_kernal_ratios}, lie between 51 and 58\%, some three times larger than for the regional load averages. Bounds of this size
are too weak to establish robustness, and we cannot conclude that the degree-one estimates are determined largely independently of the prior; the more modest variance reductions reported 
above point in the same direction. 
An identical analysis of the degree-two components, by contrast, yields relative errors of 4-6\%, so for these coefficients, which are constrained directly 
by observations rather than only through the sea-level physics, estimates are likely to be largely insensitive to the prior.
The situation at degree one may be improved either by assimilating genuinely independent observations, such as those from satellite laser ranging, or by integrating outputs from ocean bottom pressure models directly into the prior covariance structure—effectively placing the approaches of \citet{swenson2008estimating} and \citet{sun2016optimizing} within a rigorous probabilistic setting.

\subsection{Ocean altimetry and joint inversions}

\subsubsection{Formulation of the problem}

We now consider the joint inversion of GRACE/FO and ocean altimetry observations for the three-component model space introduced in Section~\ref{sec:AltimetryBias}, which comprises perturbations to sterodynamic sea-level, 
vertically averaged ocean density, and ice thickness. The construction of the joint forward problem follows readily from the earlier discussions, with the support in \texttt{pygeoinf} for direct sums and block operators making 
its numerical implementation straightforward. The data covariance is taken to be block-diagonal, so that the gravity and altimetry errors are independent, while the same model distribution is used both in generating the synthetic 
data and as the prior. Alongside the joint inversion, we can also invert each data set individually, and in this manner gauge the information gained through their combination.

\subsubsection{The surrogate Woodbury preconditioner}

Within the inversion of ocean altimetry observations, the geocentric sea-level change is sampled on a $1^{\circ} \times 1^{\circ}$ grid, leading to a data space of dimension 34\,514. 
Taken together with the GRACE/FO coefficients for $2\le l \le 100$, the data space has total dimension 44\,711. To precondition the full prior data covariance, $AQA^{*} + R$, there is no simple, 
physically motivated approach akin to the WMB preconditioner used for satellite gravity alone. Instead, we implement a matrix-free method that we term the surrogate Woodbury (SW) preconditioner. 
This could be combined with the WMB preconditioner in a block-diagonal fashion, with the SW preconditioner applied only within the altimetry sub-space, but we find no advantage in doing so: applied to the gravity 
data alone, the SW preconditioner substantially outperforms the WMB one, reducing the number of iterations per solution to between 10 and 15 for the problems considered. Its set-up cost is higher, but as this 
is incurred once and the resulting preconditioner is typically reused across many solves, the saving in iterations is decisive.

The starting point is the Woodbury matrix identity which, on the assumption that $R$ and $Q$ are invertible, allows us to write
\begin{equation}
(AQA^{*} + R)^{-1} = R^{-1} - R^{-1}A(A^{*}R^{-1}A + Q^{-1})^{-1}A^{*}R^{-1}.
\end{equation}
In this manner, the action of the inverse prior data covariance is expressed in terms of the solution of a related linear system, $(A^{*}R^{-1}A + Q^{-1})^{-1}$, posed on the model space. 
Within the exact inverse problem the model space is infinite-dimensional, while numerically it is approximated by a space whose dimension is typically higher than that of the data space. 
This identity therefore replaces the original problem with a larger one, and thus constructing the preconditioner in this way is of no benefit. Suppose, however, that we introduce a surrogate 
forward operator, $A_{s}$, acting on a significantly lower-dimensional approximation of the model space, together with surrogates $Q_{s}$ and $R_{s}$ for the prior and noise covariances. 
We can then define our preconditioner as
\begin{equation}
P = R_{s}^{-1} - R_{s}^{-1}A_{s}(A_{s}^{*}R_{s}^{-1}A_{s} + Q_{s}^{-1})^{-1}A_{s}^{*}R_{s}^{-1}.
\label{eq:swp}
\end{equation}
A surrogate noise covariance is used because the true covariance, $R$, may be either singular -- which would preclude the use of the Woodbury identity entirely -- or computationally prohibitive to invert. The surrogate model space is chosen to be small enough that 
the term $(A_{s}^{*}R_{s}^{-1}A_{s} + Q_{s}^{-1})^{-1}$ can be calculated explicitly using Cholesky factorisation or a similar direct method. The formation of the necessary dense matrix 
on this surrogate space comprises independent calculations, and hence constitutes a further instance of the embarrassingly parallel structure discussed in Section~\ref{sec:NumCon}, 
allowing the set-up costs to be minimised.

To form these surrogates for the joint problem, we make a series of simplifications. For the forward operator $A_{s}$, we reduce the 
spherical harmonic truncation degree from 256 to 32, limit the solution of the sea-level equation to a single iteration, 
and remove the gravitational, rotational, and deformational (GRD) contribution to the geocentric sea-level change. For the noise covariance, we retain the diagonal contributions but 
discard the spatially correlated component of the altimetry errors, so that $R_{s}$ is diagonal and its inverse can be computed explicitly. 
The prior covariance $Q$ is singular, owing to the spatial projections onto the oceans or ice sheets and the mass conservation constraint 
imposed on ocean dynamics; for the surrogate, we strip away these projections, as well as the mass constraint, to ensure invertibility. More generally, 
where such simple forms are unavailable, sparse or low-rank approximations can be used to arrive at a suitable form for $R_{s}^{-1}$ and $Q_{s}^{-1}$.

The result of this process is a very poor physical approximation to the true forward problem, but nevertheless a good preconditioner, whose set-up costs are modest. 
Using this approach, solutions to the joint problem are typically obtained in around 60 iterations, to the altimetry-only problem  in about 55, and to the GRACE/FO problem in 10--15.
These figures vary somewhat between individual solves, and can be reduced by increasing the order of the surrogate, though at the cost of 
additional set-up time. It is notable that the omission of the correlated component of the altimetry errors from $R_{s}$ does not appreciably degrade this performance.
That the joint problem costs only a few more iterations than the altimetry problem alone reflects the fact that the surrogate is built from a single forward operator acting 
on the shared model space, so that the coupling between the two data types is retained, if only approximately. A block-diagonal preconditioner discards this coupling entirely, 
and performs appreciably worse as a result.

\begin{figure*}
    \centering
    \includegraphics[width=0.8\textwidth]{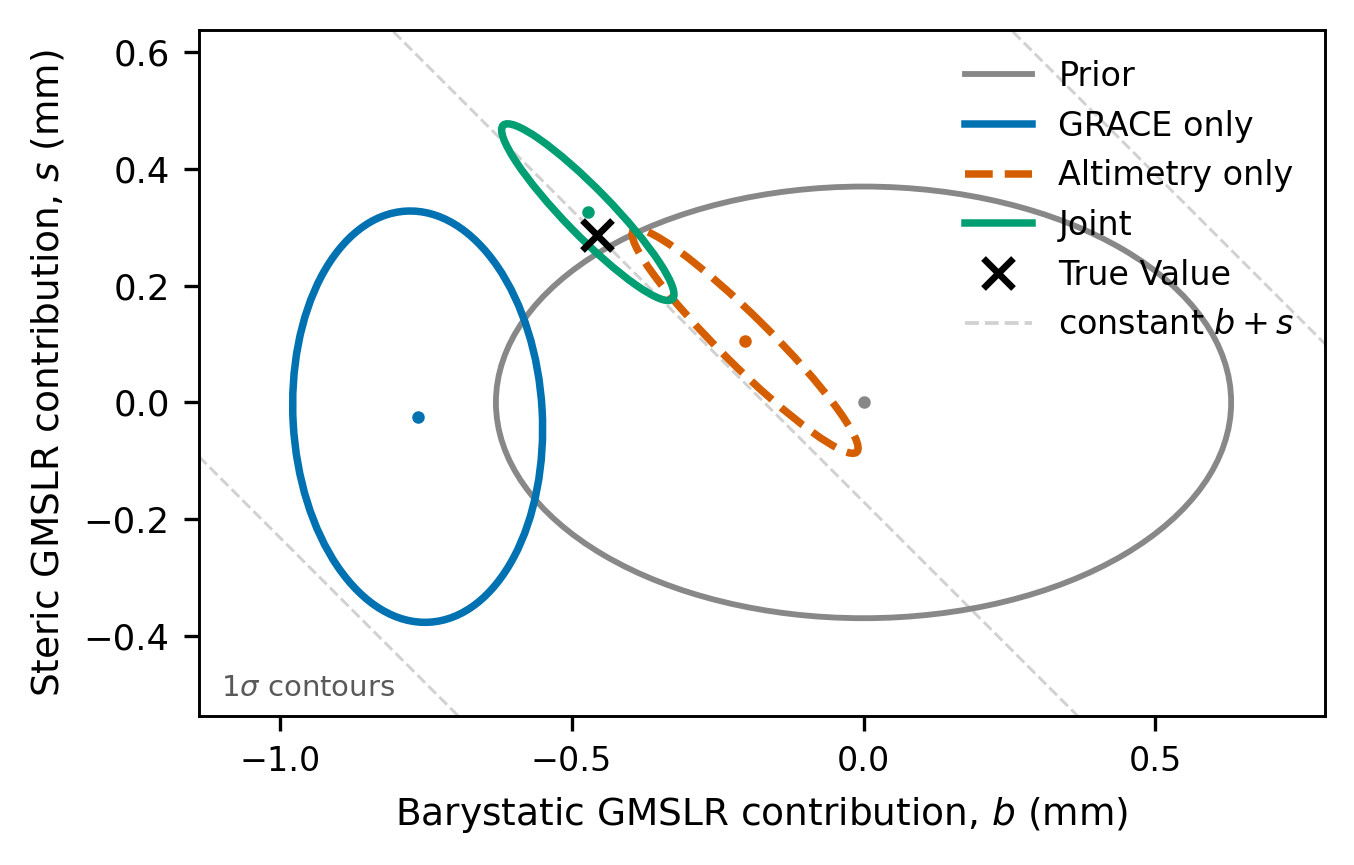}
    \caption{One-standard-deviation contours in the plane of the barystatic ($b$) and steric ($s$) contributions
    to GMSLR, for the prior and the three posteriors, with posterior expectations marked by dots and the true
    value by a cross; dashed lines indicate constant $b + s$. The altimetry-only contour is elongated along
    these lines: the total is tightly constrained while its partition is not, with a correlation of $-0.95$.
    The GRACE-only contour is narrow in $b$ alone; because GRACE does not directly observe the steric effect, 
    any narrowing in $s$ occurs only indirectly through correlated priors. The joint posterior is, in effect, their intersection,
    though it retains a correlation of $-0.93$, the sum remaining far better determined than the difference.}
    \label{fig:joint_gmslr_split}
\end{figure*}

\subsubsection{Decomposition of the sea-level field}
\label{sec:sl_decomp}

Before presenting the results, we define the quantities of interest onto which the prior and posterior distributions will be projected. We first note that the total relative sea-level change within the model is given by the sum $\eta_{d} + \xi$. Following \cite{gregory2019concepts}, we separate this total change into a steric part (associated with density variations at a fixed mass) and a manometric part (associated with changes in the mass of the water column per unit area). The steric sea-level change (SSLC) is defined as

\begin{equation}
    \eta_{s} = -\frac{\eta_{0}}{\rho_{w}}\, 1_{\mathcal{O}}\, \delta\rho_{w}.
    \label{eq:steric_sl}
\end{equation}

This scaling ensures that the mass of each water column remains pointwise unchanged, meaning $\rho_{w}\eta_{s} + \eta_{0}\,\delta\rho_{w} = 0$ over the oceans; consequently, the steric contribution applies no direct load and induces no GRD response. The ocean component of the direct load in eq.~(\ref{eq:alt_load}) is therefore carried entirely by the manometric remainder of the sterodynamic field: $1_{\mathcal{O}}(\rho_{w}\eta_{d} + \eta_{0}\,\delta\rho_{w}) = \rho_{w} 1_{\mathcal{O}}(\eta_{d} - \eta_{s})$. 

Next, we express the relative sea-level change generated by a given direct load, $\zeta$, as the linear mapping $\xi[\zeta]$. By partitioning the total load from eq.~(\ref{eq:alt_load}) into its independent ice and ocean components ($\zeta = \zeta_{i} + \zeta_{o}$), such that $\xi = \xi[\zeta_{i}] + \xi[\zeta_{o}]$, we can define the constituent sea-level terms:

\begin{equation}
    \mathrm{SSLC} = \eta_{s}, \quad
    \mathrm{DMSLC} = (\eta_{d} - \eta_{s}) + \xi[\zeta_{o}], \quad
    \mathrm{BMSLC} = \xi[\zeta_{i}].
    \label{eq:sl_decomp}
\end{equation}

By construction, these components sum exactly to the total relative change, such that $\mathrm{RSLC} = \mathrm{SSLC} + \mathrm{DMSLC} + \mathrm{BMSLC}$. The \textit{dynamic} manometric sea-level change (DMSLC) comprises the redistribution of the ocean's own mass together with the GRD response to that redistributed load, while the \textit{barystatic} manometric sea-level change (BMSLC) is the sea-level response to the ice load, including its GRD fingerprint. By introducing these two terms, we extend the terminology of \cite{gregory2019concepts}, explicitly separating the manometric change based on whether the loading originates internally within the oceans or is driven by the transfer of mass from the land.

When averaged over the oceans, this decomposition conveniently separates the physical drivers of global-mean sea-level rise (GMSLR). Let us denote the spatial average of a generic field $f$ over the oceans by $\langle f \rangle_{\mathcal{O}} = A_{\mathcal{O}}^{-1}\int_{\partial M} 1_{\mathcal{O}}\, f \,\dS$. The constraint of eq.~(\ref{eq:ocean_mass}) 
gives $\langle \eta_{d} - \eta_{s}\rangle_{\mathcal{O}} = 0$, while conservation of mass within the sea-level equation fixes the ocean average of $\xi[\zeta]$ through the net mass of the load, so that $\langle \xi[\zeta_{o}]\rangle_{\mathcal{O}}$ 
also vanishes. Hence $\langle \mathrm{DMSLC}\rangle_{\mathcal{O}} = 0$: ocean dynamics redistribute sea level but cannot alter its global mean.

With the dynamic contribution vanishing globally, GMSLR reduces strictly to its steric and barystatic components. The ocean average of the steric change yields the global-mean steric rise, $\langle \mathrm{SSLC}\rangle_{\mathcal{O}} = s$, while the average of the barystatic change yields the total barystatic rise:

\begin{equation}
\langle \mathrm{BMSLC}\rangle_{\mathcal{O}} = b = -\frac{\rho_{i}}{\rho_{w} A_{\mathcal{O}}} \int_{\partial M} (1 - 1_{\mathcal{O}})\, \eta_{i} \,\dS.
\label{eq:bary}
\end{equation}

It immediately follows that $\mathrm{GMSLR} = b + s$ \citep[cf.][eq.~43]{gregory2019concepts}. In practice, we evaluate $b$ directly through eq.~(\ref{eq:bary}) and determine $s$ as the residual $\mathrm{GMSLR} - b$. Crucially for the inversion process, every quantity defined above -- whether pointwise, regionally averaged, or global -- is a continuous linear functional of the model vector, and so prior and posterior push-forwards onto any collection of them are Gaussian, 
with moments obtained through a small number of covariance actions in the manner of Section~\ref{sec:bayes}.

\begin{figure*}
    \centering
    \includegraphics[width=0.8\textwidth]{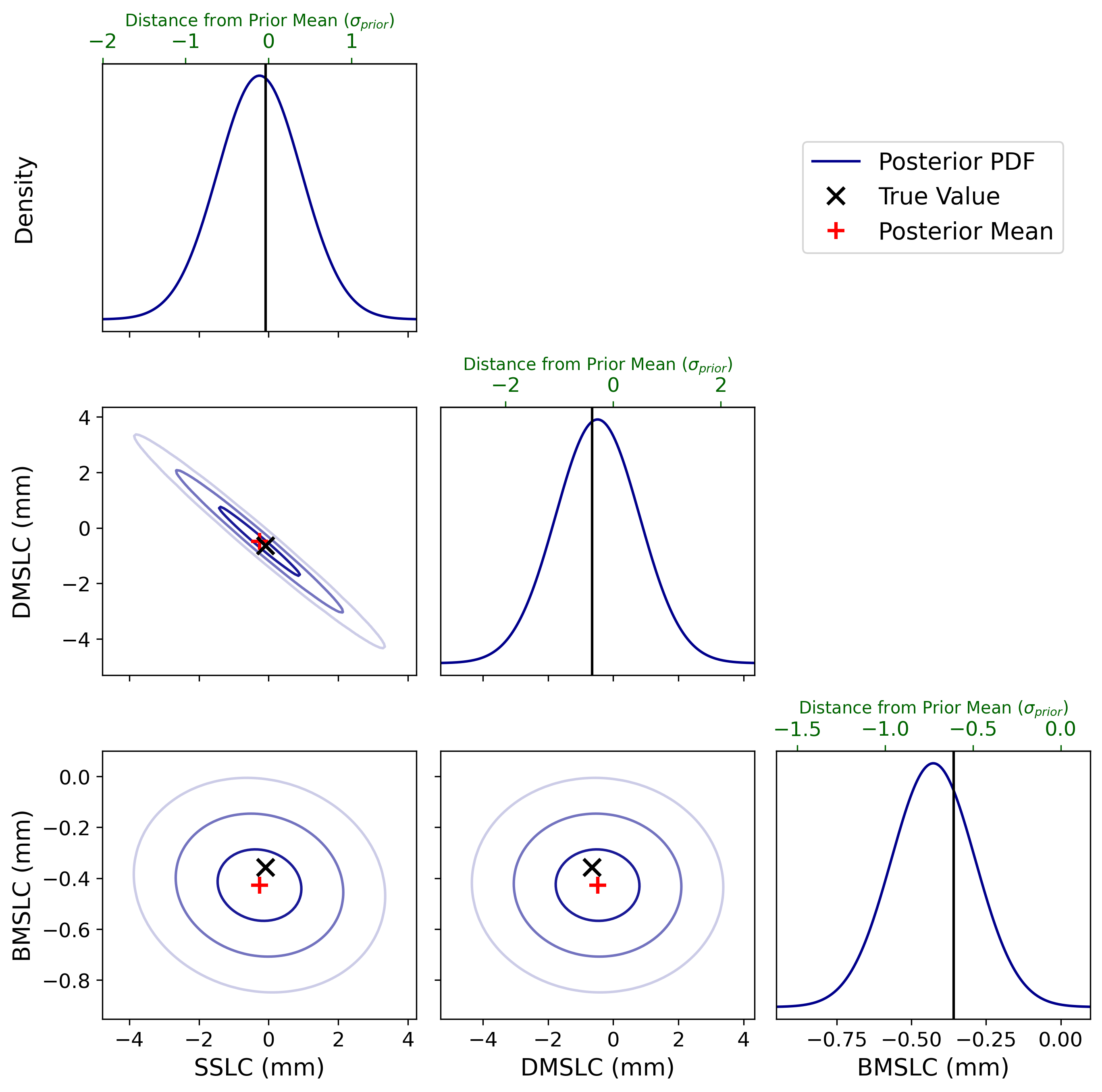}
    \caption{Corner plot for the joint-inversion posterior on the Tasman Sea averages of steric (SSLC), dynamic manometric (DMSLC), and barystatic manometric (BMSLC) 
    sea-level changes, with true values (black crosses), posterior expectations (red crosses), and upper axes in units of the prior
    standard deviation. SSLC and DMSLC are strongly anti-correlated, their sum being pinned by the altimetry
    without the partition being resolved, while BMSLC is essentially decoupled from both, reflecting the direct
    constraint on the ice load supplied by the gravity data.}
    \label{fig:altimetry_regional}
\end{figure*}

\subsubsection{Inversion results}

Fig.~\ref{fig:joint_posterior_maps} compares the true model state with the posterior expectation from the joint inversion, alongside maps of the pointwise posterior standard deviation. The sterodynamic sea-level change is 
recovered in close visual agreement with the truth, its standard deviation reduced from the prior value of 4\,mm to a few tenths of a millimetre over the well-observed ocean, and rising back towards the prior poleward of 
the $\pm66^{\circ}$ altimetry limit. The ice-thickness change is fairly well recovered, with the largest uncertainties over the East Antarctic interior. The vertically averaged density change is the least constrained 
field, its posterior standard deviation lying only modestly below the prior level over much of the ocean. This ordering reflects the fundamental sensitivities of the observing system. Altimetry provides a direct and powerful constraint on geocentric sea-level change, which is dominated by the sterodynamic contribution. Gravity observations, in contrast, sense only mass redistribution. Disentangling the remaining fields, and especially isolating the mass-conserving density variations, relies on subtle physical and spatial differences, leaving the vertically averaged density as the most poorly constrained property.

How the two data types act is clearest in the spatial covariance functions. Fig.~\ref{fig:cov_ocean} shows the covariance of the sterodynamic sea-level change at a reference point in the North Atlantic with all three model fields, 
under the prior and each of the three posteriors. In the case of the prior, the field is uncorrelated with the ice thickness, while its local anti-correlation with the density field is the imposed long-wavelength coupling described in 
Section~\ref{sec:AltimetryNum}, supplemented weakly by the mass-conservation conditioning. The GRACE/FO data alone leave these functions close to their prior forms: satellite gravity says little about the ocean fields at a point. 
The altimetry data, by contrast, collapse the variance by nearly two orders of magnitude while inducing long-range structure that couples the reference point to the ocean fields globally and, more notably, to the ice sheets. This coupling enters through the sea-level physics embedded in the forward operator. Addition of the gravity data within the joint inversion acts to reduce the magnitude of the covariance between the 
oceans and the ice sheets, but it is not removed completely. 
Fig.~\ref{fig:cov_ice} shows the corresponding functions for the ice thickness at a point on the West Antarctic Ice Sheet. Altimetry alone 
constrains this quantity only through its sea-level fingerprint, and the covariance amplitude falls modestly from its prior value while acquiring global structure within the ocean fields. The gravity data act far more directly: 
the amplitude falls, and a slight negative covariance appears with the East Antarctic Ice  Sheet, expressing the freedom to redistribute a well-constrained total mass. The joint covariance function for the ice thickness 
shows a mildly reduced variance at the observation point compared to the GRACE/FO-only case, but there are more pronounced negative correlations to distant regions of the ice sheets. As with Fig.~\ref{fig:cov_ocean}, 
we also see a reduction in the covariance between the ice and ocean components through the use of both data types.

\begin{figure*}
    \centering
    \includegraphics[width=0.8\textwidth]{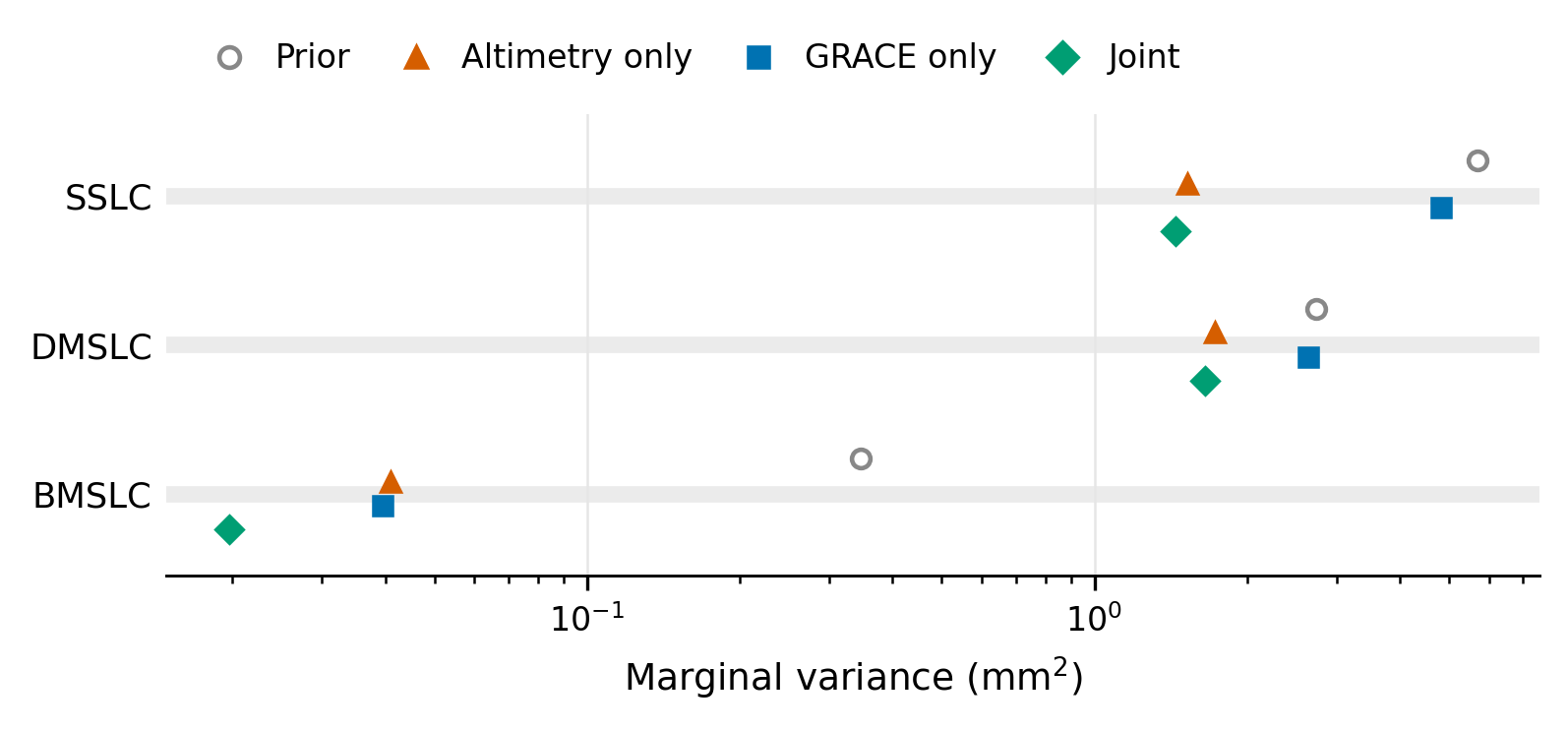}
    \caption{Marginal variances of the steric (SSLC), dynamic manometric (DMSLC), and barystatic manometric
    (BMSLC) sea-level changes of eq.~(\ref{eq:sl_decomp}), averaged over the Tasman Sea, under the prior and the
    altimetry-only, GRACE-only, and joint posteriors (note the logarithmic scale). BMSLC is well constrained by
    either data type individually and best by both together; SSLC is constrained almost entirely by the
    altimetry; DMSLC remains largely prior-limited throughout.}
    \label{fig:regional_variance}
\end{figure*}

At the global scale, Fig.~\ref{fig:joint_gmsl} presents probability density functions for GMSLR. The latitude-weighted average is overconfident: its expectation is displaced from the true value by a few times its nominal standard deviation, 
consistent with Section~\ref{sec:AltimetryBias} where the nominal error of this method was found to understate the true error by a factor of around three. The altimetry-only posterior bounds the truth comfortably, with a variance reduction 
of 99.3\% relative to the prior, and the joint inversion improves this only marginally, to 99.4\%. The GRACE-only posterior instead achieves 69.9\%, its KL divergence of 0.83 nats falling below the order-$n$ benchmark. This is to say that satellite gravity alone 
offers only a modest constraint on the total; the reason emerges upon splitting GMSLR into its parts.

Fig.~\ref{fig:joint_gmslr_split} displays one-standard-deviation contours in the $(b, s)$ plane for the prior, within which the two coordinates are uncorrelated, and for the three posteriors. We can see here that $b$ and $s$ are close to complementary; the 
altimetry-only ellipse is elongated along lines of constant $b + s$, with a posterior correlation of $-0.95$. This is because the data constrain the total tightly while leaving its partition between mass and density largely to the prior. The GRACE-only ellipse 
is instead narrow in $b$ alone, with the gravity data reducing the barystatic variance by 88.4\% but the steric by only 9.2\%. It should be noted that though the GRACE/FO data is completely insensitive to steric sea-level variations, the 
slight reduction seen in the posterior for this component arises due to the correlated nature of the prior distribution which acts to link steric and manometric sea-level changes. 
The joint posterior is, in effect, the intersection of the individual components, with variance reductions of 94.5\% in $b$ and 83.2\% in $s$, and a 
KL divergence of 3.13 nats against 2.30 for altimetry alone. A strong anti-correlation of $-0.93$ nevertheless remains, because the sum is still known far better than the difference: under the joint posterior, $\sigma_{b+s} \approx 0.06$\,mm while
$\sigma_{b-s} \approx 0.29$\,mm. 

The same complementarity holds at regional scale, where the decomposition of eq.~(\ref{eq:sl_decomp}) applies directly. In contrast to joint inversions that project the signals onto a limited set of predefined basis patterns 
\citep[e.g.][]{rietbroek2016revisiting, willen2026improving}, the present approach lets the data determine the spatial structure while carrying full uncertainties. Fig.~\ref{fig:regional_variance} summarises the marginal variances 
of SSLC, DMSLC and BMSLC averaged over the Tasman Sea under the prior and the three posteriors, and Fig.~\ref{fig:altimetry_regional} shows the corner plot for the joint inversion. The barystatic component is well constrained by either 
data type individually (variance reductions of 88.1 and 88.5\% for altimetry and gravity, respectively) and most comprehensively by both together (94.3\%). This is due to the far-field fingerprint of the ice load being smooth and visible to each sensor. The steric 
component is essentially an altimetry quantity, its reduction of 73.2\% under altimetry alone barely altered by the addition of gravity, but the dynamic manometric component resists both, with a joint reduction of only 39.4\%. 
The corner plot reveals why the joint posterior is nevertheless highly informative. Because altimetry tightly pins the sum of SSLC and DMSLC (which represents the regional sea-level change less its barystatic part), it induces a strong anti-correlation between the two without resolving their partition. Meanwhile, the barystatic manometric component remains essentially decoupled from both.
Accordingly, the generalised variance (the determinant of the $3\times3$ covariance) falls by a factor of almost $10^{4}$ under the joint inversion even though the total variance (the trace) 
falls by only 64\%: the data determine particular combinations of the components far more tightly than any component individually. The joint KL divergence of 3.76 nats, above the order-$n$ benchmark for $n = 3$, summarises the overall information gain.

\section{Conclusions}

The accurate inference of modern-day sea-level change, and the attribution of its driving mechanisms, require methods that respect the underlying physics. Building on earlier studies 
\citep[e.g.][]{riva2010sea, sterenborg2013bias, al2024reciprocity, coulson2026inverting}, we have shown that the standard estimators -- the WMB method for satellite gravity and spatial 
averaging for ocean altimetry -- share two structural limitations: they do not incorporate gravitationally self-consistent sea-level physics, and their stated uncertainties derive solely 
from the propagation of observational noise. Within our idealised but broadly representative experiments, these limitations produce systematic biases and understated uncertainties; for 
the spatial averaging of altimetry data, for example, the nominal error understates the true error by a factor of around three. While the precise numbers depend on the statistical distributions 
assumed, the underlying mechanisms do not. To address these problems, we have introduced a scalable Bayesian framework for global sea-level inversion, formulated in an infinite-dimensional setting
and implemented efficiently within our open-source Python libraries. 

The framework rests on three linked elements. First, the incorporation of adjoint sea-level theory allows the full self-gravitating elastic physics to enter both the forward and inverse calculations. Second, the 
infinite-dimensional formulation ensures that uncertainties are not artificially shaped by ad hoc spatial discretisation or truncation of the model space. Third, the computational costs 
that have historically made such inversions prohibitive are overcome through the matrix-free architecture of the open-source \texttt{pygeoinf} and \texttt{pyslfp} libraries, together with the 
WMB and surrogate Woodbury preconditioners. The basic calculations -- posterior expectations, covariance actions, and individual posterior samples -- take a few minutes on a laptop, 
while the more intensive sampling- and factorisation-based tasks decompose into independent calculations whose run times fall broadly in proportion to the number of cores available.

A series of synthetic experiments demonstrates the framework at full observational resolution and illustrates the analyses it enables. For satellite gravity, the Bayesian estimator effectively suppresses 
spectral leakage and recovers regional load averages with estimator-kernel errors roughly half those of the WMB method. Crucially, this smaller kernel difference implies a more faithful recovery for essentially any source field, not merely those drawn from the assumed prior. The framework also yields estimates, with quantified uncertainties, of the otherwise unobservable degree-one loads, 
though these are more weakly constrained than the other properties considered and are correspondingly more sensitive to the prior. For ocean altimetry, the method accounts for the restricted spatial 
sampling and for the distinction between relative and geocentric sea-level, correcting the overconfidence 
of standard averaging. The joint inversion of the two data types displays their complementarity: altimetry constrains the total sea-level change while gravimetry constrains its mass-bearing part. 
Together they reduce, though do not eliminate, the degeneracy between the steric and barystatic contributions, yielding a decomposition of regional sea-level change into steric, dynamic manometric, 
and barystatic manometric components with full uncertainties. Throughout, the same machinery supplies diagnostics of prior sensitivity (worst-case perturbation bounds, estimator-kernel comparisons, 
and direct recomputation under alternative priors) so that the robustness of any particular conclusion can be assessed rather than assumed. Because the prior and noise distributions adopted are idealised, 
the quantitative results are indicative rather than definitive; it is the qualitative structure -- which quantities the data determine, which remain prior-limited, and why -- that we expect to carry over 
to practical applications.

While the present study establishes the theoretical and computational foundations using idealised distributions, the value of the framework lies in its application to real data. Future work will move 
to the assimilation of  observational records, incorporating physically informed priors and fully correlated noise models. For the former, the evidence-maximisation 
approach of Section~\ref{sec:empirical_bayes} offers a concrete route, with priors constructed from the outputs of numerical ocean and climate models by regarding them as samples from parameterised random 
fields. The scalability of the methodology also provides a clear pathway for integrating further data sets, including tide gauge records and ice altimetry, and for extending the framework to time-dependent 
sequential assimilation and, eventually, to laterally varying viscoelastic earth models. Within such sequential applications, a low-rank factorisation of the posterior must be formed at each assimilation step, potentially 
at high resolution and correspondingly high rank. As discussed in Section~\ref{sec:NumCon}, each factorisation comprises many independent solves, with no communication between tasks and modest memory requirements 
for each, and hence distributes naturally across the hundreds of cores available within a cluster or cloud environment. A factorisation requiring several hours in serial would then complete within minutes, and 
it is this scalability, rather than the cost of any individual solve, that will make the sequential extension computationally viable.

\begin{acknowledgments}

DA-A has been supported through the Natural Environment Research Council grant NE/X013804/1. AL acknowledges support from the Vetlesen Foundation and NASA 23-MAP23-0015.

\end{acknowledgments}

\section*{Data availability statement}

There are no data associated with this paper. The \texttt{pygeoinf} and \texttt{pyslfp} libraries can be 
found at \url{https://github.com/da380/pygeoinf} and \url{https://github.com/da380/pyslfp}, while scripts used to generate all 
results within the paper are contained in \url{https://github.com/danielheathcote/heathcote2026-bayesian-sl}.

\bibliographystyle{gji}
\bibliography{refs}

\label{lastpage}
\end{document}